\documentclass[12pt,a4paper]{article}
\usepackage{xcolor}
\definecolor{darkblue}{rgb}{0,0,.6}
\usepackage[pdftex,colorlinks=true,linkcolor=darkblue, urlcolor=darkblue, citecolor=darkblue]{hyperref}
\usepackage{amsmath,caption,orcidlink}
\usepackage[top=1.5in, bottom=1.5in, left=1in, right=1in]{geometry}

\DeclareCaptionStyle{italic}[justification=centering]{labelfont={bf}, textfont={it}, labelsep=colon}
\usepackage{graphicx,psfrag,epsf}
\usepackage{enumerate,multirow}
\usepackage[round]{natbib}
\usepackage{url,doi}
\usepackage{booktabs,subfig,bm,paralist,mathpazo,tikz,longtable,microtype,dsfont,rotating}

\newcommand{\blind}{0}

\newcommand{\E}{\text{E}}

\graphicspath{{plots/}}

\newsavebox\CBox

\definecolor{a0}{rgb}{0.0, 0.5, 0.0}
\definecolor{bistre}{rgb}{0.24, 0.17, 0.12}
\definecolor{amethyst}{rgb}{0.6, 0.4, 0.8}
\definecolor{blue-violet}{rgb}{0.54, 0.17, 0.89}
\definecolor{Rcolor}{RGB}{150,160,190}
\definecolor{blush}{rgb}{0.87, 0.36, 0.51}
\definecolor{brightturquoise}{rgb}{0.03, 0.91, 0.87}
\definecolor{burntorange}{rgb}{0.8, 0.33, 0.0}
\date{}
\AtBeginDocument{}
\newcommand{\X}{\mathcal{X}}
\newcommand{\Y}{\mathcal{Y}}

\newcommand{\Rlogo}{\protect\includegraphics[height=1.8ex,keepaspectratio]{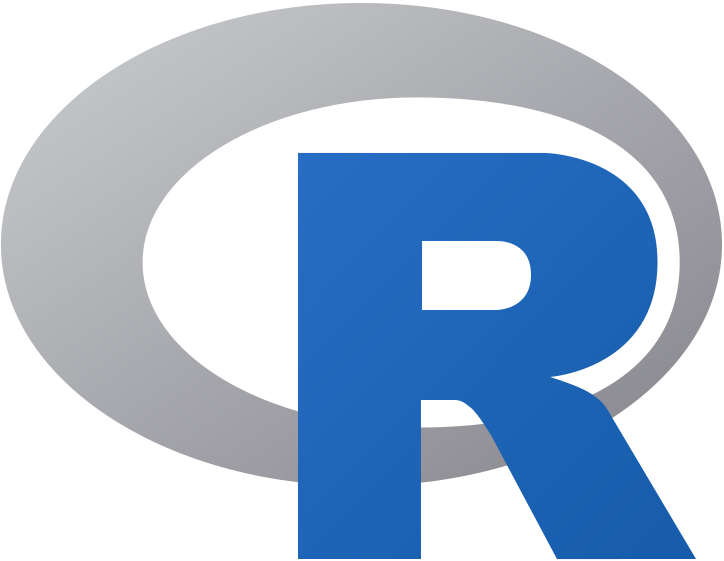}}

\begin{document}

\def\spacingset#1{\renewcommand{\baselinestretch}{#1}\small\normalsize} \spacingset{1}

\if0\blind
{
\title{\bf Age-specific demographic modeling and forecasting: Rolling window, expanding window, or both?}
}
\author
{
\normalsize Sizhe Chen \orcidlink{0009-0002-5062-2951} \qquad Han Lin Shang\footnote{Corresponding address: Department of Actuarial Studies and Business Analytics, Level 7, 4 Eastern Road, Macquarie University, Sydney, NSW 2109, Australia; Telephone: +61(2) 9850 4689; Email: hanlin.shang@mq.edu.au} \orcidlink{0000-0003-1769-6430} \\
\normalsize Department of Actuarial Studies and Business Analytics \\
\normalsize Macquarie University
}
\maketitle
\fi

\if1\blind
{
\title{\bf Age-specific demographic modeling and forecasting: Rolling window, expanding window, or both?}
} \fi

\maketitle

\begin{abstract}
Rolling and expanding windows are widely used in age-specific demographic modeling and forecasting. Building on these approaches, we propose a simple combination method that assigns equal weight to the forecasts from both schemes. Our focus is on evaluating and comparing the forecast accuracy of the two window types in modeling age-specific mortality and fertility. Based on the multi-country comparison, the superior performance of one method often persists across different forecast horizons. In the absence of prior information, our combined approach offers a robust and practical alternative.

\vspace{.1in}
\end{abstract}
	
\newpage	
\spacingset{1.55} 

\section{Introduction}\label{sec:1}

Modeling and forecasting mortality and fertility are essential for understanding population structure and its dynamics, as well as for informing data-driven policy decisions. Accurate mortality forecasts help governments and insurance companies plan for aging populations, hospital facilities, pension liabilities, and sustainable retirement. Fertility forecasts are crucial for anticipating future workforce needs, school infrastructure requirements, and family-related services. Together with migration data, these demographic components shape the size and profile of populations over time, influencing economic growth, social planning, and resource allocation.

Statistical methods for modeling and forecasting mortality and fertility have evolved significantly to address the complex patterns and uncertainties inherent in various models used to generate forecasts. For modeling mortality, popular approaches include \citeauthor{LC92}'s \citeyearpar{LC92} model and its extensions, which decompose age-specific mortality rates into age-, period-, and cohort-components to capture long-term trends and temporal dynamics \citep[see, e.g.,][]{RH06, HR09}. Some extensions include functional data-analytic methods and Bayesian modeling \citep[for comprehensive reviews, refer to][]{BHT+06, Booth06, BT08, BCB23}. For modeling fertility, models should account for fluctuation and postponement behaviors using the Lee-Carter model and its extension, such as the Bayesian model, as in \cite{Lee93} and \cite{SZG+14}.

Despite the increasing number of statistical methods proposed, it is often necessary to compare forecast accuracy against benchmarks. In these accuracy comparisons, researchers often face the choice between forecast schemes: a rolling window and an expanding window. In a rolling-window scheme, the model is repeatedly re-estimated using a fixed-size window that moves forward through the data, discarding older observations as new ones become available. This forecasting scheme focuses on recent information, which is particularly useful when the underlying process undergoes a smooth or sudden structural change. In contrast, an expanding-window scheme incorporates all available past data up to the most recent data, gradually increasing the training sample size. This forecasting scheme uses the entire history of the series, making it more suitable when the underlying process remains stable over time.

A fundamental decision in the two forecasting schemes is the fitting period used for parameter estimation. Despite its practical importance, only a few studies have investigated the effects of the fitting period on forecast accuracy; and we aim to address this issue in the paper. The typical approach has been to randomly choose a fitting period, which often leads to a misrepresentation of future mortality trends \citep{JK07}. If a mean change exists, only the most recent data stretch should be used. In the Lee-Carter framework, \cite{BMS02} considers a change point detection method to determine the optimal fitting period. The idea of a change point was also considered in \cite{SX22} for analyzing excess deaths during the COVID-19 era. An implicit issue with this method is that any misestimation of the change-point location can alter forecast accuracy. Additionally, truncating the data after the change point results in a reduction in the number of samples. When the relationship between the out-of-sample forecast error and the length of the fitting period is non-monotonic, selecting the optimal fitting period becomes difficult \citep{MUS20}. For a data set with long series, data from the distant past is not very useful for forecasting. This idea motivates the work of \citet{HS09} and \cite{SH24} to incorporate a set of geometrically decaying weights into the estimation of age-specific mortality. The performance of a finite sample is crucially dependent on the determination of the weight parameter.

Instead of optimally selecting a tuning parameter, we propose a simple approach that averages forecasts across different fitting periods. Using the same time-series forecasting method, we evaluate the effectiveness of this combination approach and compare its performance with rolling and expanding windows. Using the 23 countries from \cite{HMD23} in Section~\ref{sec:2.1}, the rolling-window forecast scheme tends to be advantageous with the smallest forecast errors in almost all horizons. Using the 16 countries from \cite{HFD23} in Section~\ref{sec:2.2}, the expanding-window forecast scheme yields the smallest forecast errors. Using the same functional time-series forecasting method as in Section~\ref{sec:3}, this discrepancy in accuracy highlights the importance of forecast schemes and our combined approach, which examines point and interval forecasts from both schemes in Sections~\ref{sec:4} and~\ref{sec:5}, respectively. The conclusion is presented in Section~\ref{sec:6}, along with some ideas on how the methodology can be further extended.

\section{Age-specific mortality and fertility rates}\label{sec:2}

\subsection{Australian and Canadian age-specific mortality rates}\label{sec:2.1}

We demonstrate our forecasting approach using Australia's age and sex-specific mortality data from 1921 to 2021, sourced from \cite{HMD23}. The original data set captures the mortality rates for both sexes. For analytical refinement, age groups have been aggregated to span from 0 to 94 in single years, with the last age group being 95+. This aggregation helps generate a comprehensive overview of mortality trends. By scrutinizing changes in the mortality rate as a function of age and year, it becomes evident that mortality rates have varied significantly over the centuries. 

\begin{figure}[!htb]
\centering
\subfloat[\small Australian Female Raw data]
{\includegraphics[width=8.7cm]{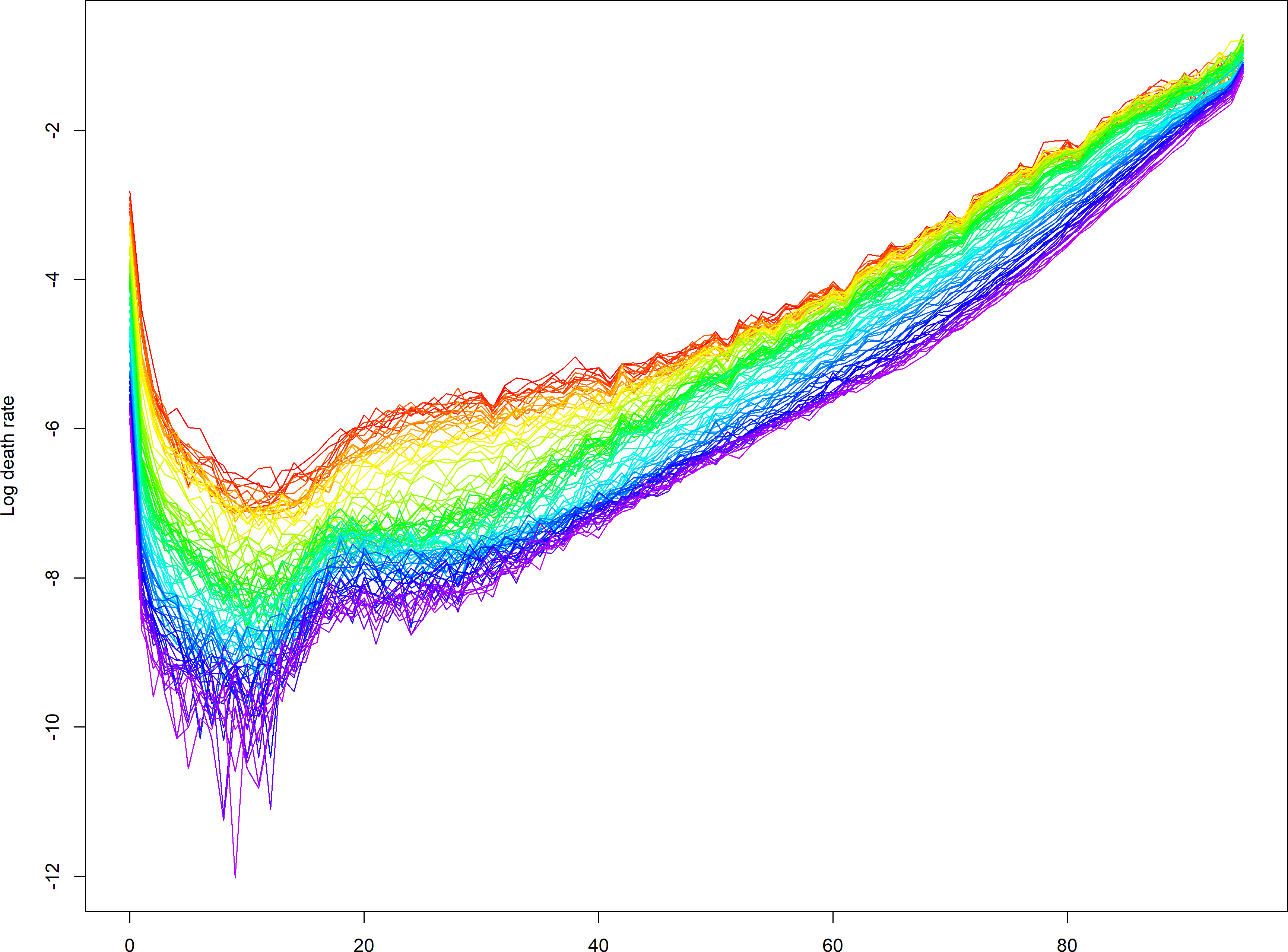}}
\quad
\subfloat[\small Australian Male Raw data]
{\includegraphics[width=8.4cm]{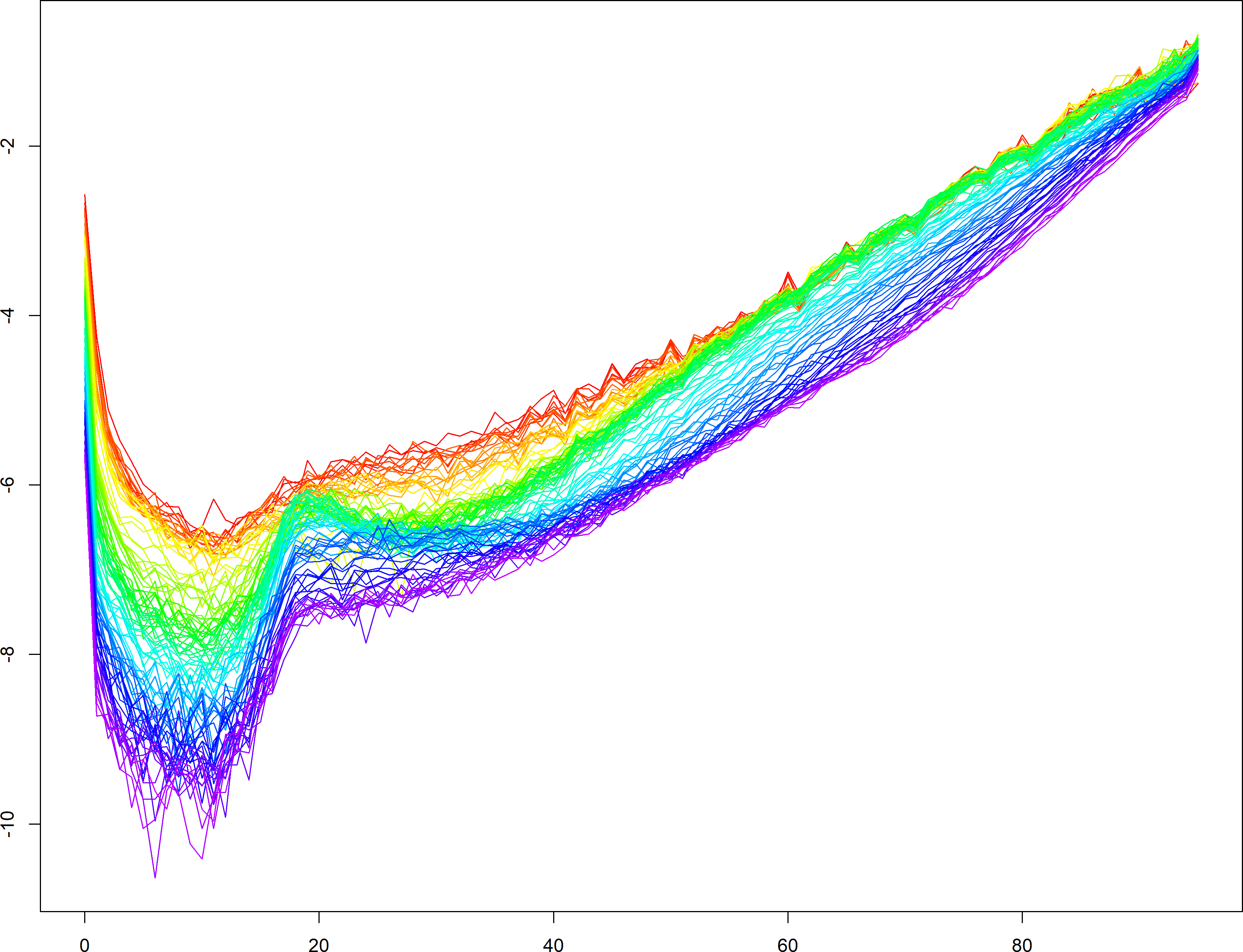}}
\\
\subfloat[\small Australian Female Smoothed data]
{\includegraphics[width=8.7cm]{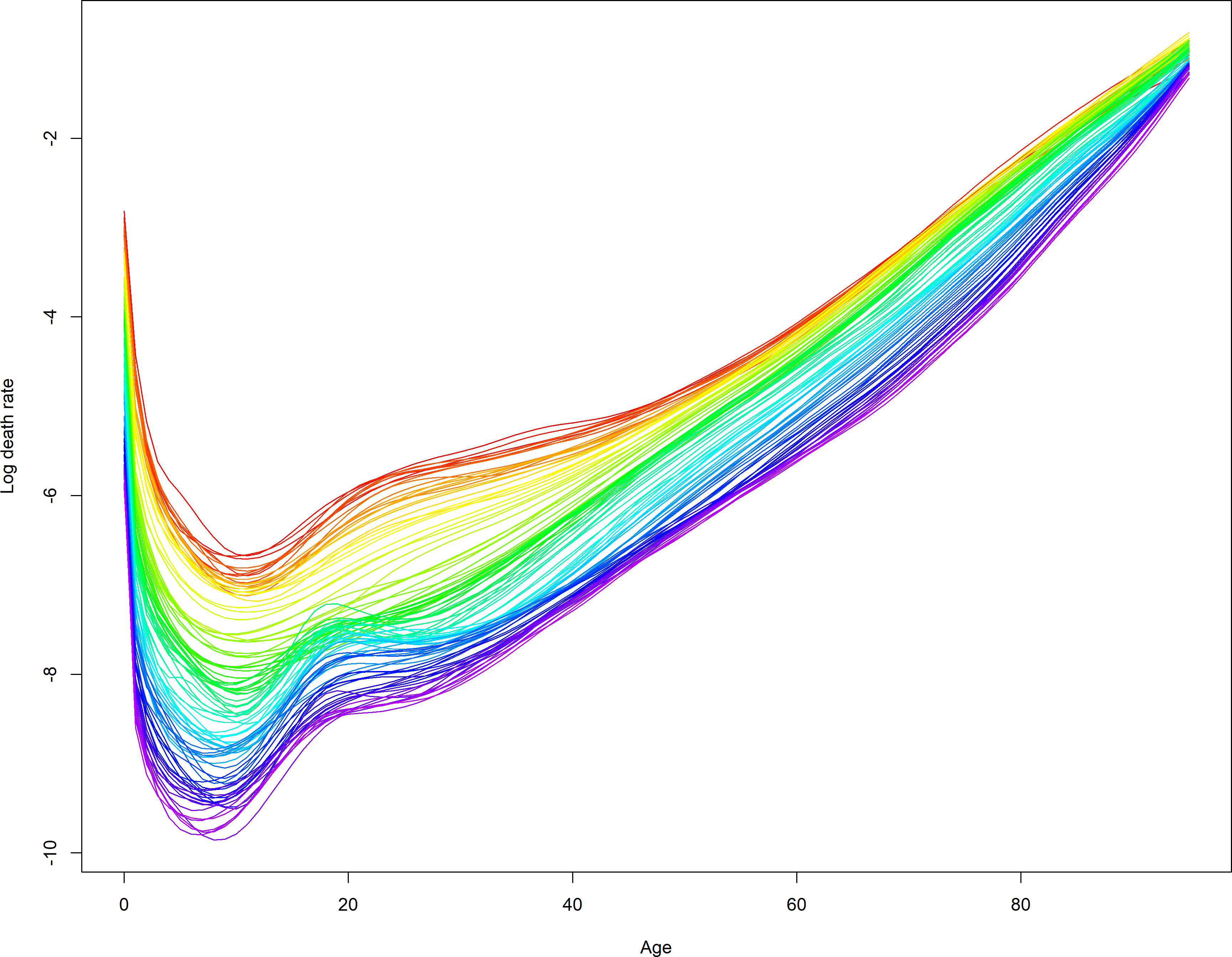}}
\quad
\subfloat[\small Australian Male Smoothed data]
{\includegraphics[width=8.4cm]{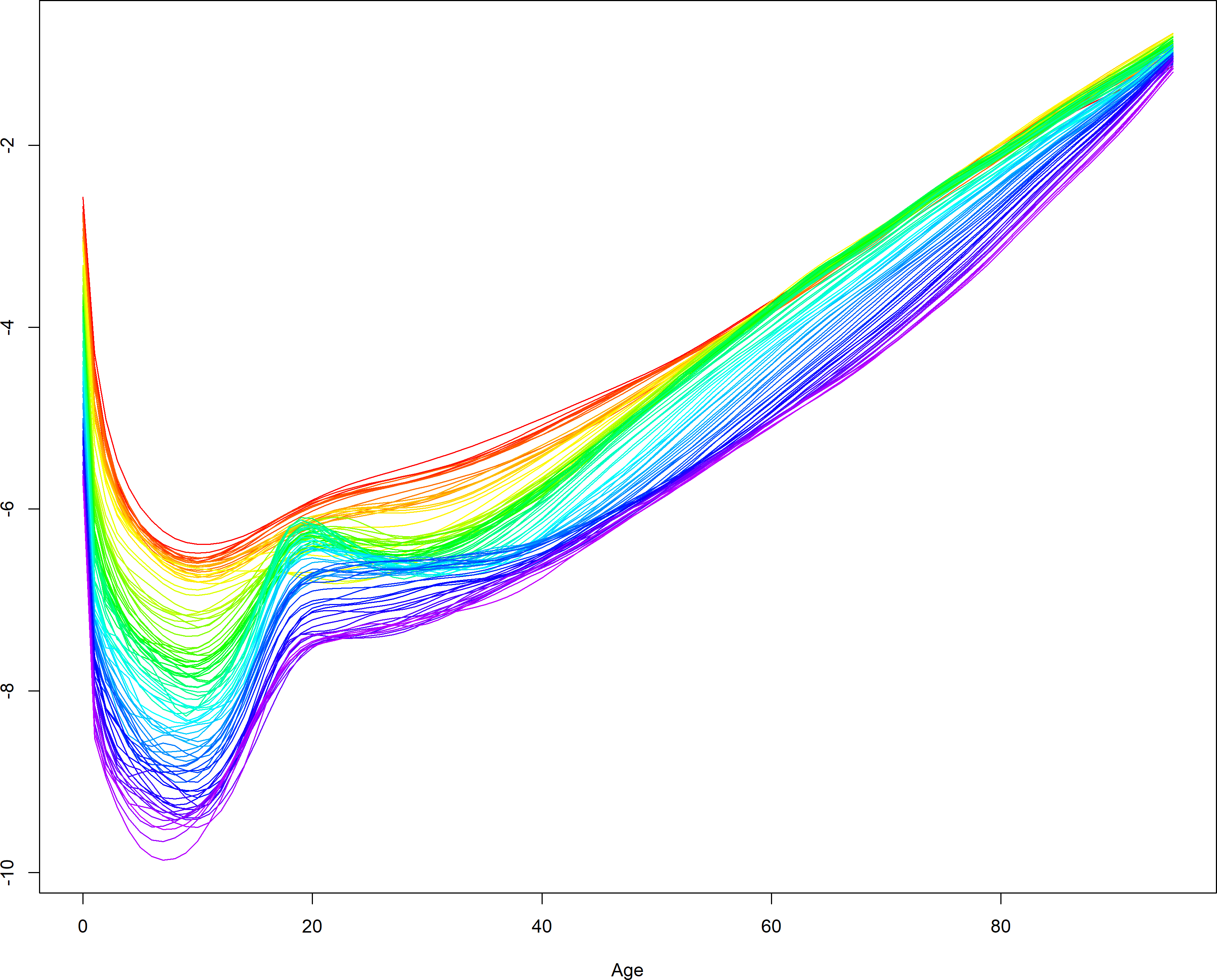}}
\caption{\small{Rainbow Plots: Australian age-specific female and male mortality rates from 1921 to 2021 with and without smoothing by single years of age from 0 to 94 and the last age group of 95+. The data from the distant past are shown in red, while the data in the most recent years are shown in purple.}}\label{fig:1.9}
\end{figure}

To illustrate this historical trajectory, we present the mortality rates using rainbow plots in Figure~\ref{fig:1.9}. Logarithmic mortality rates from the distant past are colored red, while data from the most recent years are colored purple. For modeling, we employ the weighted penalized regression splines method with a monotonicity constraint of \cite{wood1994monotonic} to smooth our mortality data \citep[see, e.g.,][]{HU07}. Smoothing strategies can improve forecast accuracy when combined with the averaging fitting method.

\begin{figure}[!htb]
\centering
\subfloat[\small Canadian Female Raw data]
{\includegraphics[width=8.7cm]{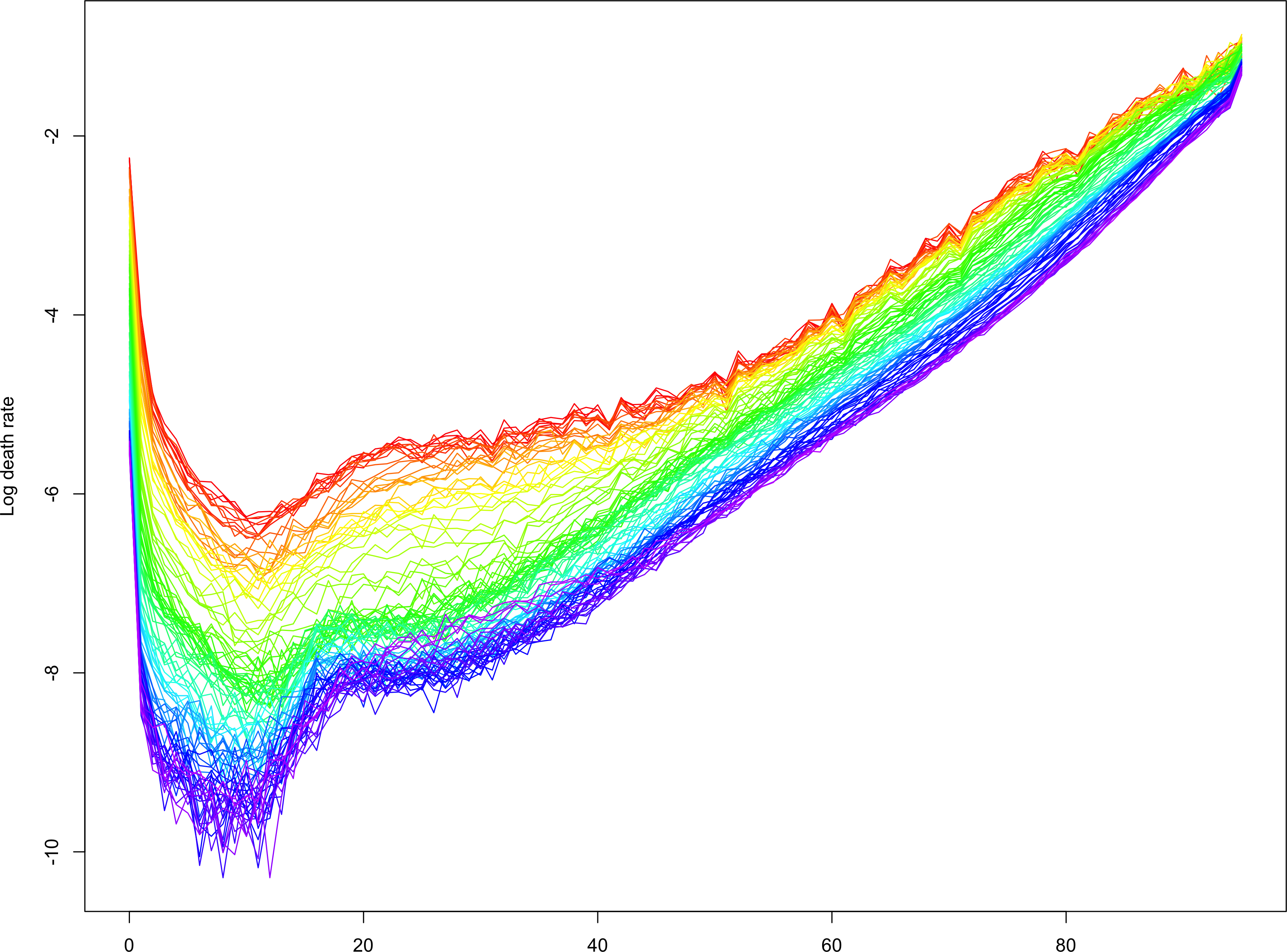}}
\quad
\subfloat[\small Canadian Male Raw data]
{\includegraphics[width=8.4cm]{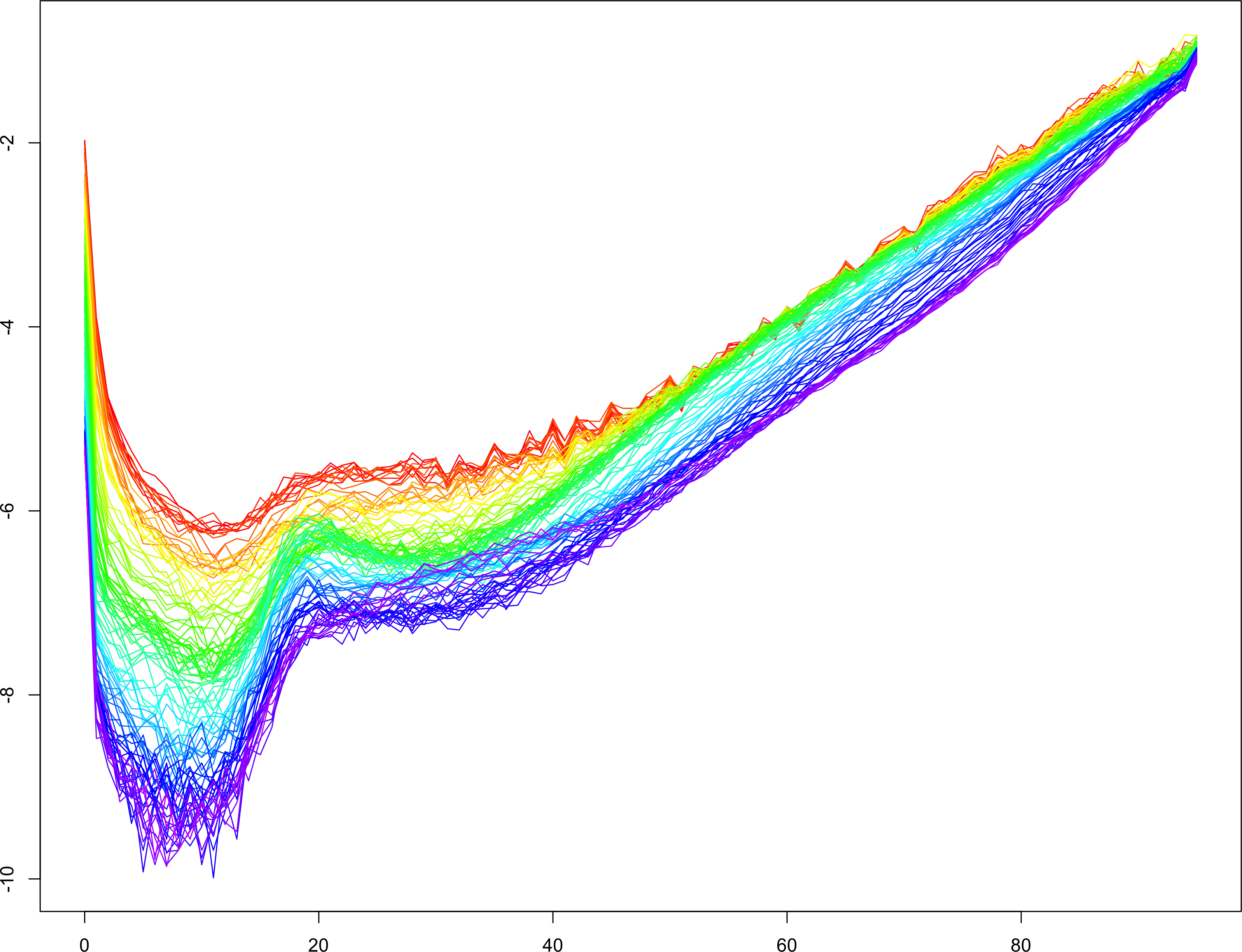}}
\\
\subfloat[\small Canadian Female Smoothed data]
{\includegraphics[width=8.7cm]{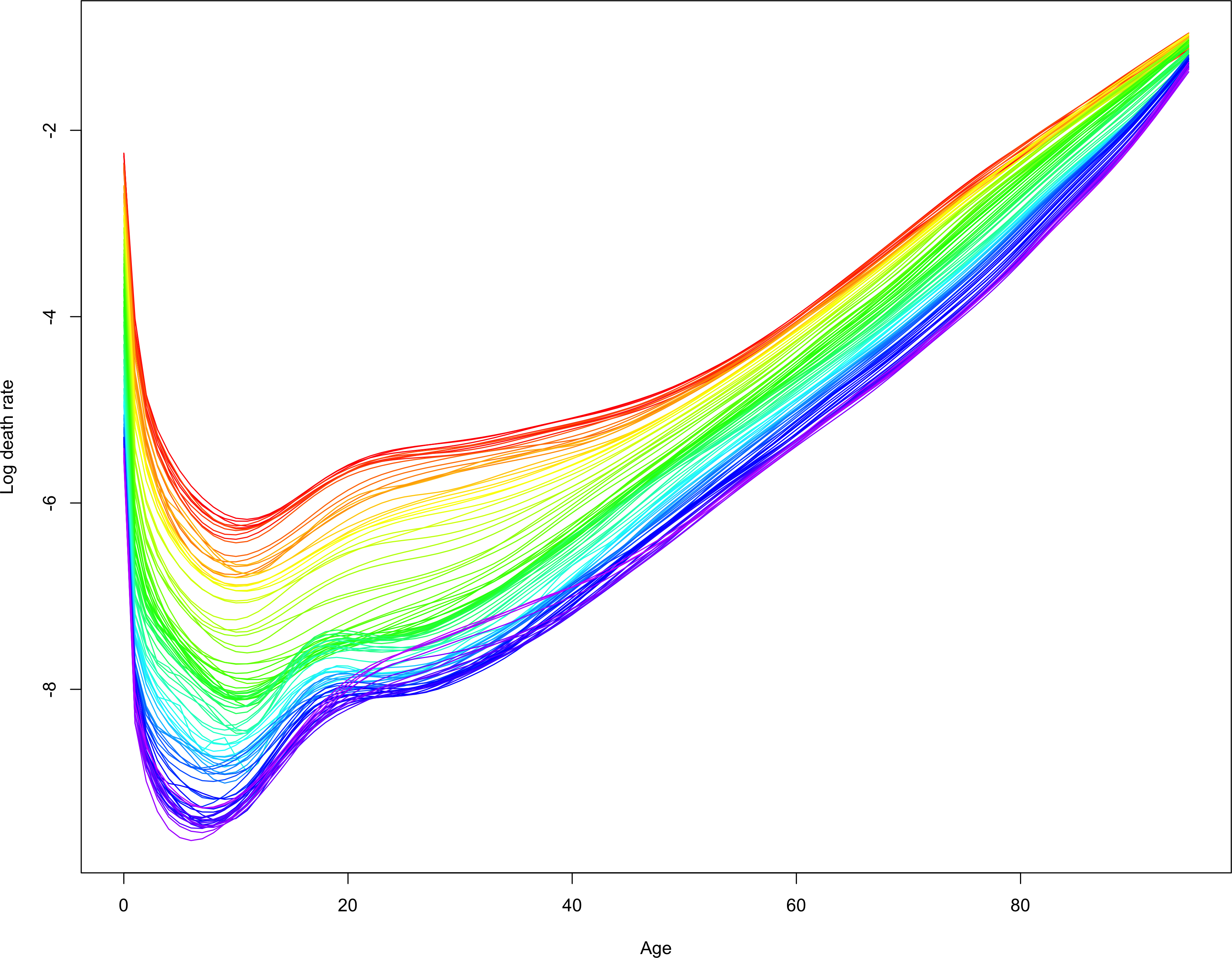}}
\quad
\subfloat[\small Canadian Male Smoothed data]
{\includegraphics[width=8.4cm]{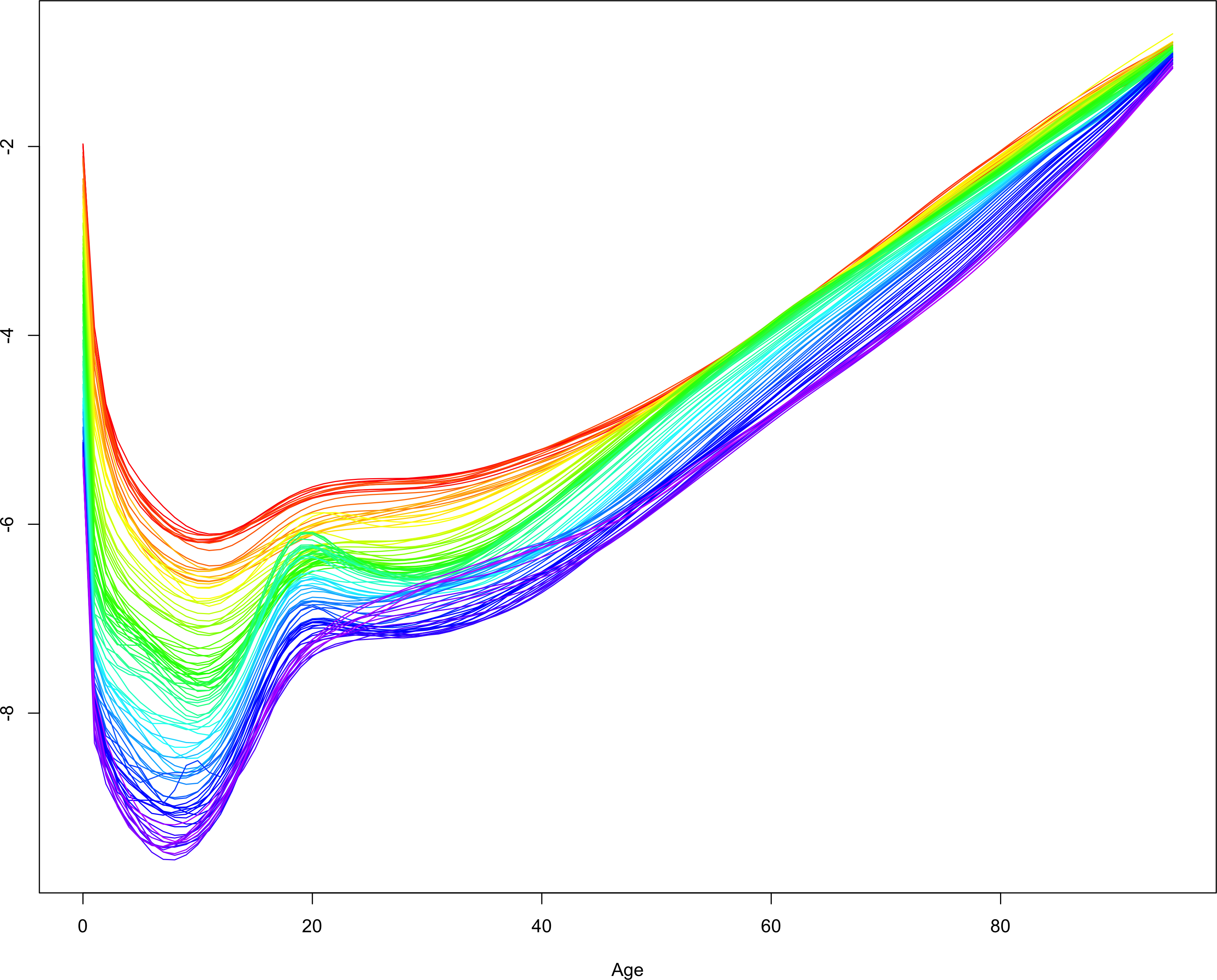}}
\caption{\small{Rainbow Plots: Canadian female and male mortality rates from 1921 to 2023 with and without smoothing by single years of age from 0 to 94 and the last age group of 95+.}}\label{fig:15}
\end{figure}

From \cite{HMD23}, we obtain a multi-country data set used in this study for comparison. It comprises age- and sex-specific mortality rates for the 23 countries listed in Table~\ref{tab:countries}, exclusively using data from years up to and including 1950. This focuses on the analysis of mortality patterns from the first half of the 20\textsuperscript{th} century and earlier, a period characterized by significant demographic and epidemiological transitions. To enable fair comparisons, we retain only countries with continuous national series through 1950 and consistent territorial coverage and documentation; countries that do not meet these criteria are excluded. For example, Belgium is excluded because the series in the \cite{HMD23} has a documented gap for 1914–1918 (no mortality statistics for World War I years), which breaks the continuity needed for pre-1950 cross-country comparisons. 
\begin{table}[!htb]
\centering
\tabcolsep 0.13in
\caption{Countries and their respective sample periods for mortality.}\label{tab:countries}
\begin{tabular}{@{}lll@{\hspace{2em}}lll@{}}
\toprule
\textbf{Country} & \textbf{Code} & \textbf{Period} & \textbf{Country} & \textbf{Code} & \textbf{Period} \\
\midrule
Australia  				& AUS     		& 1921--2021 & Austria                 	& AUT     		& 1947--2023 \\
Bulgaria   				& BGR     		& 1947--2021 & Canada                  	& CAN     		& 1921--2023 \\
Czech      				& CZE     		& 1950--2021 & Denmark                 	& DNK     		& 1835--2024 \\
Finland    				& FIN     		& 1878--2024 & France                  	& FRATNP  	& 1816--2023 \\
Hungary    			& HUN     		& 1950--2020 & Iceland                 	& ISL     		& 1838--2023 \\
Ireland    				& IRL     		& 1950--2022 & Italy                   	& ITA     		& 1872--2022 \\
Japan      				& JPN     		& 1947--2023 & Netherlands             & NLD     		& 1850--2022 \\
New Zealand			& NZL\_NP 	& 1948--2021 & Norway                  	& NOR     		& 1846--2023 \\
Portugal   				& PRT     		& 1940--2023 & Slovakia                	& SVK     		& 1950--2019 \\
Spain      				& ESP     		& 1908--2023 & Sweden                  	& SWE     		& 1751--2024 \\
Switzerland			& CHE     		& 1876--2023 & United Kingdom	& GBR\_NP 	& 1922--2022 \\
United States of America 	& USA 		& 1933--2023 & 				& 			& \\
\bottomrule
\end{tabular}
\end{table}

For mortality, we demonstrate using Australia and Canada as illustrative countries in Figures~\ref{fig:1.9} and~\ref{fig:15}. All three of our panel countries are OECD members operating in comparable developed-economy contexts, which helps maintain consistent background conditions as we examine methodological contrasts. Australia and Canada combine long and uninterrupted series in the \cite{HMD23} with large populations and stable registration systems, yielding smooth horizon-wise error profiles in both point and interval evaluations \citep{HMD23}. 

At the same time, their mortality histories differ in ways that are directly relevant for method comparison, particularly for males: the timing and intensity of the cohort smoking epidemic differ in high-income countries and have been shown to account for large cross-national gaps in adult mortality \citep{preston2010contribution}, and secular declines in external causes (e.g., transport injuries) have progressed at different paces between OECD members \citep{OECDHealth2023}. This ``similar macro-context but different epidemiological timing'' pairing yields clear, legible trajectories and contrasting method rankings across horizons and sexes in our figures, demonstrating that the relative performance of rolling, expanding, and their equal-weight combination reflects genuine differences in how recent information versus full history should be weighted, rather than data artifacts.

\subsection{Canadian and Japanese age-specific fertility rates}\label{sec:2.2}

\begin{figure}[!htb]
\centering
\subfloat[\small Canadian Raw Data]
{\includegraphics[width=8.07cm]{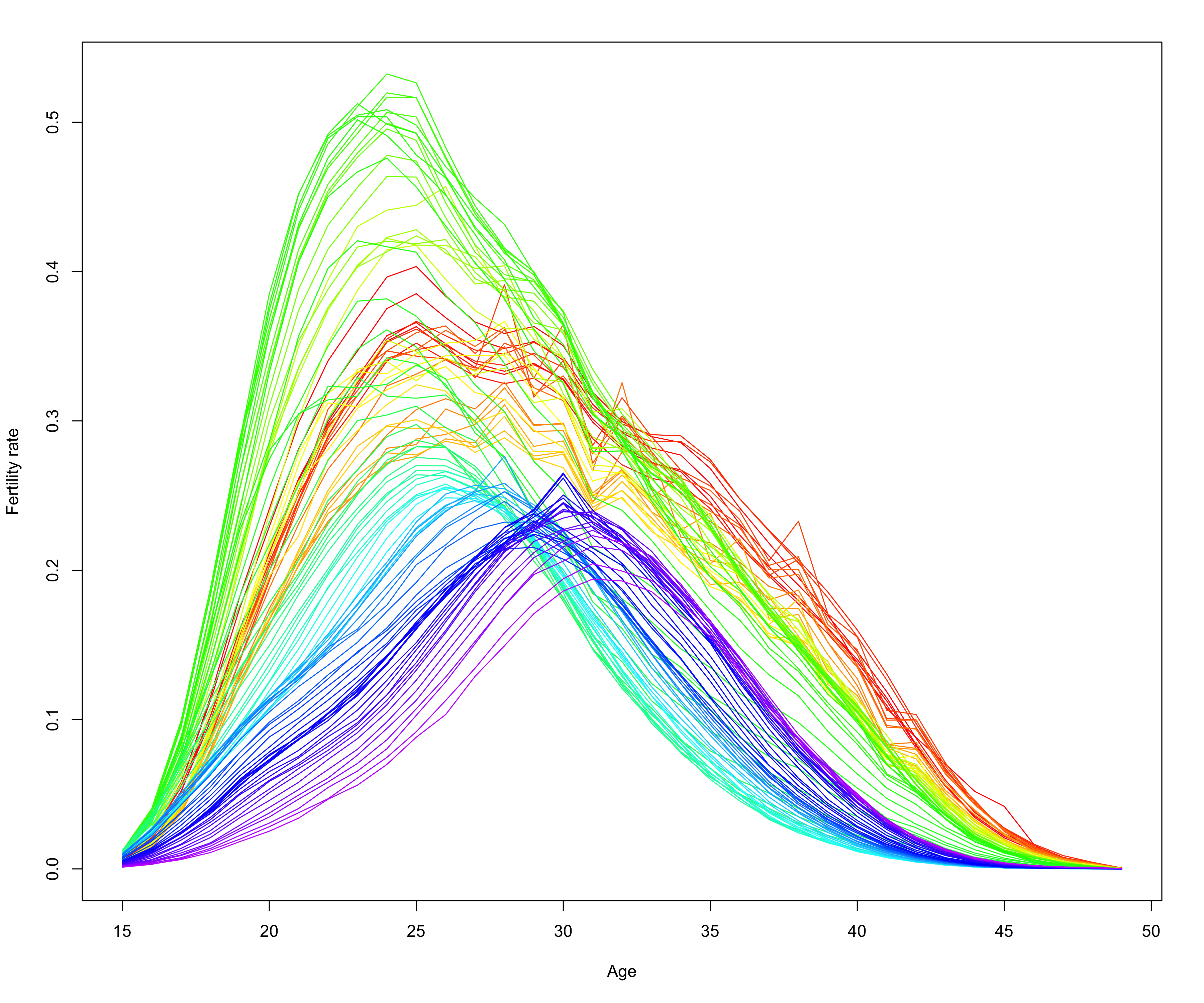}}
\qquad
\subfloat[\small Canadian Smoothed Data]
{\includegraphics[width=8.07cm]{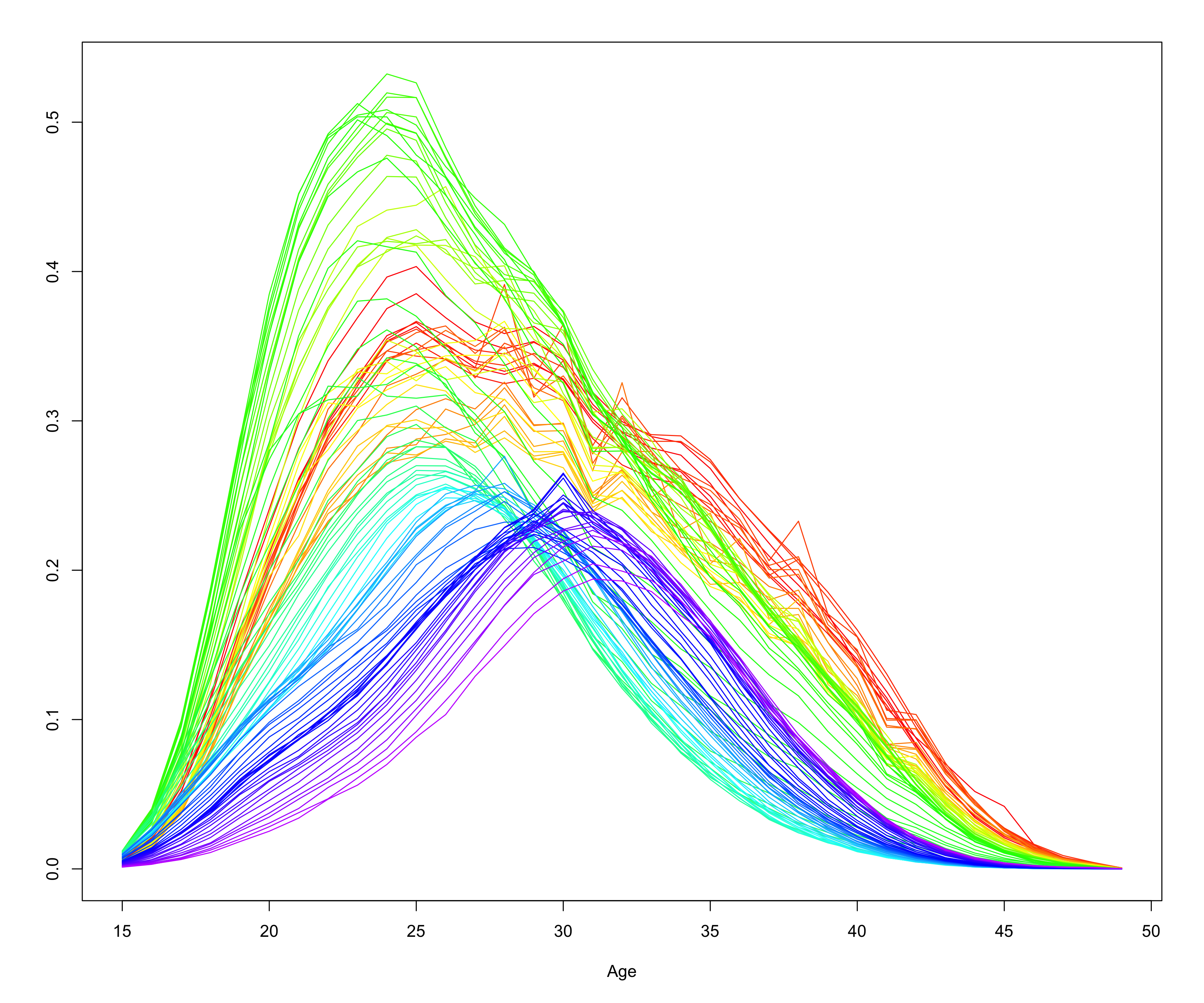}}
\\
\subfloat[\small Japanese Raw Data]
{\includegraphics[width=8.07cm]{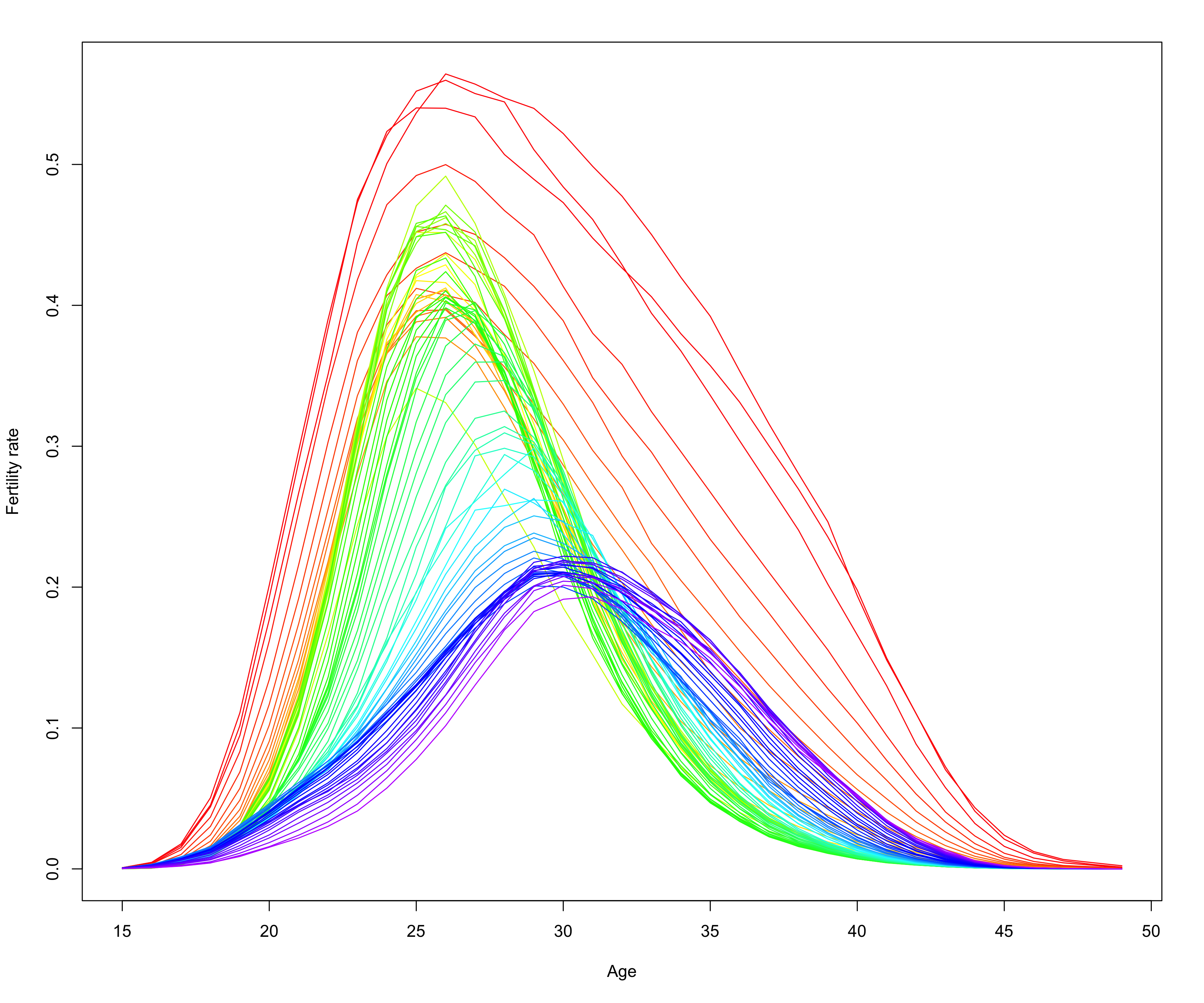}}
\qquad
\subfloat[\small Japanese Smoothed Data]
{\includegraphics[width=8.07cm]{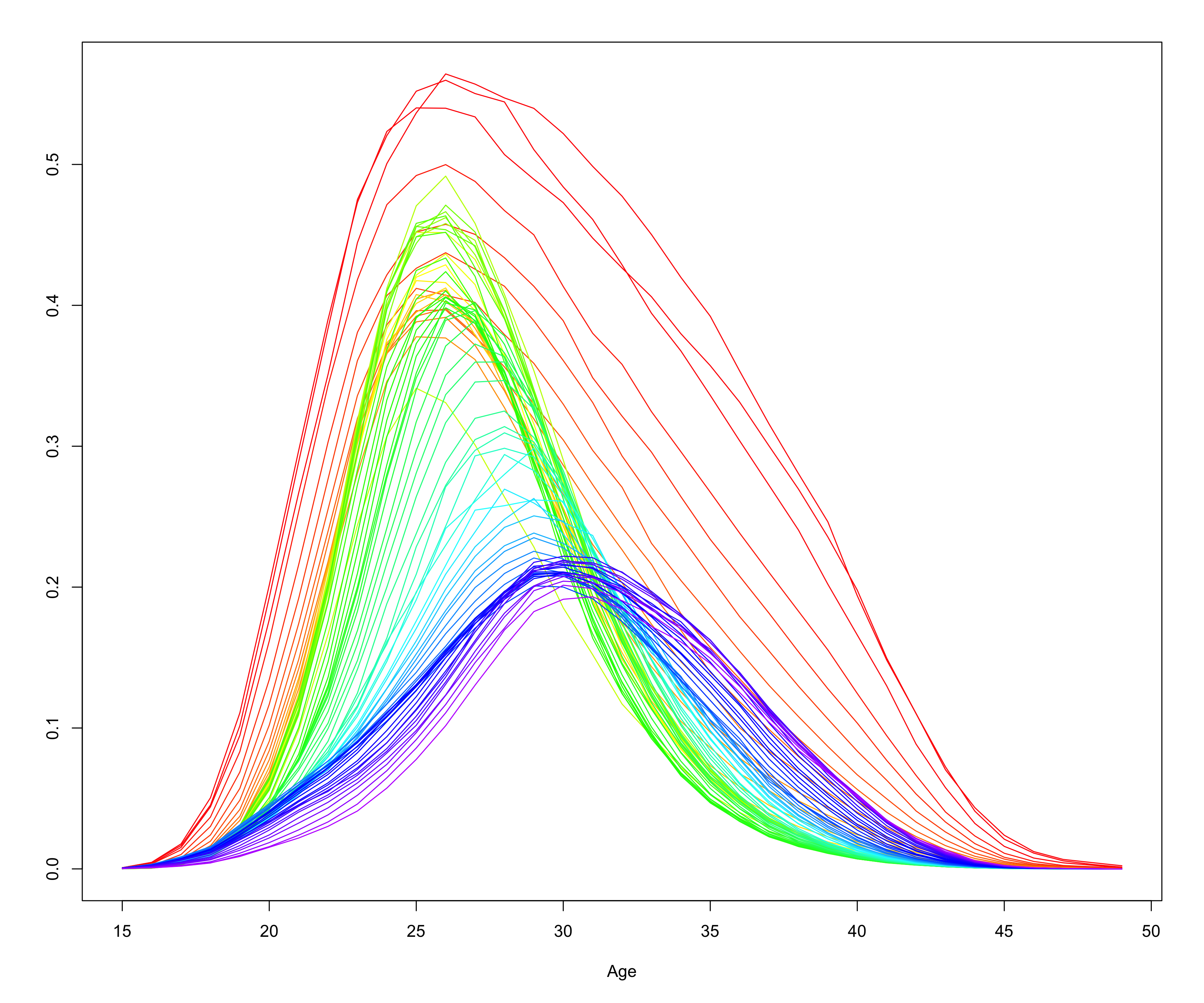}}
\caption{\small{Rainbow plots: Canadian fertility rates (Ages 15-49+) from 1921 to 2023 and Japanese fertility rates (Ages 15-49+) from 1947 to 2023.}}\label{fig:1}
\end{figure}

We illustrate our forecasting method using Canada's age-specific fertility rate (ASFR) data from 1921 to 2023, sourced from \cite{HFD23}, and initially capture fertility rates for women aged $12-$ to $55+$. For analytical purposes, the age groups 12-15 and 49-55 have been aggregated to $15-$ and $49+$, respectively, emphasizing the fertility trends within the primary working age group of $15-$ to $49+$. This aggregation helps us understand fertility patterns relevant to Canadian and Japanese workforce demographics. By analyzing changes in fertility rates by age and year, we observe a decreasing trend over time. To visualize this evolution, we present fertility rates as rainbow plots in Figure~\ref{fig:1}, with the distant past shown in red and the most recent years in purple. 

We utilize the weighted regression B-spline method with a concavity constraint, drawing on the methodology of \cite{he1999cobs} to enhance fertility data by smoothing. Generally, the use of smoothing techniques aims to improve forecast accuracy. This approach tends to yield superior predictive results when combined with the averaging fitting method. From \cite{HFD23}, we obtain the multi-country datasets used in this study for comparison. It comprises age-specific fertility rates for the 16 countries listed in Table~\ref{tab:countries_asfr}, exclusively using data from years up to and including 1950. 
\begin{table}[!htb]
\tabcolsep 0.18in
\centering
\caption{Countries and their ASFR sample periods (latest \cite{HFD23} vintage).}\label{tab:countries_asfr}
\begin{tabular}{@{}lll@{\hspace{2em}}lll@{}}
\toprule
\textbf{Country} & \textbf{Code} & \textbf{Period} & \textbf{Country} & \textbf{Code} & \textbf{Period} \\
\midrule
Belgium 	& BEL 		& 1947--2022 & Japan 				& JPN 	& 1947--2023 \\
Bulgaria 	& BGR 		& 1947--2021 & Netherlands 			& NLD 	& 1950--2022 \\
Canada 	& CAN 		& 1921--2023 & Portugal 				& PRT 	& 1940--2023 \\
Czechia 	& CZE 		& 1950--2021 & Slovakia 				& SVK 	& 1950--2014 \\
Denmark 	& DNK 		& 1916--2024 & Spain 				& ESP 	& 1922--2023 \\
Finland 	& FIN 		& 1939--2024 & Sweden 				& SWE 	& 1891--2024 \\
France  	& FRATNP 	& 1946--2022 & Switzerland 			& CHE 	& 1932--2023 \\
Hungary 	& HUN 		& 1950--2020 & United States of America & USA 	& 1933--2023 \\
\bottomrule
\end{tabular}
\end{table}

For fertility, we demonstrate using Canada and Japan as illustrative countries in Figure~\ref{fig:1}. Australia is not included for ASFR panels in our study because it is not covered in \cite{HFD23}. Instead, Japan is selected because it is an OECD member with well-documented postponement and very low fertility, providing an informative contrast to Canada’s more moderate levels and smoother timing changes \citep{frejka2010east, raymo2015marriage}. This substitution keeps the examples within a comparable developed-country context while supplying the substantive variation, strong tempo shifts versus smoother dynamics, which reveals how methods behave across horizons when underlying fertility regimes differ. 

\section{Functional time-series forecasting method}\label{sec:3}

We consider a functional time-series method of \cite{HU07} to produce $h$-step-ahead forecasts. The method combines the ideas of nonparametric smoothing, functional principal component regression, and functional data analysis to model age-specific mortality and fertility rates. Logarithmic mortality rates are first smoothed using penalized regression splines with a partial monotonic constraint for ages 65 and above. We can express
\begin{equation}
\Y_t(u_i) = \X_t(u_i) + \sigma_t(u_i)\varepsilon_{t,i}, \quad i=1, 2, \dots,U, \quad t=1, 2, \dots,n,\label{eq:1}
\end{equation}
where $\Y_t(u_i)$ denotes the log of the observed mortality rate for age $u_i$ in year $t$, and $U$ denotes the total number of ages; $\sigma_t(u_i)$ allows the amount of noise to vary with age $u_i$ in year $t$, and $\varepsilon_{t, i}$ represents an independent and identically distributed standard normal random variable. Based on the sample covariance function of the log (smoothed) mortality rates, we compute the estimated functional principal components and their associated scores,
\begin{equation}
\X_t(u) = \mu(u) + \sum^K_{k=1}\beta_{t,k}\phi_{k}(u) + e_t(u),\label{eq:2}
\end{equation}
where $\X_t(u)$ denotes the log of the observed mortality rate for age $u$ in year $t$, $\mu(u)$ can be estimated by $\overline{\X}(u) = \frac{1}{n}\sum_{t=1}^n\X_t(u)$, $\{\beta_{t,1},\dots,\beta_{t,K}\}$ denotes a set of the first $K$ functional principal component scores in time $t$, $\{\phi_{1}(u),\dots,\phi_K(u)\}$ denotes a set of the first $K$ functional principal components, $e_t(u)$ denotes the residual function with a mean of zero, and $K<n$ denotes the number of retained functional principal components. Following \cite{HBY13}, we choose $K=6$.

By conditioning on the observed data $\bm{\Y}(u) = \{\Y_1(u),\dots, \Y_n(u)\}$ and the set of estimated functional principal components $\bm{\Phi} = [\phi_1(u),\dots,\phi_K(u)]$, the $h$-step-ahead forecast of $\Y_{n+h}(u)$ can be obtained by
\begin{align*}
\widehat{\Y}_{n+h|n}(u) &= \E[\Y_{n+h}(u)|\bm{\X}(u),\bm{\Phi}] \\
&= \overline{\X}(u)+\sum^K_{k=1}\widehat{\beta}_{n+h|n,k}\phi_k(u),
\end{align*}
where $\widehat{\beta}_{n+h|n,k}$ denotes the $h$-step-ahead forecast of $\beta_{n+h,k}$ using a univariate time series model, such as the exponential smoothing state space model of \cite{HKO+08}. 

The smoothing step in~\eqref{eq:1} does not affect the mean forecast, but it is included in the bootstrap sample. Using the univariate time-series model, we obtain multi-step-ahead forecasts for the principal component scores, $\{\widehat{\beta}_{1,k},\dots,\widehat{\beta}_{n,k}\}$. Let the $h$-step-ahead forecast errors be given by
\begin{equation*}
\widehat{\xi}_{\omega,k} = \widehat{\beta}_{\omega,k} - \widehat{\beta}_{\omega|\omega-h,k}, \qquad \omega = h+1,\dots,n.
\end{equation*}
These can be sampled with replacement to produce a bootstrap sample of $\beta_{n+h,k}$:
\begin{equation*}
\widehat{\beta}_{n+h|n,k}^{(b)} = \widehat{\beta}_{n+h|n,k} + \widehat{\xi}^{(b)}_{*,k},\qquad b=1,\dots,B,
\end{equation*}
where $\widehat{\xi}^{(b)}_{*,k}$ are sampled with replacement from $\{\widehat{\xi}_{h+1,k},\dots,\widehat{\xi}_{n,k}\}$, and $B$ denotes the number of bootstrap samples.

Assuming that the first $K$ functional principal components approximate the data well, the residual of the model should be random noise. Thus, we can bootstrap the residuals of the model in~\eqref{eq:2} by sampling with replacement from the residuals $\{e_1(u), e_2(u),\dots,e_n(u)\}$. Similarly, we can also bootstrap the smoothing error $\bm{\varepsilon}_{n+h}^{(b)}$ by randomly sampling with replacement $\{\varepsilon_{1}, \varepsilon_{2}, \dots,\varepsilon_{n}\}$. 

Adding all sources of variability, we obtain $B$ variants for $\Y^{(b)}_{n+h|n}(u)$:
\begin{equation*}
\widehat{\Y}_{n+h|n}^{(b)}(u) = \overline{\X}(u) + \sum^K_{k=1}\widehat{\beta}_{n+h|n,k}^{(b)}\phi_k(u) + \widehat{\epsilon}_{n+h|n}^{(b)}(u) + \widehat{\sigma}_{n+h}(u)\widehat{\varepsilon}_{n+h}^{(b)}.
\end{equation*}

\section{Comparison of point forecast accuracy}\label{sec:4}

\subsection{Point forecast errors}\label{sec:4.1}

We consider the mean absolute forecast error (MAFE) and the root mean square forecast error (RMSFE) to evaluate the performance of the forecasts on the holdout test data. Let $\Y_{\eta}$ represent the actual observed value for the $\eta$\textsuperscript{th} observation, let $\widehat{\Y}_{\eta}$ represent the corresponding forecast value, and let $\eta=1, 2,\dots,n_{\text{test}}$ be a subscript, where $n_{\text{test}}$ is the total number of data points being evaluated. To compare the accuracy of point forecasts, we reserve the last 20 calendar years (i.e., $n_{\text{test}}=20$ for each country) as the holdout test set to evaluate forecast error, and use all previous years for training.

The MAFE measures the average absolute error between predicted and observed values, with all individual differences in age given equal weight. Evaluating on the original scale, it can be written as
\begin{equation*}
\text{MAFE} = \frac{1}{n_{\text{test}}} \sum_{\eta=1}^{n_{\text{test}}} \int_{u}\left|\exp^{\Y_{\eta}(u)} - \exp^{\widehat{\Y}_{\eta}(u)}\right|du,
\end{equation*}
where $\Y_{\eta}(u)$ denotes the holdout data in the test set and $\widehat{\Y}_{\eta}(u)$ denotes the forecast.

The RMSE is a quadratic scoring rule that measures the average magnitude of the error. It is the square root of the average of the squared differences between the prediction and the actual observation. Because errors are squared before averaging, RMSE places relatively greater weight on large errors. It is expressed as
\begin{equation*}
\text{RMSFE} = \sqrt{\frac{1}{n_{\text{test}}} \sum_{\eta=1}^{n_{\text{test}}} \int_{u}\left[\exp^{\Y_{\eta}(u)} - \exp^{\widehat{\Y}_{\eta}(u)}\right]^2du}.
\end{equation*}

\subsection{Point forecast results for age-specific mortality rates}\label{sec:4.2}

Our combination approach begins by systematically comparing and combining two distinct modeling approaches: rolling-window and expanding-window forecasts, both obtained from the same method. The objective is to enhance predictive accuracy while avoiding additional tuning complexity. In addition, we consider modeling the logarithm of raw or smoothed data sets. Using the functional time-series method, we form a tuning-free, equal-weight ensemble
\begin{equation}
\widehat{\Y}_{\eta}^{\text{Combined}} = \tfrac{1}{2}\,\widehat{\Y}_{\eta}^{\text{Rolling}} + \tfrac{1}{2}\,\widehat{\Y}_{\eta}^{\text{Expanding}}.
\label{eq:equal-combination}
\end{equation}
All models are estimated in a common training span and evaluated in a holdout set. Point-forecast accuracy is assessed using RMSFE and MAFE across horizons $h=1, 2, \dots,20$.

\begin{figure}[!htb]
\centering
{\includegraphics[width=\textwidth]{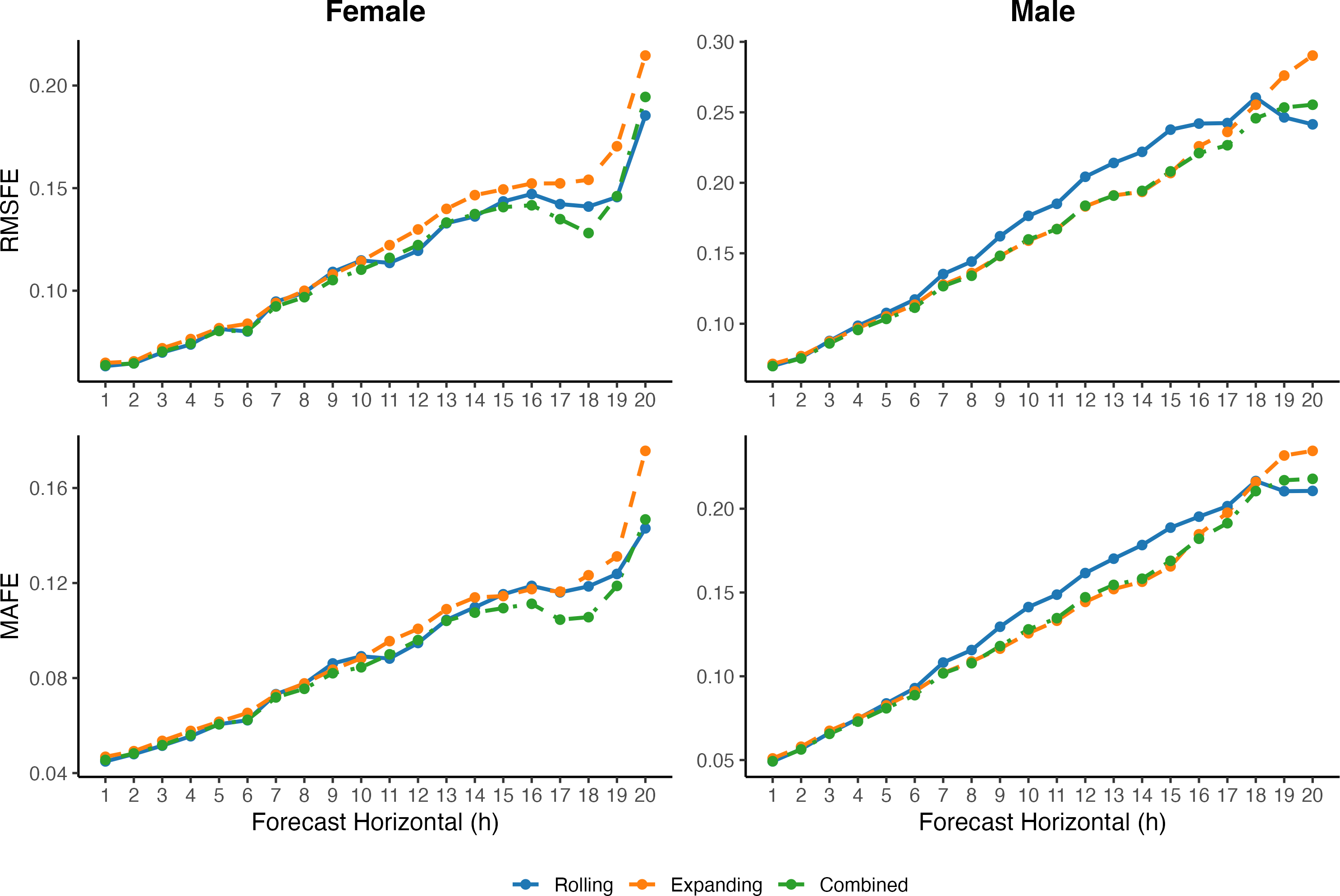}}
\caption{\small{\textcolor{purple}{AUS mortality — point forecast errors (RMSFE/MAFE vs forecast horizon ($h$)).}}}\label{fig:7}
\end{figure}

\begin{figure}[!htb]
\centering
{\includegraphics[width=\textwidth]{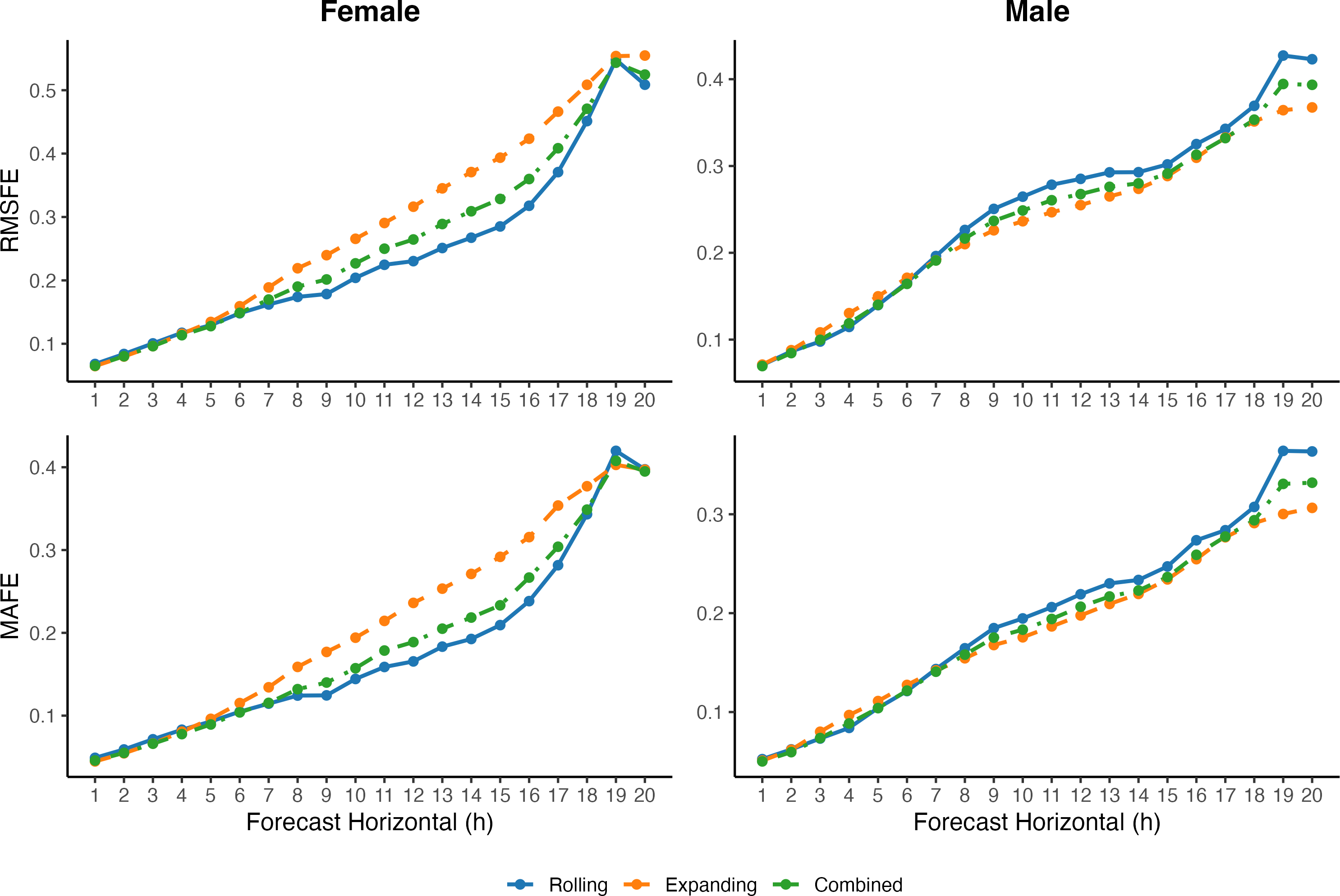}}
\caption{\small{\textcolor{purple}{CAN mortality — point forecast errors (RMSFE/MAFE vs forecast horizon ($h$)).}}}\label{fig:8}
\end{figure}

To ensure interpretability and comparability across all figures, we adopt fixed visual encodings that are used consistently for every metric plot and heatmap. In the line graphs of Figures~\ref{fig:7} and~\ref{fig:8}, each forecast method is represented by a unique color-line type pair that does not change between countries, sexes, or metrics: \texttt{Rolling} is blue with a \emph{solid} line, \texttt{Expanding} is orange with a \emph{dashed} line, and the combined forecast of equal-weight is green with a \emph{dot–dash} line. This redundant encoding (color and line type) preserves legibility in grayscale reproductions and under common forms of color-vision deficiency. The horizontal axis is the forecast horizon $h$, the vertical axis is the error (RMSFE or MAFE), and the lower curves indicate better accuracy at a given horizon. A grid-free base theme is used to emphasize the relative performance of methods and avoid visual clutter.

\begin{figure}[!htb]
\centering
{\includegraphics[width=\textwidth]{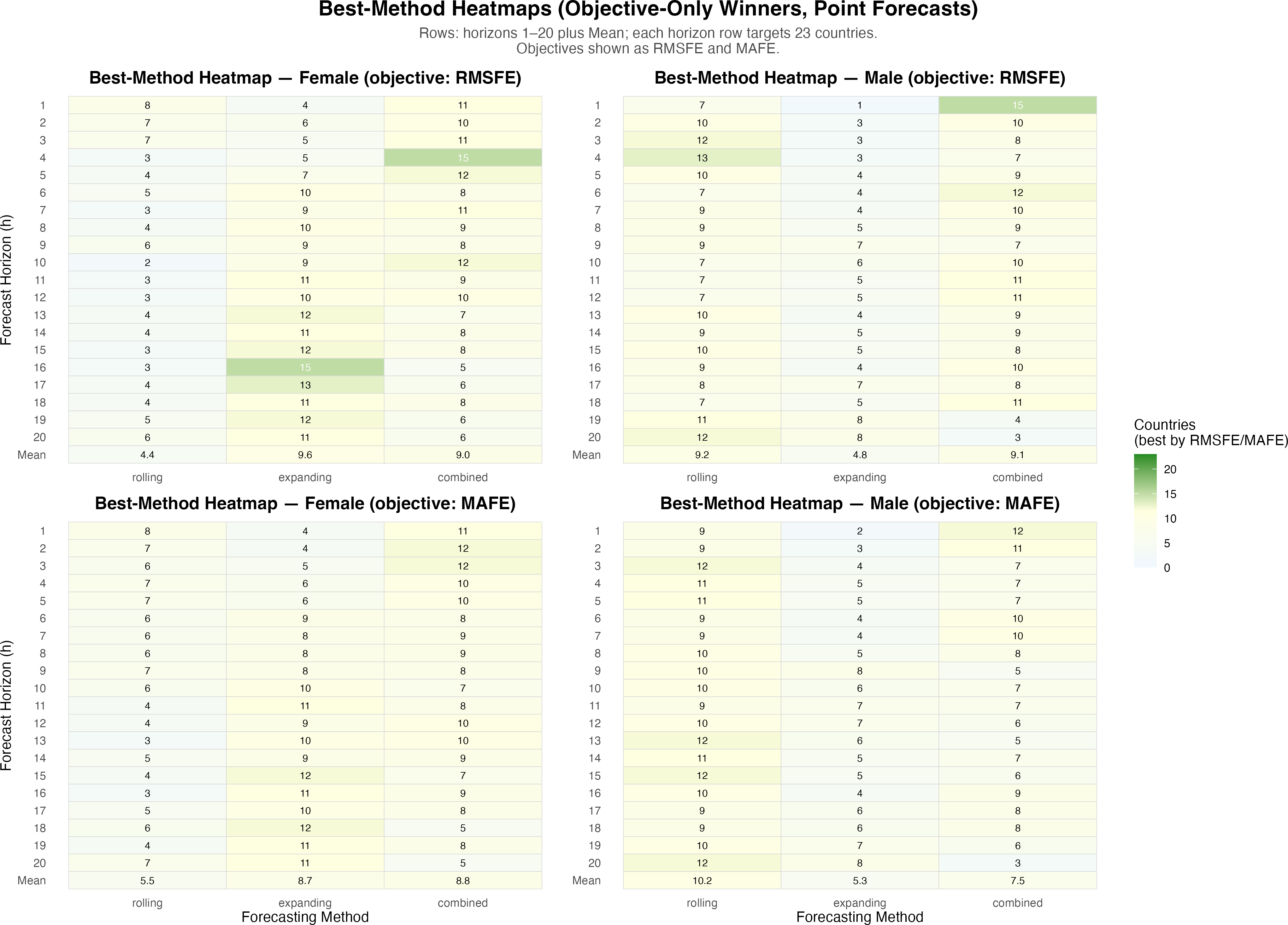}}
\caption{\small{Horizon-specific ranking heatmaps for the female and male mortality, measured by the MAFE and RMSFE, for the 23 countries considered.}}\label{fig:3}
\end{figure}

In Figure~\ref{fig:3}, the horizon-specific heatmaps synthesize the performance between countries at each horizon. Formally, for each method–horizon cell, the fill color encodes the \emph{count of countries} in which that method attains the lowest error (``wins'') at that horizon. The divergent fill scale maps low counts to light blue (``aliceblue''), mid-range counts to yellow (``lightyellow''), and high counts to dark green (``forestgreen''); the midpoint is set to half of the observed maximum count to enhance contrast around typical values, and the scale is bounded between 0 and the maximum observed across methods and horizons. 

Each tile also displays its win count as an overlaid label, with the text color chosen for legibility against the local background. The $x$-axis lists the forecasting methods; the $y$-axis lists horizons $h=1,\dots,20$ and includes a terminal “Mean” row that averages performance across the horizons. Consequently, darker green tiles indicate horizons (or the mean row) where a method dominates a larger share of countries, while pale tiles indicate few or no wins. Titles, axes, and typography are centered and simplified to keep attention on the method–horizon structure.

Figure~\ref{fig:3} summarizes the horizon-wise rankings for \emph{female} and \emph{male} mortality under RMSFE and MAFE. A clear pattern emerges: for \emph{female} mortality, \texttt{Expanding} and \texttt{Combined} win most often across horizons; for \emph{male} mortality, \texttt{Rolling} and \texttt{Combined} account for most wins. \texttt{Combined} is consistently competitive, typically among the top two, and often the winner. As the forecast horizon $h$ increases, the sex-specific contrast strengthens: \texttt{Expanding} gains share for \emph{females}, while \texttt{Rolling} gains share for \emph{males}; the terminal mean row confirms these tendencies.

The RMSFE/MAFE split does not overturn these patterns: RMSFE (which emphasizes large misses) magnifies the advantage of the sex-aligned scheme, whereas MAFE narrows gaps; \texttt{Combined} remains robust in both cases. Cross-country heatmaps show a concentration of wins in the \texttt{Expanding} column for \emph{females} and the \texttt{Rolling} column for \emph{males}, indicating systematic rather than idiosyncratic country effects. Because leadership can change with $h$, horizon-specific reporting should accompany averages.

If a single scheme must be chosen beforehand, the defaults are \texttt{Expanding} for females and \texttt{Rolling} for males. However, the \texttt{Combined} scheme serves as a practical, \textbf{tuning-free hedge} that tracks the better of the two schemes over \textbf{medium- to long-horizon} periods while damping volatility and imposing little short-term cost.

The alignment between the \emph{female}–\texttt{Expanding} and \emph{male}–\texttt{Rolling} patterns and well-known demographic regularities reinforces the heatmap evidence \citep{preston2006sex, wensink2020progression}. \emph{Female} mortality typically evolves smoothly, rewarding the use of the full historical record; \emph{male} mortality shows stronger cohort and external-cause imprints, so responsiveness to recent data is beneficial. Because \texttt{Combined} performs competitively in both settings, it is a pragmatic default when prior knowledge about stability versus recent change is weak, delivering reliable accuracy across horizons and metrics.

Country-level examples further illustrate these cross-sectional patterns. For Australia (AUS; Figure~\ref{fig:7}), the female series shows \texttt{Combined} as the lowest or near-lowest over medium to long horizons in both RMSFE and MAFE; \texttt{Expanding} degrades slightly at the longest horizons, and \texttt{Rolling} remains close but is typically worse around $h\approx 12$–$18$. For Australian male mortality, \texttt{Combined} and \texttt{Expanding} generally outperform \texttt{Rolling} across many horizons, and averaging reduces late-horizon variability in the single-scheme curves.

For Canada (CAN, Figure~\ref{fig:8}), the female series favors \texttt{Rolling} across most horizons for both metrics. \texttt{Expanding} is generally the weakest. The \texttt{Combined} path inherits much of \texttt{Rolling}'s advantage and smooths the late-horizon end. By contrast, male mortality typically favors \texttt{Expanding} at medium and long horizons, with \texttt{Rolling} showing the poorest performance. \texttt{Combined} again closely tracks the stronger scheme, underscoring its hedging role.

Taken together, the evidence indicates that no single fitting scheme dominates uniformly across sex, country, and horizon. The equal-weight combination serves as a robust default: it avoids the need for additional tuning, is never systematically inferior, frequently matches or exceeds the best single scheme, and attenuates end-of-horizon volatility. In circumstances where prior knowledge about structural changes is limited or varies among populations, this tuning-free combination offers a practical approach. It provides a dependable method for forecasting age-specific mortality rates.

\subsection{Point forecast results for age-specific fertility rates}\label{sec:4.3}

We also apply the framework to \emph{age-specific fertility rate} (ASFR) data from the \cite{HFD23}. Sixteen countries with long, high-quality series are included (Table~\ref{tab:countries_asfr}), and we analyze observations through 2023 to capture major demographic transitions while avoiding pandemic-period distortions. As in the mortality analysis, we compare \texttt{rolling} and \texttt{expanding} fitting schemes within the same functional time-series method and use the same tuning-free, equal-weight ensemble (see Equation~\eqref{eq:equal-combination}). All models are trained on a common span and evaluated on held-out data. Point-forecast accuracy is assessed using RMSFE and MAFE across horizons $h=1,2,\dots,20$.

\begin{figure}[!htb]
\centering
{\includegraphics[width=18cm]{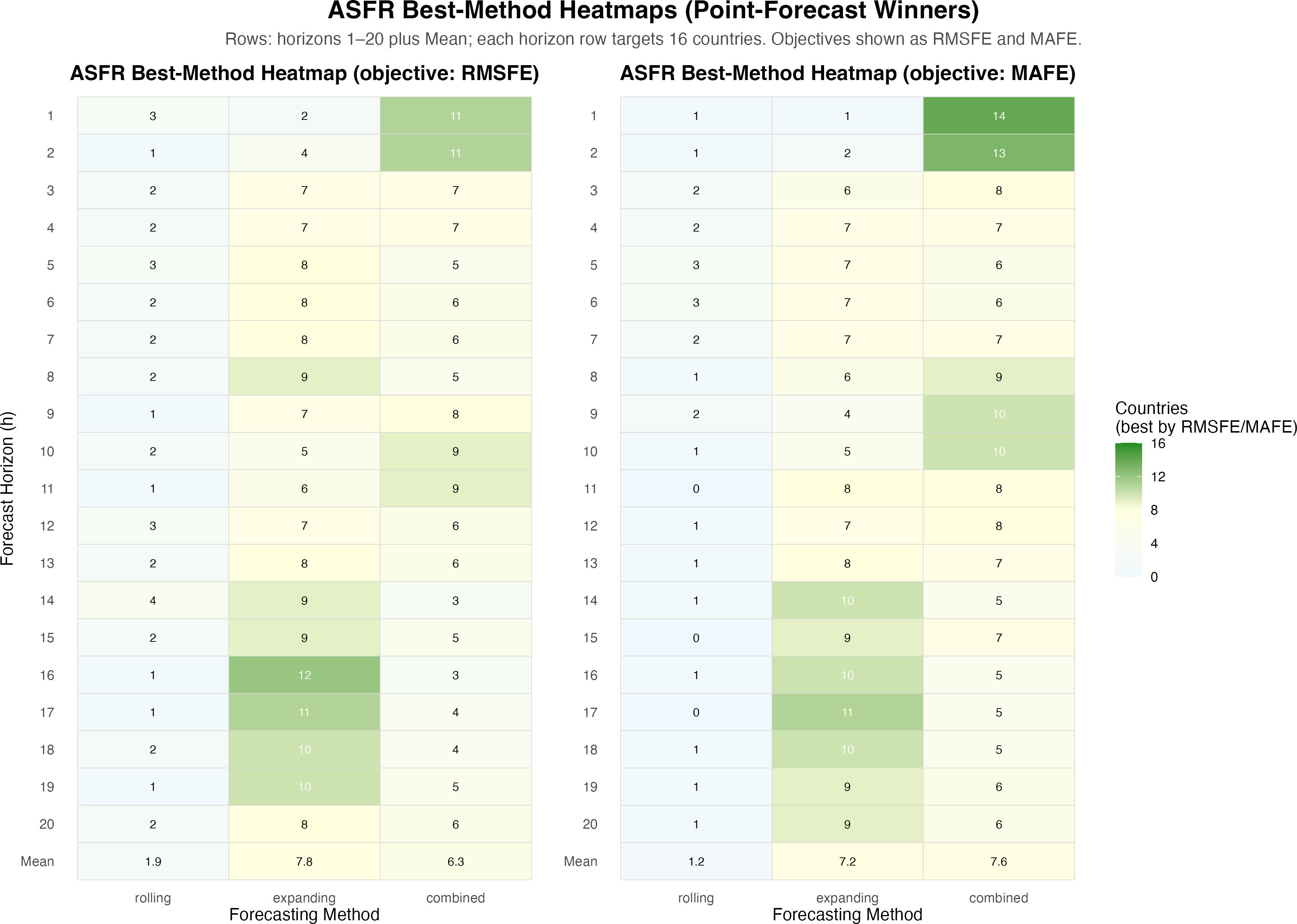}}
\caption{\small{\textcolor{purple}{Horizon-specific best-method heatmaps for ASFR point forecasts, measured by RMSFE and MAFE, for the 16 countries considered.}}}\label{fig:4}
\end{figure}

Figure~\ref{fig:4} summarizes horizon-specific rankings across countries, revealing a consistent pattern. For RMSFE (left panel), \texttt{expanding} achieves the largest mean win count, with \texttt{combined} close behind and \texttt{rolling} rarely best. For MAFE (right panel), \texttt{combined} edges \texttt{expanding} on average wins, again with \texttt{rolling} seldom dominant. The heatmaps also clarify how the ranking evolves with the horizon. At short horizons, the three methods are often close, but from medium horizons onward, the advantage of using the full historical record becomes more apparent: the concentration of darker tiles in the \texttt{expanding} and \texttt{combined} columns increases as $h$ lengthens. The mean row confirms that these horizon-wise gains accumulate across countries. The split between RMSFE and MAFE does not overturn the conclusion; RMSFE, which accentuates larger errors, favors \texttt{expanding} slightly more, whereas MAFE, which downweights occasional large misses, favors \texttt{combined}. In practical terms, for ASFR, the evidence supports exploiting the full history or averaging with the recent-information scheme; the equal-weight combination is reliably competitive and often the top performer, especially at medium and long horizons where volatility in single-scheme forecasts tends to be larger.

\begin{figure}[!htb]
\centering
{\includegraphics[width=18cm]{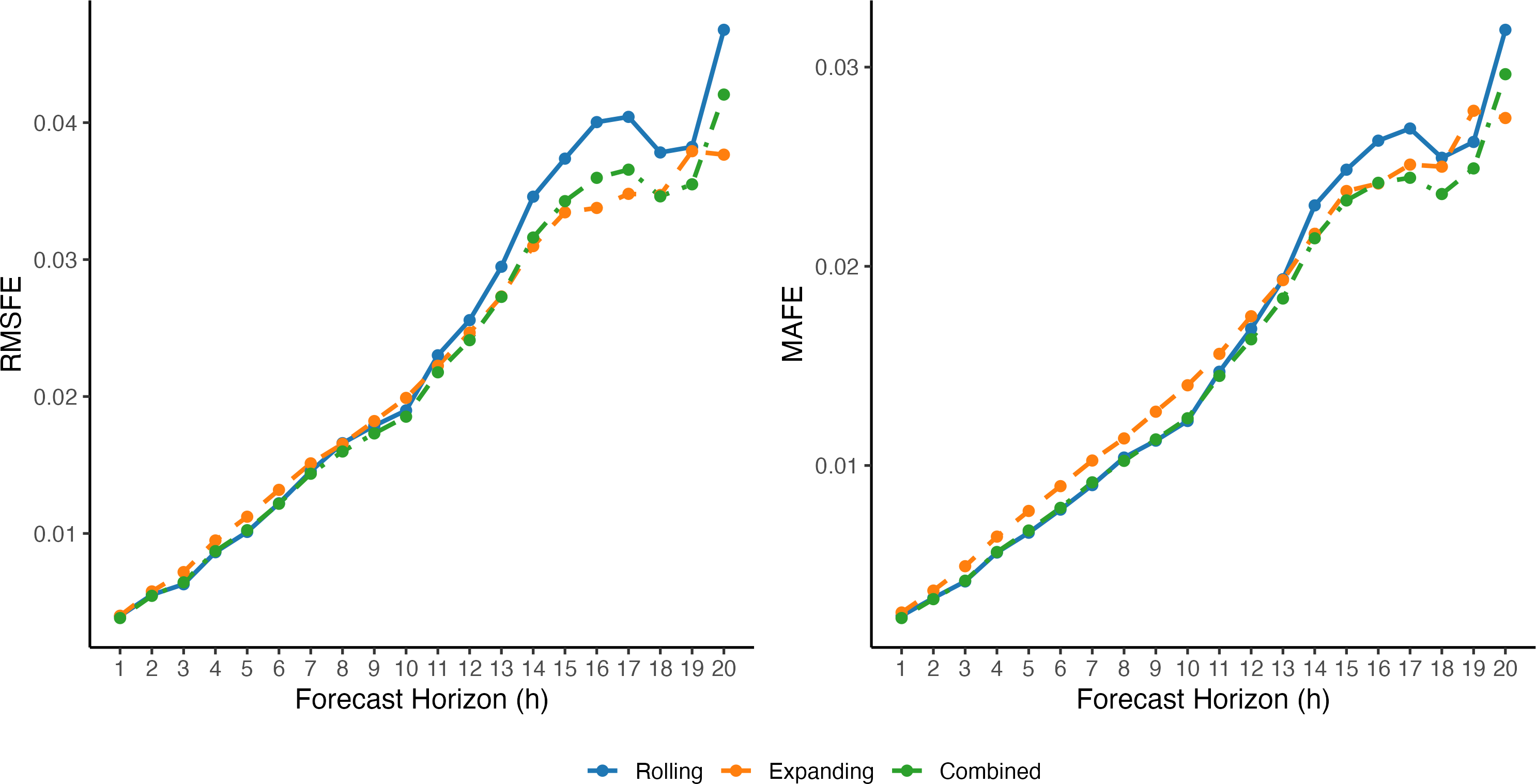}}
\caption{\small{\textcolor{purple}{CAN ASFR — point forecast errors (RMSFE/MAFE vs forecast horizon ($h$)).}}}\label{fig:9}
\end{figure}

Country-level evidence from Canada (CAN; Figure~\ref{fig:9}) illustrates these cross-sectional patterns. At short horizons ($h\!\leq\!10$), all methods are similar. From medium to long horizons ($h\!\ge\!12$), \texttt{rolling} develops a pronounced late-horizon spike and becomes clearly inferior, whereas \texttt{expanding} and \texttt{combined} remain competitive. For RMSFE, \texttt{combined} closely tracks or slightly improves on \texttt{expanding} through $h\!\approx\!14$–$19$; for MAFE, \texttt{combined} is typically the lowest at the longer horizons, consistent with the heatmap’s average-wins advantage. The equal-weight average, therefore, inherits much of the stronger single scheme while attenuating end-of-horizon variability.

\begin{figure}[!htb]
\centering
{\includegraphics[width=18cm]{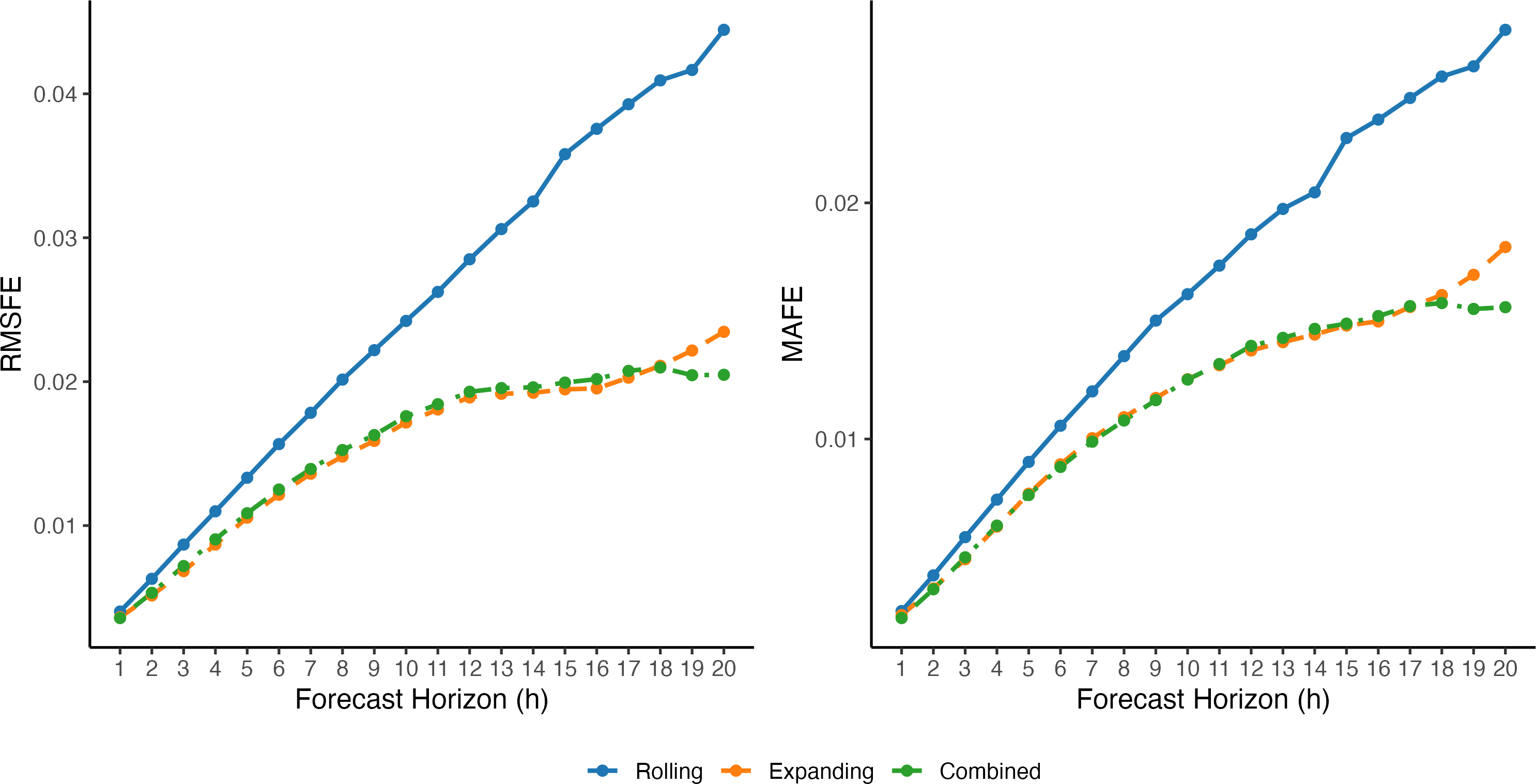}}
\caption{\small{\textcolor{purple}{JPN ASFR — point forecast errors (RMSFE/MAFE vs forecast horizon ($h$)).}}}\label{fig:17}
\end{figure}

Country-level evidence from Japan (JPN; Figure~\ref{fig:17}) provides a complementary example to Canada. At short horizons ($h\le 10$), the methods are close, but by medium horizons ($h\ge 12$) \texttt{rolling} becomes clearly inferior in both RMSFE and MAFE. The \texttt{expanding} and \texttt{combined} trajectories remain tightly grouped over most horizons; for RMSFE, the \texttt{combined} line closely tracks or is marginally below \texttt{expanding} through approximately $h=14$–$19$, while for MAFE, \texttt{combined} is typically the lowest at the longer horizons. The observed behavior aligns with the heatmap evidence. In scenarios with significant timing shifts and consistently low fertility, utilizing either the complete historical trajectory or the equal-weight average produces the most accurate point forecasts. This averaging process also reduces the slight increase seen at the later horizons in the individual scheme curves.

Overall, no single fitting scheme dominates uniformly in fertility forecasting. \texttt{Expanding} is often preferred under RMSFE, while the equal-weight \texttt{combined} forecast frequently leads under MAFE and remains highly competitive under RMSFE. Given heterogeneous country dynamics and limited prior information on structural change, the tuning-free combination provides a practical default for ASFR point forecasting.

\section{Comparison of interval forecast accuracy}\label{sec:5}

\subsection{Interval forecast errors}\label{sec:5.1}

To evaluate pointwise interval forecast accuracy, we consider the empirical coverage probability (ECP) and the difference between it and the nominal coverage probability. This difference is known as the coverage probability difference (CPD). The ECP can be defined as follows:
\textcolor{purple}{\begin{equation*}
\text{ECP} = 1-\frac{1}{n_{\text{test}}\times p}\sum_{\eta=1}^{n_{\text{test}}} \sum_{i=1}^p \left[\mathds{1}\{\exp^{\Y_{\eta}(u_i)}>\exp^{\widehat{\Y}_{\eta}^{\text{ub}}(u_i)}\}+\mathds{1}\{\exp^{\Y_{\eta}(u_i)}<\exp^{\widehat{\Y}^{\text{lb}}_{\eta}(u_i)}\}\right],
\end{equation*}}
\hspace{-.075in} where $n_{\text{test}}$ denotes the number of curves in the forecasting period, $p$ denotes the number of discretized ages, $[\widehat{\Y}_{\eta}^{\text{lb}}(u), \widehat{\Y}_{\eta}^{\text{ub}}(u)]$ represents the lower and upper bounds of the constructed prediction interval, and $\mathds{1}\{\cdot\}$ denotes the binary indicator function. Pointwise CPD is defined as
\begin{equation*}
\text{CPD} = |\text{ECP} - \text{nominal coverage}|.
\end{equation*}
The lower the CPD value, the better the forecasting method's performance.

Although the CPD presents a calibration metric, it does not measure the sharpness of the constructed prediction interval. Not only do we want the prediction interval to achieve the smallest CPD, but we also want the smallest width of the prediction interval. Here, we also consider the interval score of \cite{GR07}. For each year in the test set, the $h$-step-ahead prediction intervals were calculated at the $100(1-\alpha)\%$ nominal coverage probability. The lower and upper bounds are predictive quantiles at $\alpha/2$ and $1-\alpha/2$, denoted by $\widehat{\Y}_{\eta}^{\text{lb}}(u_i)$ and $\widehat{\Y}_{\eta}^{\text{ub}}(u_i)$. The scoring rule for the interval forecast at the discretized age $u_i$ is
\textcolor{purple}{\begin{align*}
S_{\alpha, \eta}(u_i) =& \left[\exp^{\widehat{\Y}_{\eta}^{\text{ub}}(u_i)} - \exp^{\widehat{\Y}_{\eta}^{\text{lb}}(u_i)}\right] + \frac{2}{\alpha}\left[\exp^{\widehat{\Y}_{\eta}^{\text{lb}}(u_i)}-\exp^{\Y_{\eta}(u_i)}\right]\mathds{1}\left\{\exp^{\widehat{\Y}_{\eta}^{\text{lb}}(u_i)}> \exp^{\Y_{\eta}(u_i)}\right\}  \\
& \hspace{1.77in} + \frac{2}{\alpha}\left[\exp^{\Y_{\eta}(u_i)} - \exp^{\widehat{\Y}_{\eta}^{\text{ub}}(u_i)}\right]\mathds{1}\left\{\exp^{\Y_{\eta}(u_i)} >  \exp^{\widehat{\Y}_{\eta}^{\text{ub}}(u_i)}\right\},
\end{align*}}
where $\alpha$ denotes a level of significance, customarily $\alpha = 0.2$ or 0.05. We compute the mean interval score as follows.
\begin{equation*}
\overline{S}_{\alpha} = \frac{1}{n_{\text{test}}\times p}\sum^{n_{\text{test}}}_{\eta=1}\sum_{i=1}^pS_{\alpha, \eta}(u_i).
\end{equation*}
The optimal interval score is achieved when the holdout data lie between the constructed prediction interval, with the smallest interval width. For comparison of the accuracy of interval forecasts, we reserve the last 20 calendar years (i.e.,  $n_{\text{test}} = 20$) for each country as the holdout test set to evaluate forecast errors. The preceding 20 years serve as a validation set for selecting the optimal parameter~$\xi$, \textcolor{purple}{while the observations prior to the validation period form the training set, which is used for model estimation under the corresponding rolling or expanding window scheme.} As shown in Figures~\ref{fig:5} and~\ref{fig:6}, the reason for setting~$h$ to range from 1 to 19 in the heatmaps is that the SD approach of \cite{SH25} for interval forecasting requires a minimum of a two-year data set to compute the pointwise standard deviation and assess the optimal parameter $\xi$.

\subsection{Interval forecast results for age-specific mortality rates}\label{sec:5.2}

\begin{figure}[!htb]
\centering
{\includegraphics[width=17.6cm]{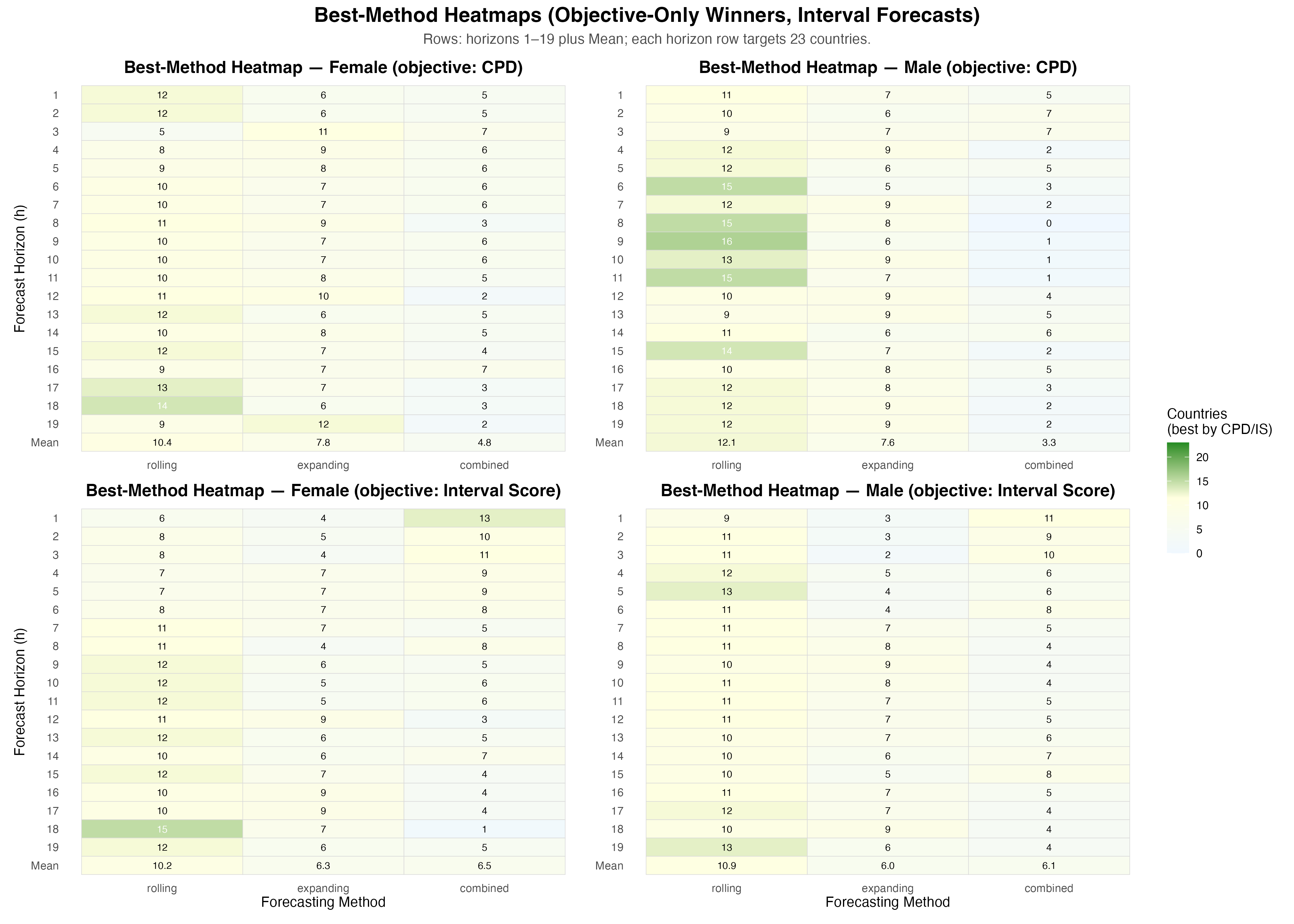}}
\caption{\small{\textcolor{purple}{Horizon-specific ranking heatmaps for the female and male mortality, measured by the CPD and score, for the 23 countries considered.}}}\label{fig:5}
\end{figure}

We evaluate interval forecasts using the calibration and sharpness criteria in Section~\ref{sec:5.1}. Across countries, the horizon-specific winners are summarized in Figure~\ref{fig:5}. For the \emph{CPD}, the darkest shading concentrates in the \texttt{rolling} column for both females and males over most horizons, and the terminal mean row confirms that \texttt{rolling} attains the highest average number of wins. The dominance is especially pronounced among males, where many horizons show \texttt{rolling} winning in most countries. For the \emph{interval score}, which evaluates the calibration–sharpness trade-off, \texttt{rolling} again leads on average for both sexes. The female panels show that \texttt{combined} is frequently competitive on short- to medium-horizon horizons, sometimes taking the largest share of wins when horizons are small, but the advantage tends to shift toward \texttt{rolling} as $h$ lengthens. The male panels display a similar ordering, with fewer horizons, in which \texttt{combined} overtakes \texttt{rolling}. The \texttt{expanding} scheme remains in contention across a subset of horizons, particularly at the lower horizon $h$, but it seldom produces the darkest tiles once the horizons are longer.

These heatmaps indicate that, when interval performance is evaluated using calibration (via CPD) and the joint calibration–sharpness criterion (interval score), emphasizing recent information with a \texttt{rolling} fitting window most often yields the best results across horizons and countries. At the same time, the equal-weight \texttt{combined} forecast remains broadly competitive, especially at shorter horizons and in the female interval-score panel, providing a simple hedge that tracks the better single scheme while reducing the risk of horizon-specific instability.

The ECP is plotted against the nominal level (horizontal dashed line); smaller absolute deviations yield a smaller CPD, indicating better calibration. Sharpness is assessed via the interval score, which penalizes both excessive width and miscoverage; lower values indicate better overall interval performance. Line styles and colors follow our global convention (blue/solid \texttt{rolling}, orange/dashed \texttt{expanding}, green/dot–dash \texttt{combined}).
\begin{figure}[!htb]
\centering
{\includegraphics[width=16.8cm]{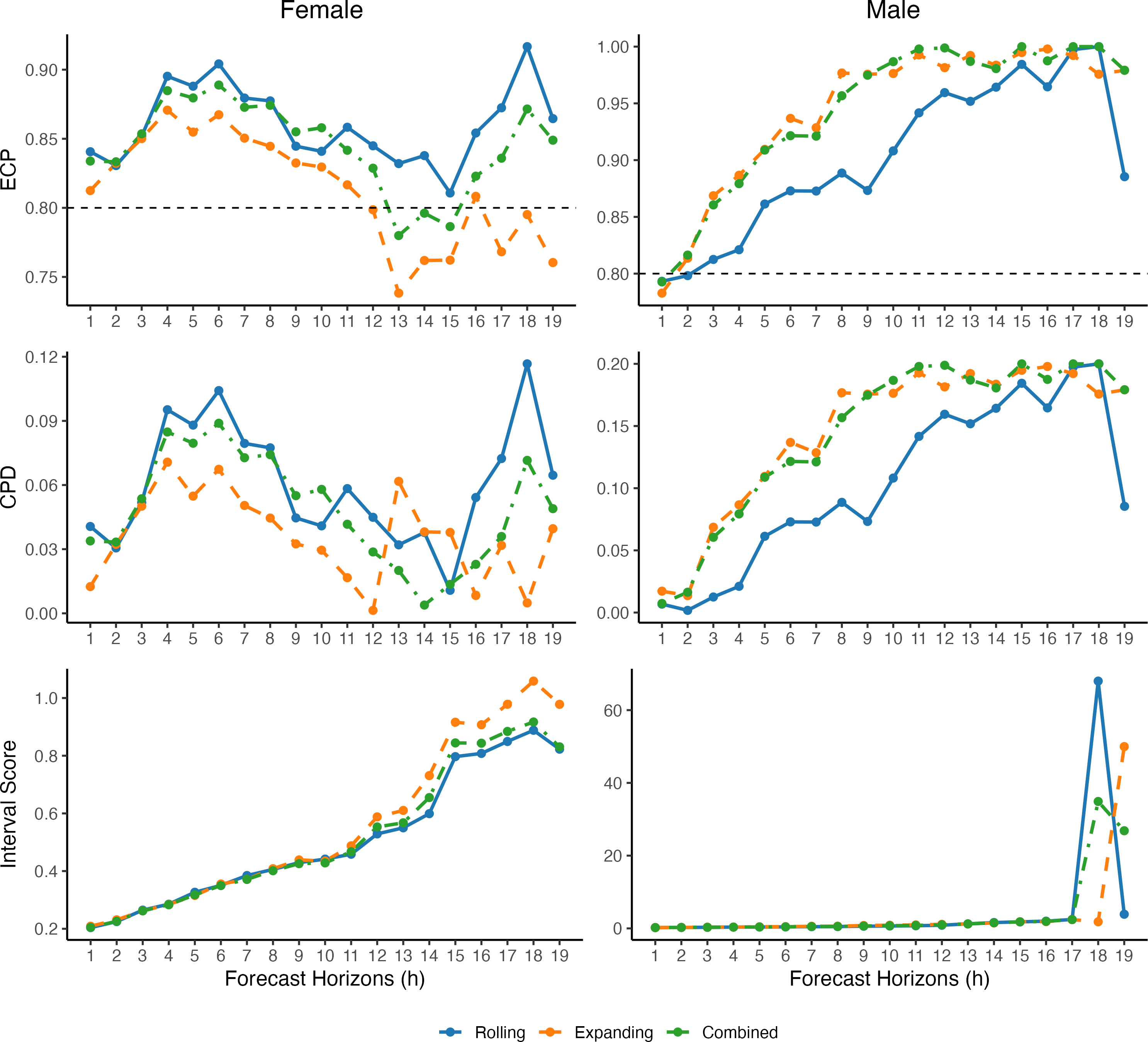}}
\caption{\small{AUS mortality — interval forecast evaluation (ECP / CPD / Interval Score by Horizon ($h$)).}}\label{fig:10}
\end{figure}

Country examples illustrate these aggregate patterns and the heterogeneity between populations. For Australia (Figure~\ref{fig:10}), the female ECP fluctuates around nominal on short horizons and then diverges; \texttt{expanding} often achieves the smallest CPD at medium horizons, but its interval scores rise later, leaving \texttt{rolling} and \texttt{combined} with lower scores at longer horizons. For Australian males, ECP increases steadily above nominal with horizon; \texttt{combined} typically delivers smaller CPD from mid to long horizons and keeps interval scores among the lowest, while \texttt{rolling} exhibits a sharper end-of-horizon rise that averaging helps dampen.
\begin{figure}[!htb]
\centering
{\includegraphics[width=19cm]{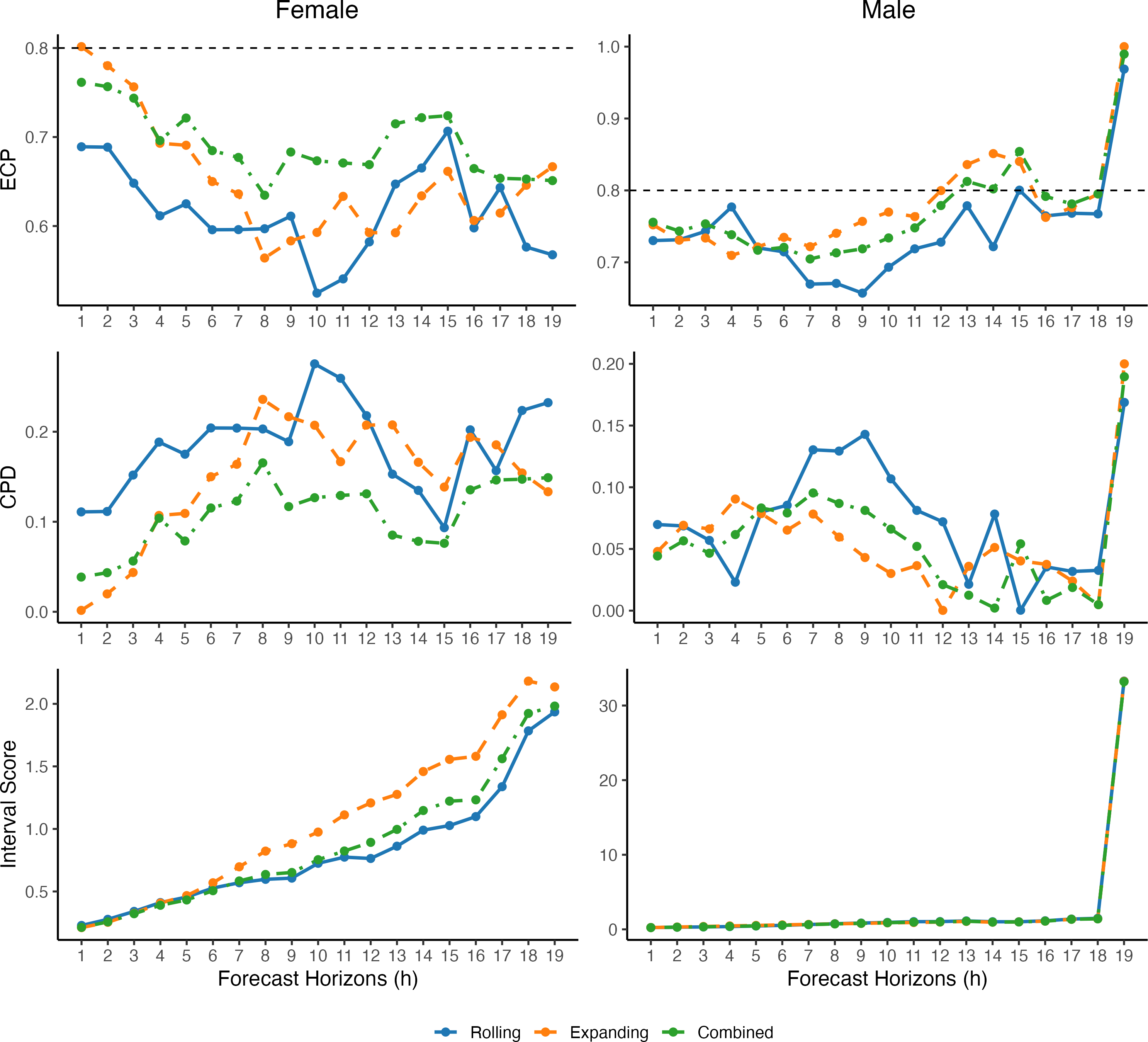}}
\caption{\small{CAN mortality — interval forecast evaluation (ECP/CPD/Interval Score by Horizon ($h$)).}}\label{fig:11}
\end{figure}

For Canada (Figure~\ref{fig:11}), female ECP shows sustained under-coverage at many horizons; here, \texttt{combined} most often attains the smallest CPD (better calibration) while \texttt{rolling} tends to produce the lowest interval scores—narrower but adequately calibrated intervals for this series. The Canadian male ECP is near nominal at short horizons but exhibits a pronounced spike at the longest horizon; \texttt{combined} mitigates, though it does not eliminate, this instability. Before the spike, \texttt{combined} and \texttt{expanding} often yield lower CPD than \texttt{rolling} with similar interval scores.

In general, interval-forecast comparisons show that \texttt{rolling} most frequently achieves the best calibration and the most favorable calibration–sharpness balance between horizons and countries, as reflected in the heatmaps. At the same time, single-country trajectories confirm that no scheme dominates uniformly: \texttt{expanding} can calibrate better at intermediate horizons in some settings, and the equal-weight \texttt{combined} method serves as a robust hedge by reducing late-horizon volatility and closely tracking the stronger single scheme. These results complement the point-forecast analysis by indicating that, for mortality, recent-information weighting (\texttt{rolling}) often yields superior interval performance, whereas averaging preserves robustness when horizon-end instability arises.

\subsection{Interval forecast results for age-specific fertility rates}\label{sec:5.3}

We evaluate ASFR interval forecasts using the same criteria as in Section~\ref{sec:5.1}: calibration by ECP relative to the nominal level (dashed line) and its absolute deviation (CPD), and overall sharpness–calibration performance by interval score (lower is better). Line styles and colors follow the global convention.

\begin{figure}[!htb]
\centering
{\includegraphics[width=19cm]{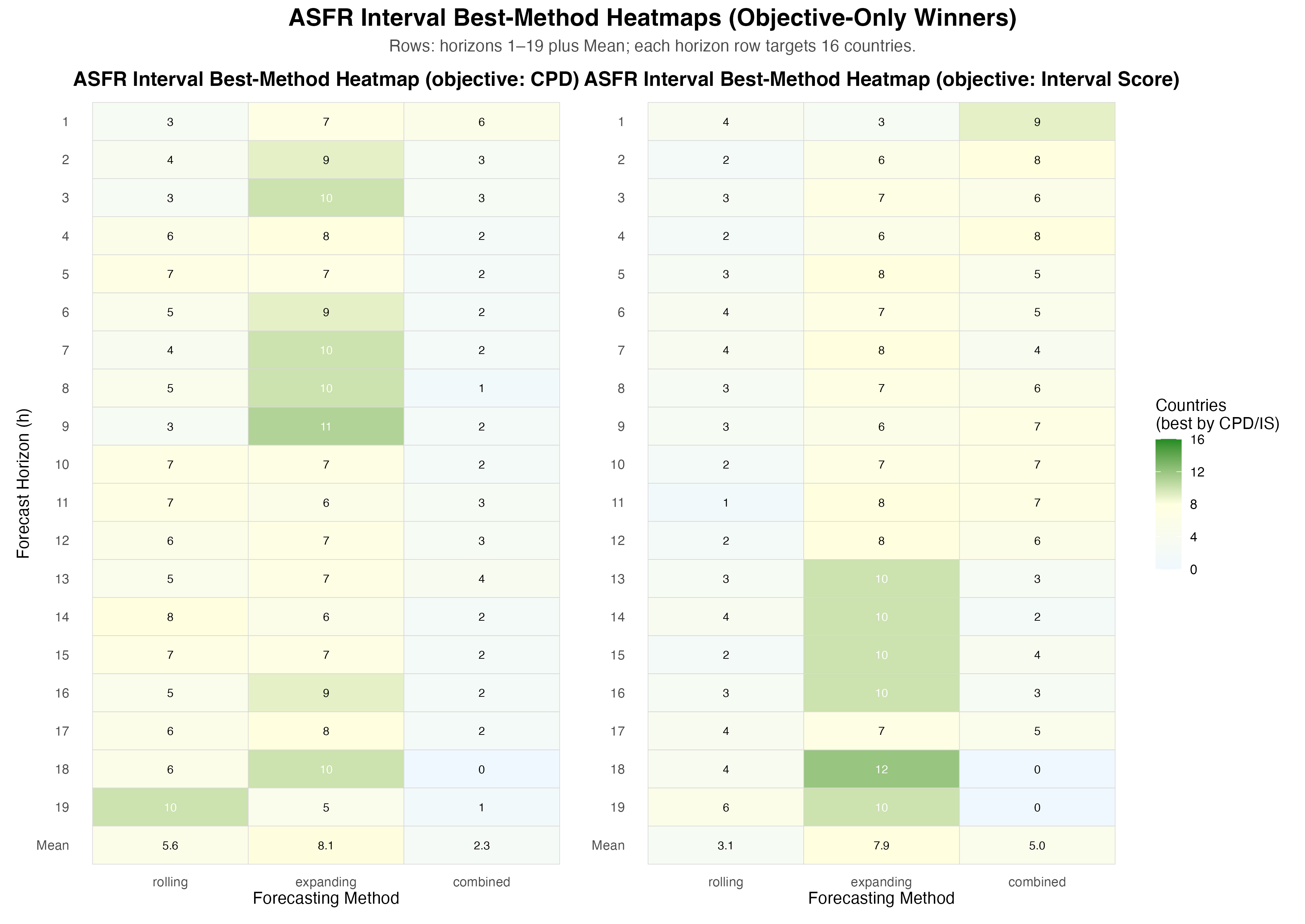}}
\caption{\small{\textcolor{purple}{Horizon-specific ranking heatmaps for the ASFR, measured by the CPD and score, for the 16 countries considered.}}}\label{fig:6}
\end{figure}

Across countries, the horizon-specific winners are summarized in Figure~\ref{fig:6}. A consistent pattern emerges in favor of \texttt{expanding}. For the \emph{CPD}, the darker shading is concentrated in the \texttt{expanding} column across a wide band of horizons, and the mean row confirms that this advantage persists when the horizon-wise evidence is averaged. For the \emph{interval score}, which captures the calibration–sharpness trade-off, the ordering is similar: as horizons lengthen, tiles increasingly darken in the \texttt{expanding} column, while \texttt{combined} remains competitive, particularly at short to medium horizons where it often matches the best single scheme. The \texttt{rolling} scheme seldom dominates once the horizons extend, indicating that intervals constructed from a short fitting period tend to either miscover or widen more than necessary in fertility series. Taken together, the heatmaps indicate that ASFR interval forecasts generally benefit from exploiting the full historical record; \texttt{expanding} most often delivers superior calibration and the most favorable calibration–sharpness balance, while the equal-weight \texttt{combined} method provides a robust alternative that closely tracks the leader and cushions horizon-end instability.

\begin{figure}[!htb]
\centering
{\includegraphics[width=12.3cm]{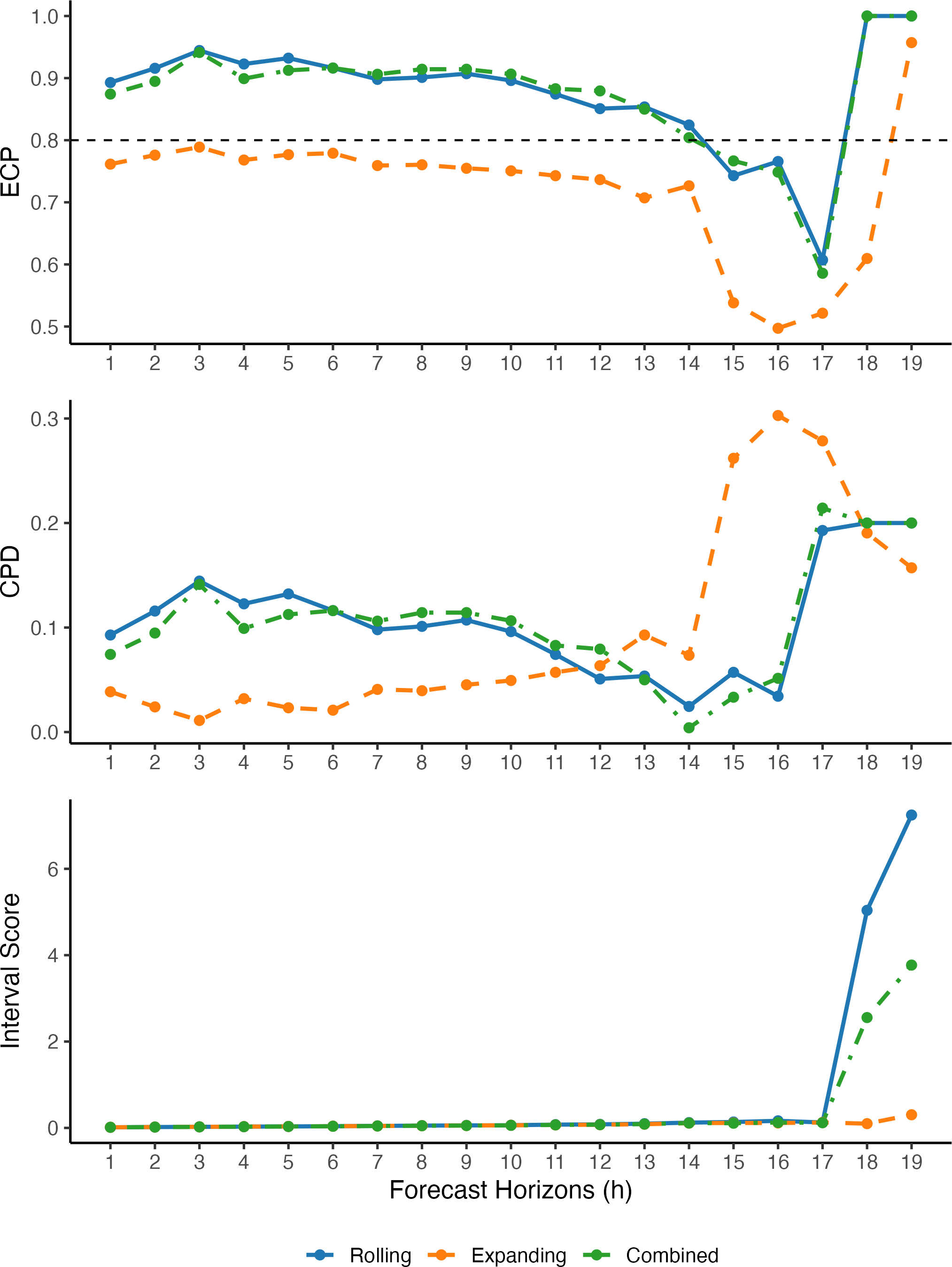}}
\caption{\small{CAN ASFR — interval forecast evaluation (ECP / CPD / Interval Score by horizon ($h$)).}}\label{fig:12}
\end{figure}

Country-level evidence from Canada (CAN; Figure~\ref{fig:12}) illustrates these aggregate patterns and their horizon dependence. In the \emph{ECP} panels, \texttt{rolling} and \texttt{combined} stay near or above the nominal level for short horizons and gradually decline toward the nominal level by \(h\approx 12\); \texttt{expanding} sits below nominal for many mid-horizons (under-coverage) before recovering at the longest horizons. The \emph{CPD} panel mirrors this: \texttt{combined} and \texttt{rolling} have a lower CPD through most short-to-mid horizons, while \texttt{expanding} exhibits a pronounced CPD spike around \(h\approx 16\!-\!17\) before improving. In the \emph{interval score} panel, the values are close to zero at short horizons for all methods; at very long horizons (\(h\ge 18\)) \texttt{rolling} experiences a sharp spike (reflecting miscoverage penalties and/or widened intervals), which the \texttt{combined} average partly dampens. \texttt{Expanding} maintains low interval scores across most horizons but pays a calibration cost (higher CPD) in the mid-range. Overall for CAN, \texttt{combined} provides the most balanced performance across horizons, while \texttt{expanding} excels in sharpness, and \texttt{rolling} occasionally attains better calibration at shorter horizons.

\begin{figure}[!htb]
\centering
{\includegraphics[width=12.3cm]{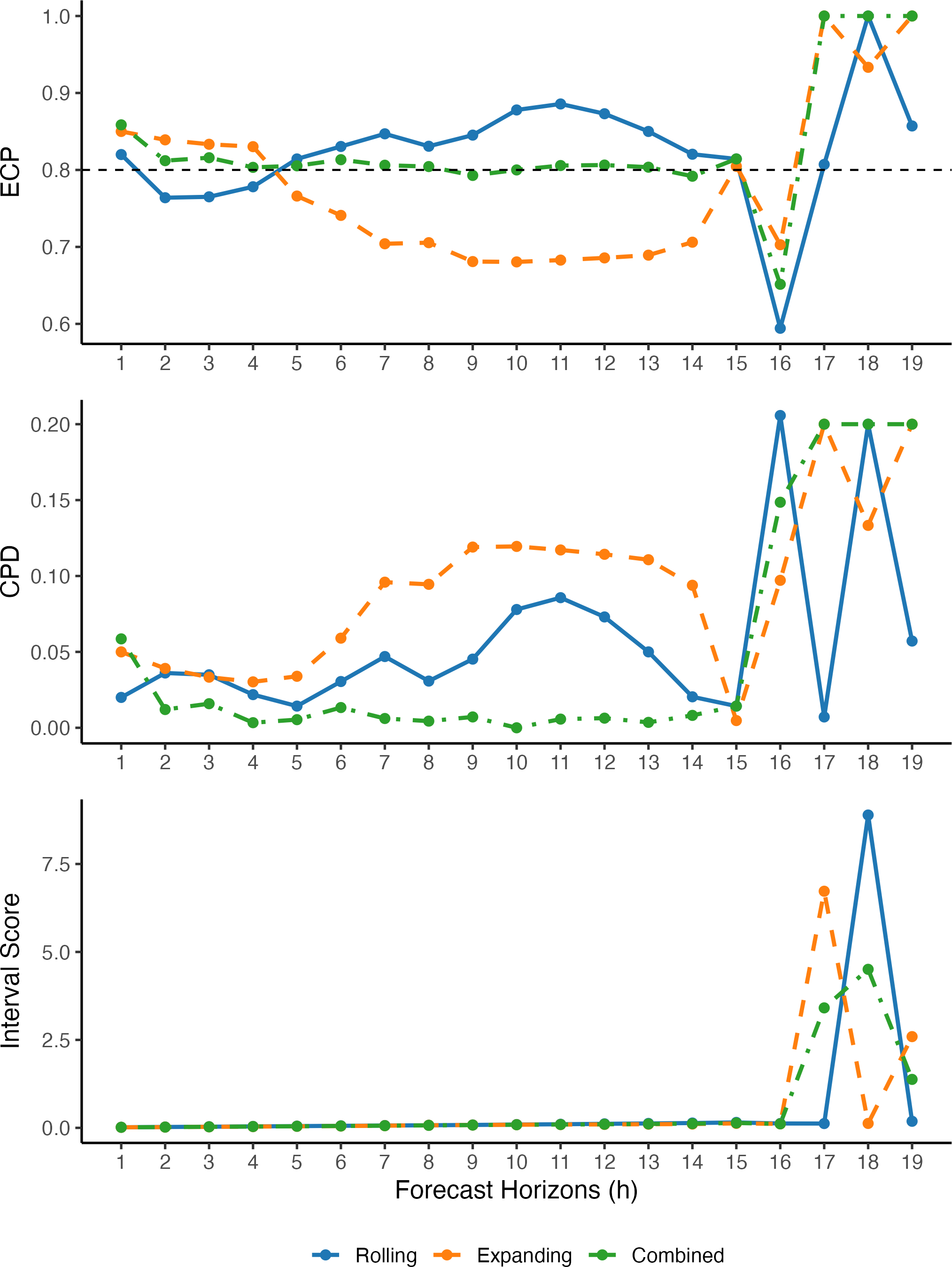}}
\caption{\small{JPN ASFR — interval forecast evaluation (ECP/CPD/Interval Score by horizon ($h$)).}}\label{fig:18}
\end{figure}

The country-level evidence from Japan (JPN; Figure~\ref{fig:18}) complements the Canadian example and shows how these cross-sectional patterns arise from horizon-wise trajectories. In the \emph{ECP} panel, \texttt{combined} remains close to the nominal level across most horizons, whereas \texttt{expanding} tends to under-cover through the middle part of the horizon range before recovering near the end, and \texttt{rolling} exhibits greater variability, including a sharp departure from nominal at the longest horizons. The \emph{CPD} panel mirrors this behavior: \texttt{combined} attains the smallest deviations from nominal over most horizons, \texttt{expanding} incurs higher CPD in the mid-range, and \texttt{rolling} shows pronounced late-horizon spikes. In the \emph{interval score} panel, all methods yield very low values at short horizons, but large penalties appear at the longest horizons; the increase is most severe for \texttt{rolling}, moderate for \texttt{expanding}, and smallest for \texttt{combined}, indicating that averaging dampens the volatility of single-scheme intervals without sacrificing earlier-horizon sharpness. Overall, for Japan, \texttt{combined} offers the most balanced performance across horizons; \texttt{expanding} provides sharp intervals but incurs a mid-horizon calibration cost; and \texttt{rolling} is comparatively unstable at the horizon end.

The ASFR results indicate that no single scheme dominates uniformly across horizons and countries. Nevertheless, \texttt{expanding} most often wins on the CPD and interval score in cross-sectional summaries, consistent with fertility dynamics benefiting from a longer historical context. The equal-weight \texttt{combined} method remains a robust hedge, tracking the stronger single scheme while reducing horizon-end volatility and mitigating extreme spikes that can arise within single-scheme intervals.

\section{Conclusion}\label{sec:6}

We address a straightforward and practical question posed in demographic forecasting: should models be estimated using a rolling window, an expanding window, or both? We explored this question within a unified functional time-series framework, focusing on two key components of population dynamics: age-specific mortality and fertility. The empirical design included multiple countries from high-quality databases, considered both sexes in mortality analyses, and assessed forecasts over a wide range of time horizons. Point forecasts were evaluated using RMSFE and MAFE, while interval forecasts were assessed via ECP and its deviation from the nominal level (CPD), along with a proper interval score to address the trade-off between calibration and sharpness. Our analysis emphasized \textbf{horizon-wise} evidence, aggregated cross-country patterns through heatmaps, and representative country panels to link aggregate findings with the behavior of individual series.

The results indicate that no single fitting scheme consistently outperforms the others in all contexts. For mortality point forecasts, expanding windows and an \textit{equal-weight combination} are often preferred for females, while rolling windows and the equal weight combination tend to be favored for males. This pattern aligns with the notion that female mortality trends are relatively stable and therefore benefit from a longer historical context. In contrast, males tend to reward responsiveness to recent information more often. In terms of mortality interval forecasts, cross-country summaries show that rolling windows frequently achieve the best calibration and lowest interval scores across horizons, although the combination method is close behind. Individual-country examples suggest that averaging can be particularly beneficial when single-scheme intervals are unstable at longer horizons; in such cases, the combination approach leverages the strengths of the more robust scheme while mitigating spikes at the end of the horizon.

In the realm of fertility, the evidence is more definitive. For point forecasting, expanding windows yield the highest mean win counts under RMSFE, whereas the equal-weight combination often slightly outperforms expanding under MAFE; rolling windows rarely excel. In interval forecasting, expanding windows typically provide the best calibration and the most favorable balance between calibration and sharpness across various horizons and countries. These findings imply that fertility dynamics develop over extended demographic transitions and reflect cohort postponement and tempo effects. Using the complete historical record benefits this analysis, and the combination method offers stable performance across horizons.

In summary, these results provide a clear answer to the question posed in the title. When there is little prior knowledge about structural stability or recent structural breaks, it is advisable to use both approaches via an equal-weight combination. This combination method is tune-free, easy to implement, seldom inferior, and often among the best performers across different horizons and datasets. The analysis provides clear guidance based on component and sex. For mortality, with a focus on point accuracy, females should prioritize expanding or combined methods, while males should focus on rolling or combined methods. When assessing mortality interval accuracy, rolling generally offers the best calibration and overall performance, and combined serves as a useful stabilizer. For fertility forecasts, expanding should be the default choice for both point and interval forecasts, and combining should be used to minimize volatility at the end of the forecast period.

The practical implications of these findings are significant for policy and planning. Mortality and fertility forecasts directly influence indicators such as life expectancy, old-age dependency ratios, school-age cohort sizes, and labor force projections. Because the most effective method can vary according to the forecast horizon, sex, and component, horizon-wise reporting and cross-country comparisons should accompany the headline results. The combined strategy is particularly beneficial for forecasts that inform decisions sensitive to tail risk or calibration, such as hospital capacity planning, pension sustainability assessments, and education infrastructure scheduling, as it minimizes the risk of horizon-specific failures without adding complexity to the tuning process.

Building on the findings of the current study, several key areas warrant further investigation; we outline four below.
\begin{inparaenum}[1)]
\item Transitioning from descriptive win heatmaps to formal horizon-wise tests of equal predictive ability would help quantify sampling uncertainty in cross-country comparisons. 
\item Rather than focusing solely on a single scheme (either rolling or expanding), future research should explicitly include the equal-weight combination as a baseline. 
\item It should explore regime-switching mechanisms that can alternate between rolling and expanding methods during structural breaks. 
\item All innovations should be compared against the established, tuning-free equal-weight baseline to ensure operational robustness and transparency.
\end{inparaenum}

In conclusion, the evidence suggests that the choice between rolling and expanding windows depends on the context. However, a simple combination of the two methods offers a robust and operationally attractive default option. This conclusion aligns with the broader goal of demographic forecasting for social planning: to create reliable, transparent, and well-calibrated projections that are stable across different horizons, populations, and indicators. At the same time, these projections must allow enough adaptability to accommodate varying temporal dynamics. For reproducibility, the computer \Rlogo \ code is available at an \href{https://github.com/csz16/Combined_window}{Github repository}.





\newpage

\section*{\textcolor{purple}{Appendix: Sensitivity analysis}}\label{sec:appendix}

This appendix reports a sensitivity analysis using two classical demographic forecasting models: the Lee--Carter (LC) model and an age--period--cohort (APC) extension. The objective is to examine whether the relative performance of \texttt{Rolling}, \texttt{Expanding}, and the equal-weight \texttt{Combined} scheme in Section~\ref{sec:4} is robust to replacing the functional time-series model in Section~\ref{sec:3} with lower-dimensional demographic benchmarks. 

Let $\Y_t(u_i)$ denote the transformed age-specific mortality or fertility rate at age $u_i$ and calendar year $t$, using the same preprocessing, training/test split, and accuracy criteria as described in Sections~\ref{sec:2} and~\ref{sec:4}. The LC model represents the age-specific surface by one age profile and one period index,
\begin{equation}
\Y_t(u_i) = a(u_i) + b(u_i)\kappa_t + \varepsilon_t(u_i),
\label{eq:app-lc}
\end{equation}
where $a(u_i)$ is the average age pattern, $b(u_i)$ measures the sensitivity of age $u_i$ to the period index $\kappa_t$, and $\varepsilon_t(u_i)$ is the residual term. The usual identification constraints, such as $\sum_i b(u_i)=1$ and $\sum_t \kappa_t=0$, are imposed to make the decomposition unique. The LC model was introduced for mortality forecasting by \citet{LC92} and was later adapted to fertility forecasting by \citet{Lee93}; see also \citet{SZG+14} for fertility applications.

The APC model augments the LC structure with a cohort component. It can be written as
\begin{equation}
\Y_t(u_i) = a(u_i) + b(u_i)\kappa_t + d(u_i)\gamma_{t-u_i} + \varepsilon_t(u_i),
\label{eq:app-apc}
\end{equation}
where $\gamma_{t-u_i}$ denotes the cohort index and $d(u_i)$ measures the age-specific loading of the cohort effect. This extension allows the fitted surface to capture persistent cohort-driven deviations from the period trend, which is important when the observed age profiles are affected by birth-cohort or death-cohort regularities. The APC specification follows the Lee--Carter cohort-extension literature, particularly \citet{RH06} and related developments reviewed in \citet{HR09}. For both LC and APC, the period and cohort indexes are extrapolated by univariate time-series models to obtain $h$-step-ahead forecasts.

For each benchmark model, we apply the same three fitting schemes used in the main analysis. The \texttt{Rolling} scheme estimates the model on a fixed moving fitting window, the \texttt{Expanding} scheme estimates the model on all data available up to the forecast origin, and the \texttt{Combined} scheme uses the equal-weight average in Equation~\eqref{eq:equal-combination}. Point forecast accuracy is evaluated by RMSFE and MAFE for horizons $h=1,2,\dots,20$, as defined in Section~\ref{sec:4.1}. In the heatmaps below, each cell reports the number of countries for which a method attains the lowest error at a given horizon. Hence, mortality rows sum to 23 countries, while ASFR rows sum to 16 countries.

\subsection*{Comparison of point forecast accuracy}\label{sec:appendix_1}

\begin{figure}[!htb]
\centering
{\includegraphics[width=0.95\textwidth]{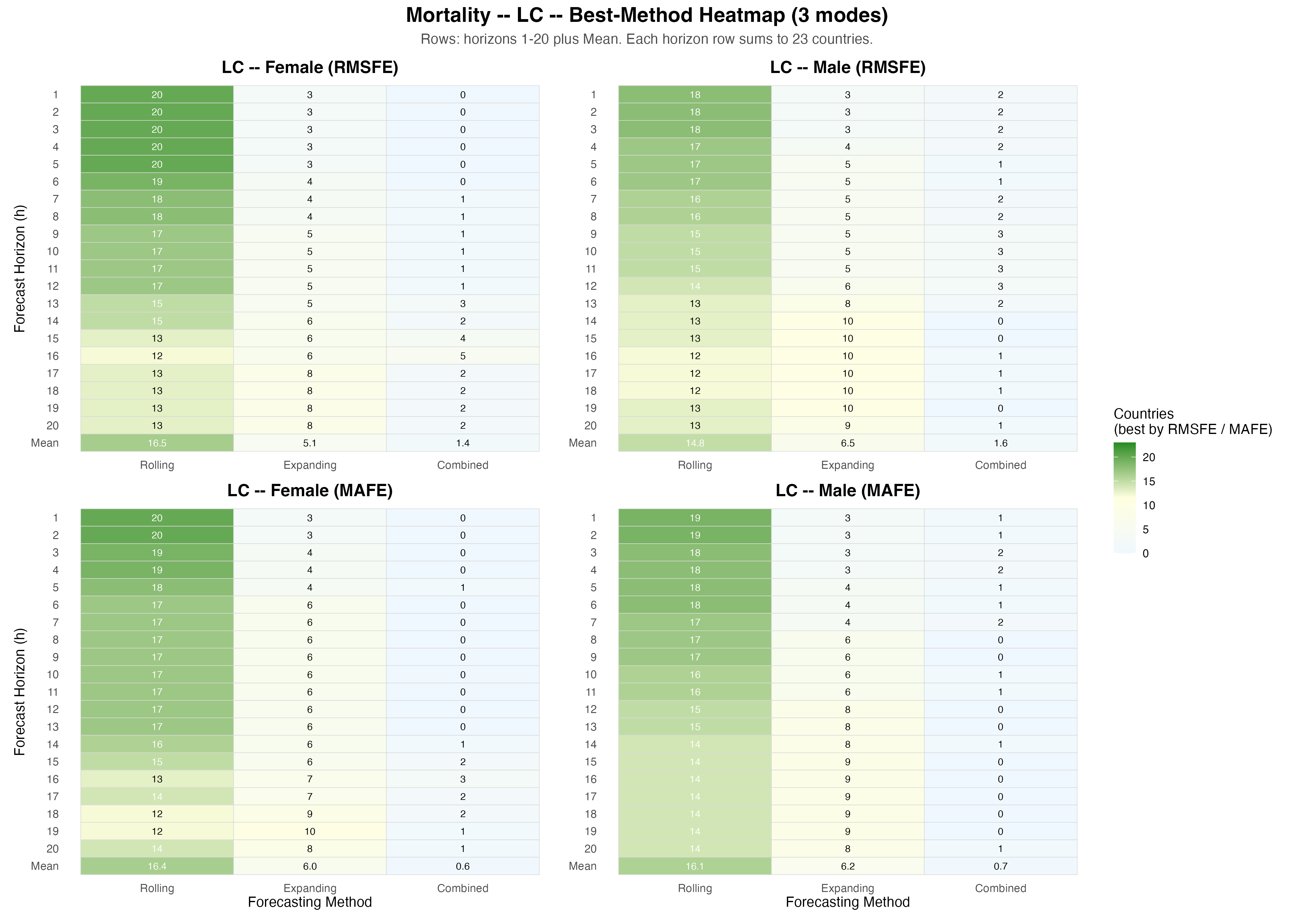}}
\caption{\small{Horizon-specific ranking heatmaps for the female and male mortality under the LC model, measured by the RMSFE and MAFE, for the 23 countries considered.}}\label{fig:app-mortality-lc-heatmap}
\end{figure}

\begin{figure}[!htb]
\centering
{\includegraphics[width=0.86\textwidth]{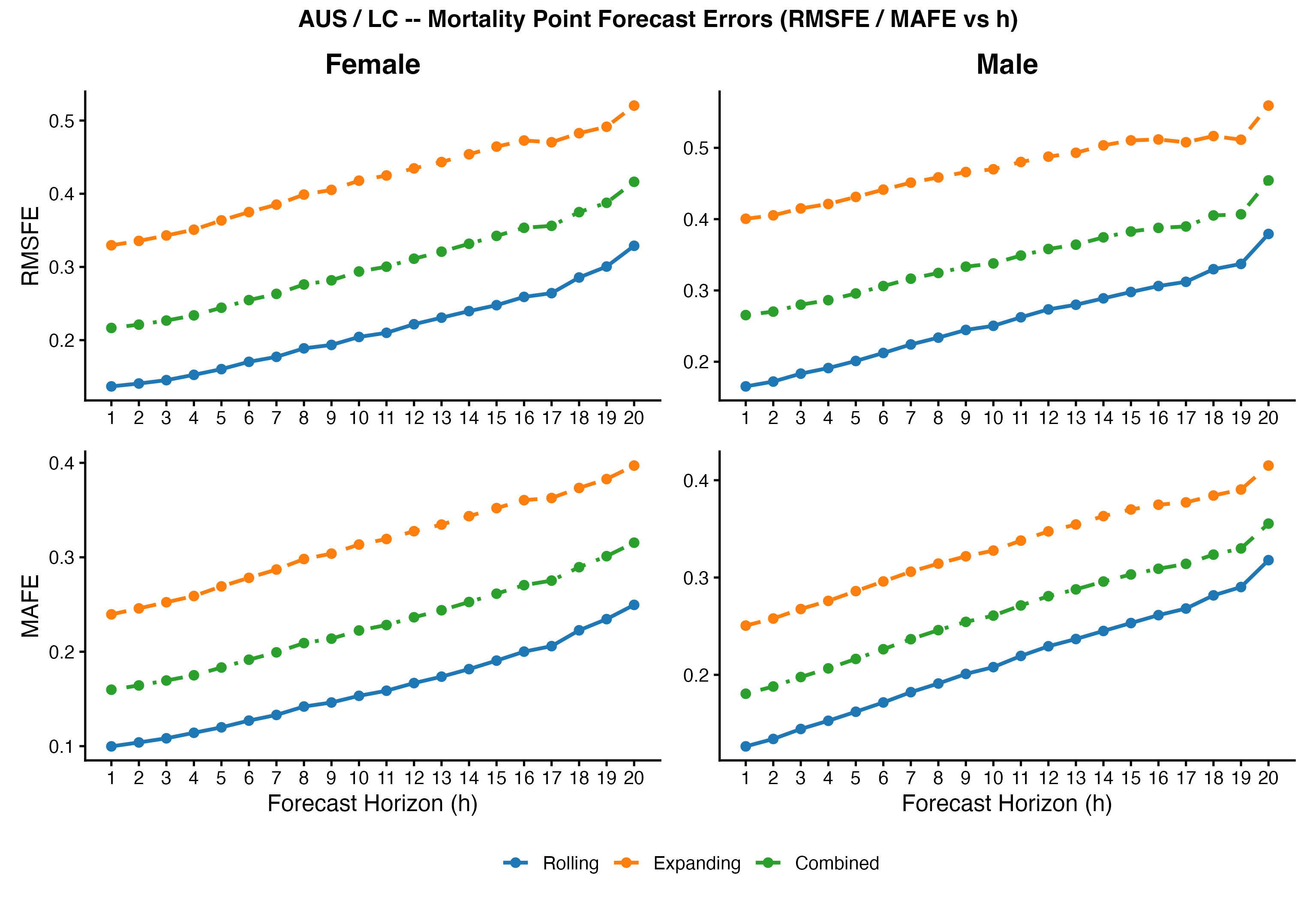}}
\caption{\small{AUS mortality — LC Point Forecast Errors (RMSFE/MAFE vs Forecast Horizon ($h$)).}}\label{fig:app-mortality-lc-aus}
\end{figure}

\begin{figure}[!htb]
\centering
{\includegraphics[width=0.86\textwidth]{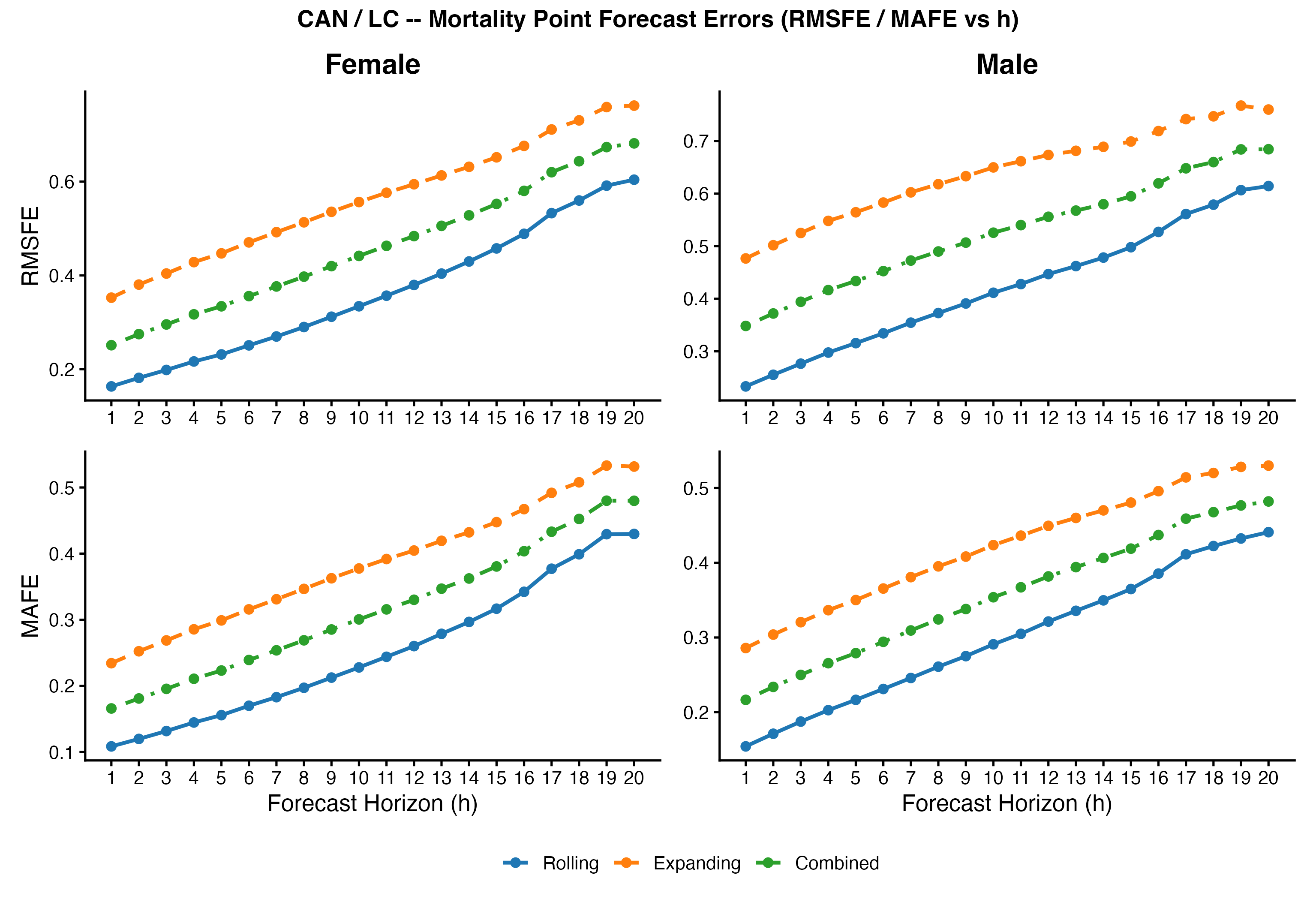}}
\caption{\small{CAN mortality — LC point forecast errors (RMSFE/MAFE vs forecast horizon ($h$)).}}\label{fig:app-mortality-lc-can}
\end{figure}

Figures~\ref{fig:app-mortality-lc-heatmap}--\ref{fig:app-mortality-lc-can} report the LC mortality sensitivity results. The heatmaps indicate a strong overall advantage for \texttt{Rolling}. For female mortality, \texttt{Rolling} wins on average in 16.5 countries under RMSFE and 16.4 countries under MAFE. For male mortality, the corresponding average win counts are 14.8 and 16.1. \texttt{Expanding} is the main competitor, particularly at some medium and longer horizons, but its average win counts remain well below those of \texttt{Rolling}. \texttt{Combined} rarely wins the heat-map rankings, with mean counts ranging from 0.6 to 1.6 depending on sex and metric.

The Australian and Canadian LC panels support the conclusion from heatmaps. Across both sexes and both accuracy measures, \texttt{Rolling} is the lowest or nearly lowest curve over the full horizon range. \texttt{Expanding} is usually the highest-error curve, and \texttt{Combined} occupies an intermediate position. The separation is especially clear in the Australian panels and in Canadian female mortality. These results suggest that the LC mortality benchmark benefits from recent-information weighting, even though the expanding window remains competitive in a subset of cross-country long-horizon comparisons.

\begin{figure}[!htb]
\centering
{\includegraphics[width=\textwidth]{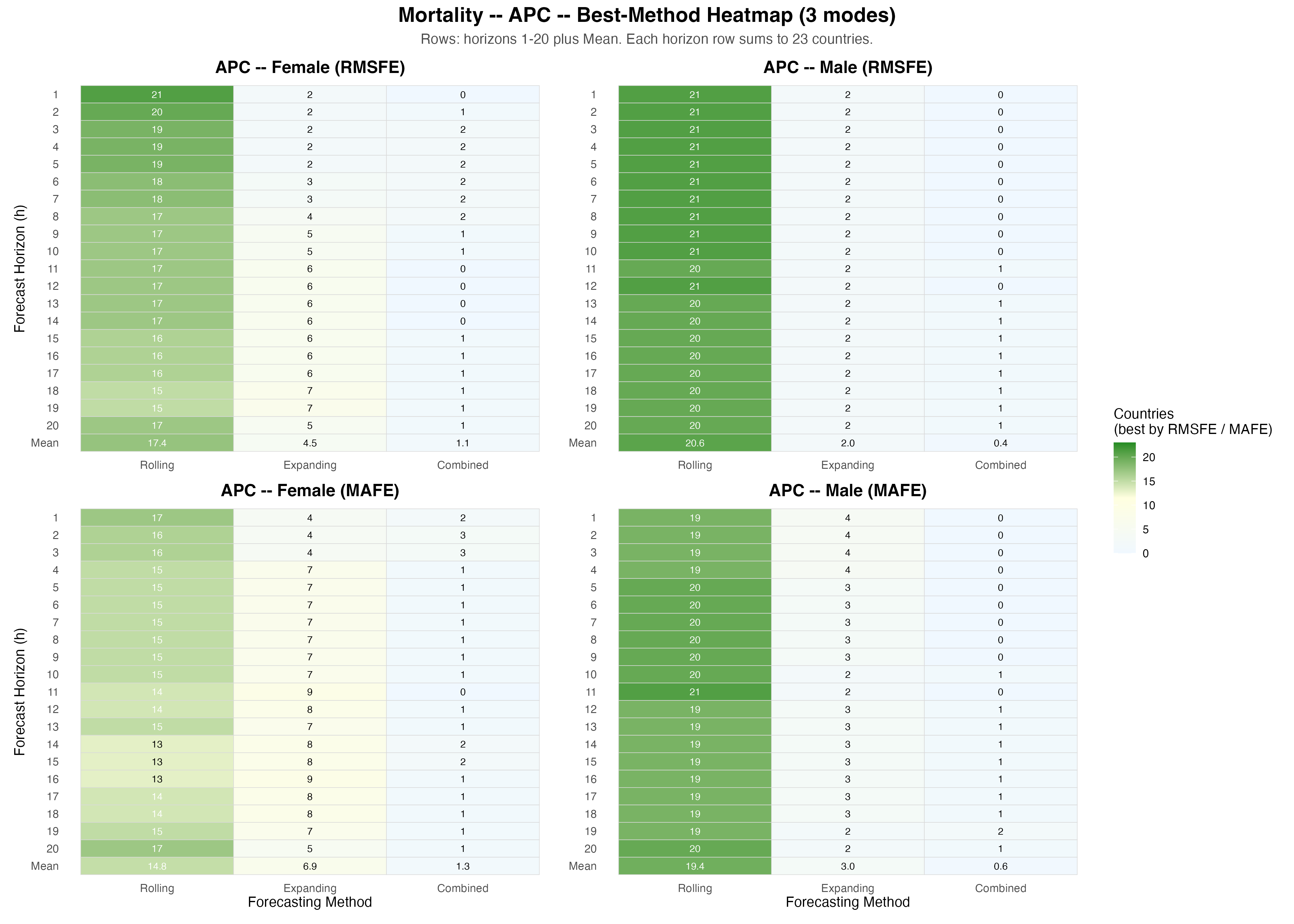}}
\caption{\small{Horizon-specific ranking heatmaps for the female and male mortality under the APC model, measured by the RMSFE and MAFE, for the 23 countries considered.}}\label{fig:app-mortality-apc-heatmap}
\end{figure}

\begin{figure}[!htb]
\centering
{\includegraphics[width=0.86\textwidth]{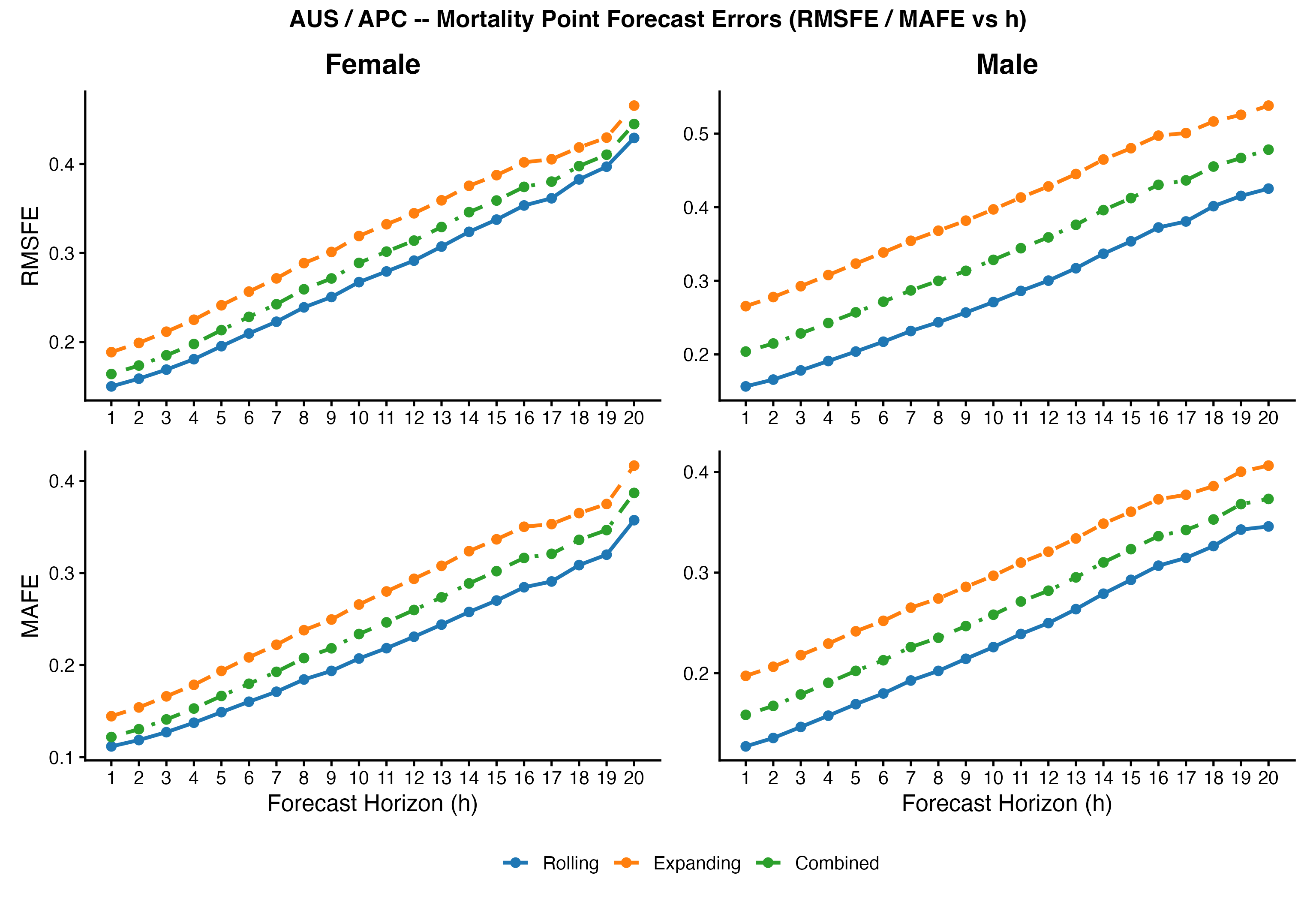}}
\caption{\small{AUS mortality — APC point forecast errors (RMSFE/MAFE vs forecast horizon ($h$)).}}\label{fig:app-mortality-apc-aus}
\end{figure}

\begin{figure}[!htb]
\centering
{\includegraphics[width=0.86\textwidth]{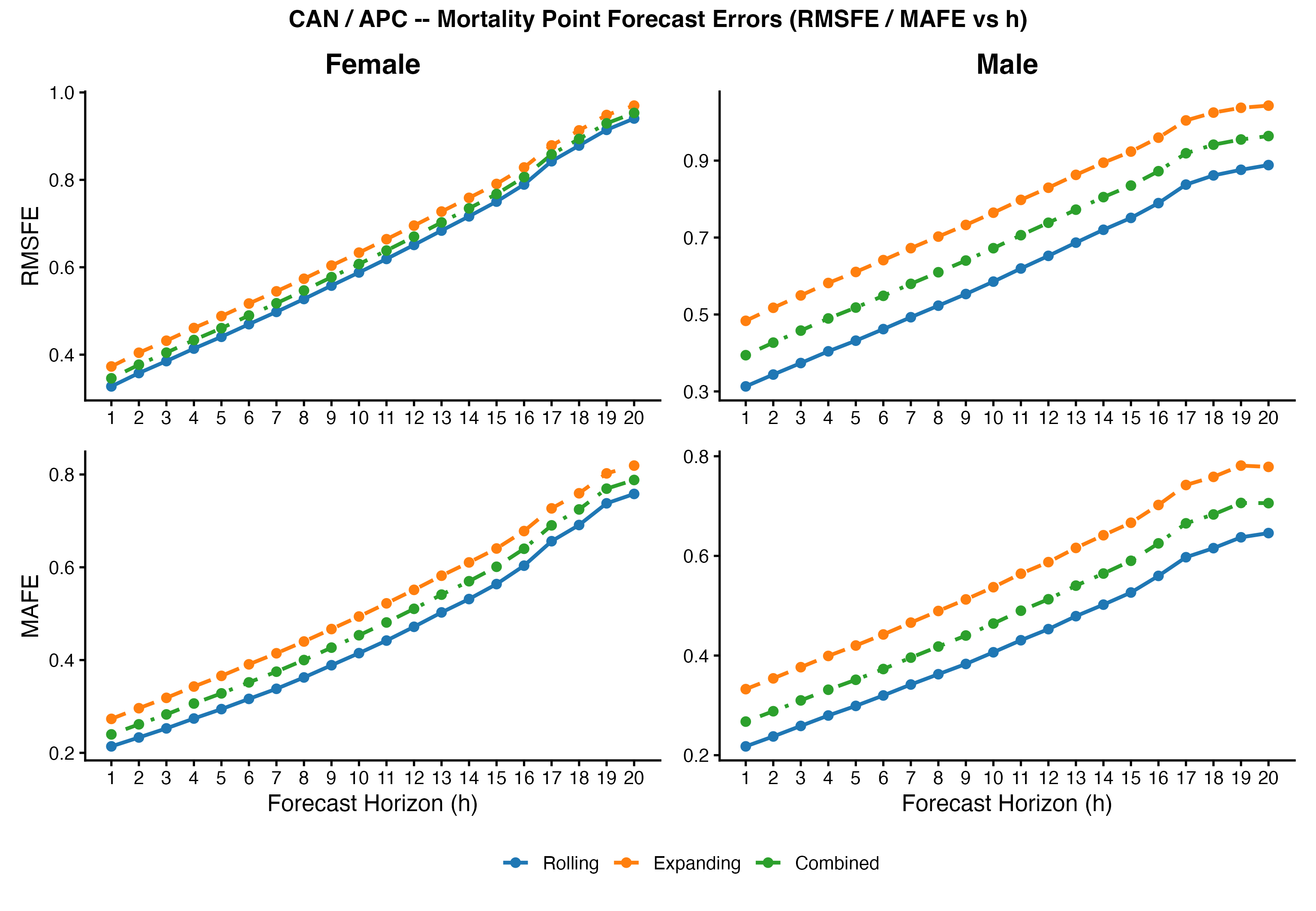}}
\caption{\small{CAN mortality — APC point forecast errors (RMSFE/MAFE vs forecast horizon ($h$))}}
\label{fig:app-mortality-apc-can}
\end{figure}

Figures~\ref{fig:app-mortality-apc-heatmap}--\ref{fig:app-mortality-apc-can} demonstrate the mortality results from the APC model. The dominance of \texttt{Rolling} becomes even more pronounced for mortality under APC. For RMSFE, \texttt{Rolling} wins on average in 17.4 countries for females and 20.6 countries for males. For MAFE, it wins in 14.8 countries for females and 19.4 countries for males. \texttt{Expanding} remains a secondary competitor, especially in the female MAFE panel, but it does not challenge the overall ranking. The \texttt{Combined} method records very few wins across panels.

The Australian and Canadian APC line plots are consistent with this cross-country ranking. In both countries, \texttt{Rolling} has the lowest error curves for most horizons, \texttt{Expanding} has the highest, and \texttt{Combined} generally tracks the midpoint between the two single-window methods. The Canadian female APC curves are comparatively close, but the ordering remains stable. This pattern indicates that, after adding a cohort term, the mortality forecasts place even greater value on the most recent fitting window, particularly for male mortality.

\begin{figure}[!htb]
\centering
{\includegraphics[width=0.95\textwidth]{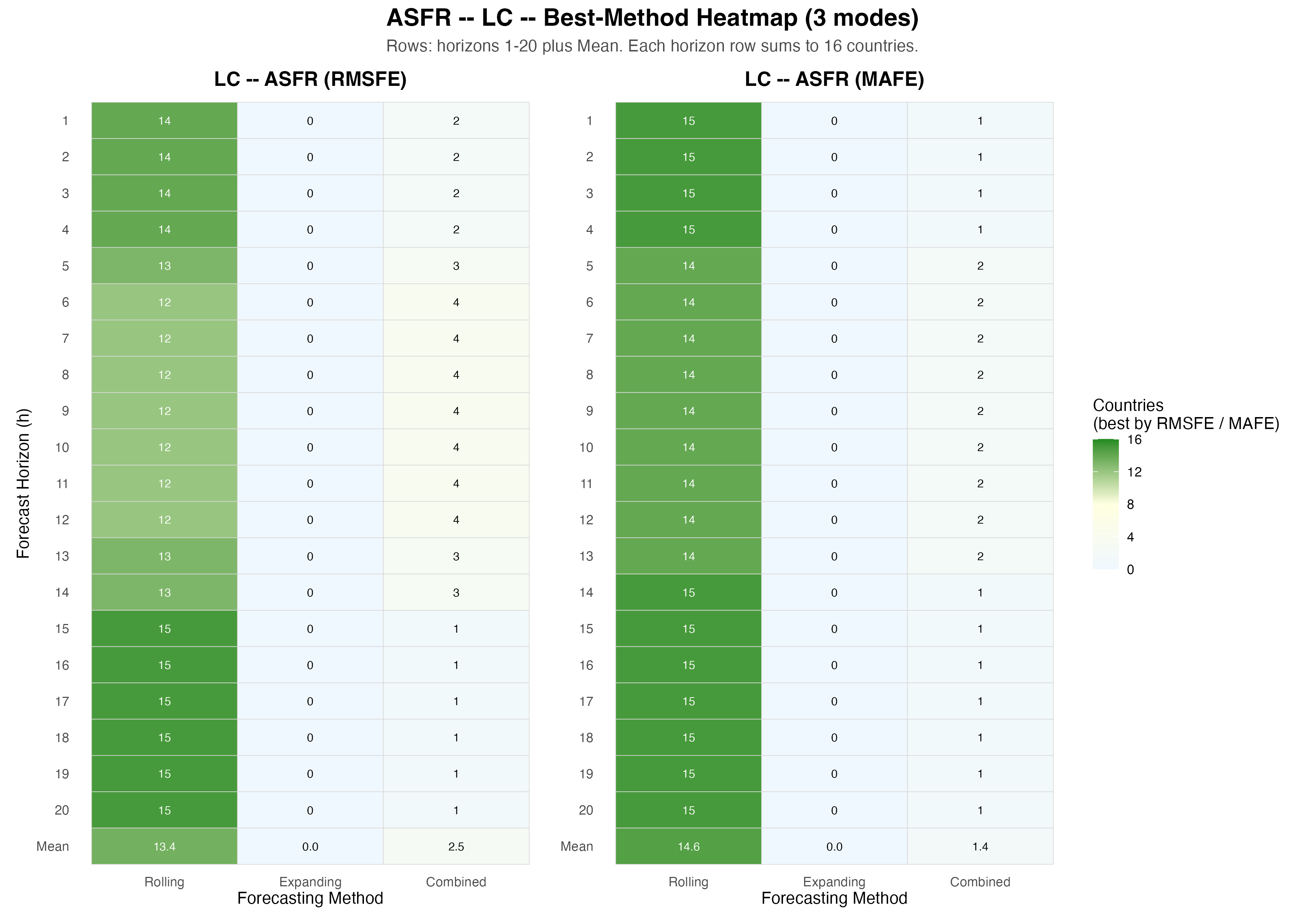}}
\caption{\small{Horizon-specific ranking heatmaps for the ASFR under the LC model, measured by the RMSFE and MAFE, for the 16 countries considered.}}
\label{fig:app-asfr-lc-heatmap}
\end{figure}

\begin{figure}[!htb]
\centering
{\includegraphics[width=\textwidth]{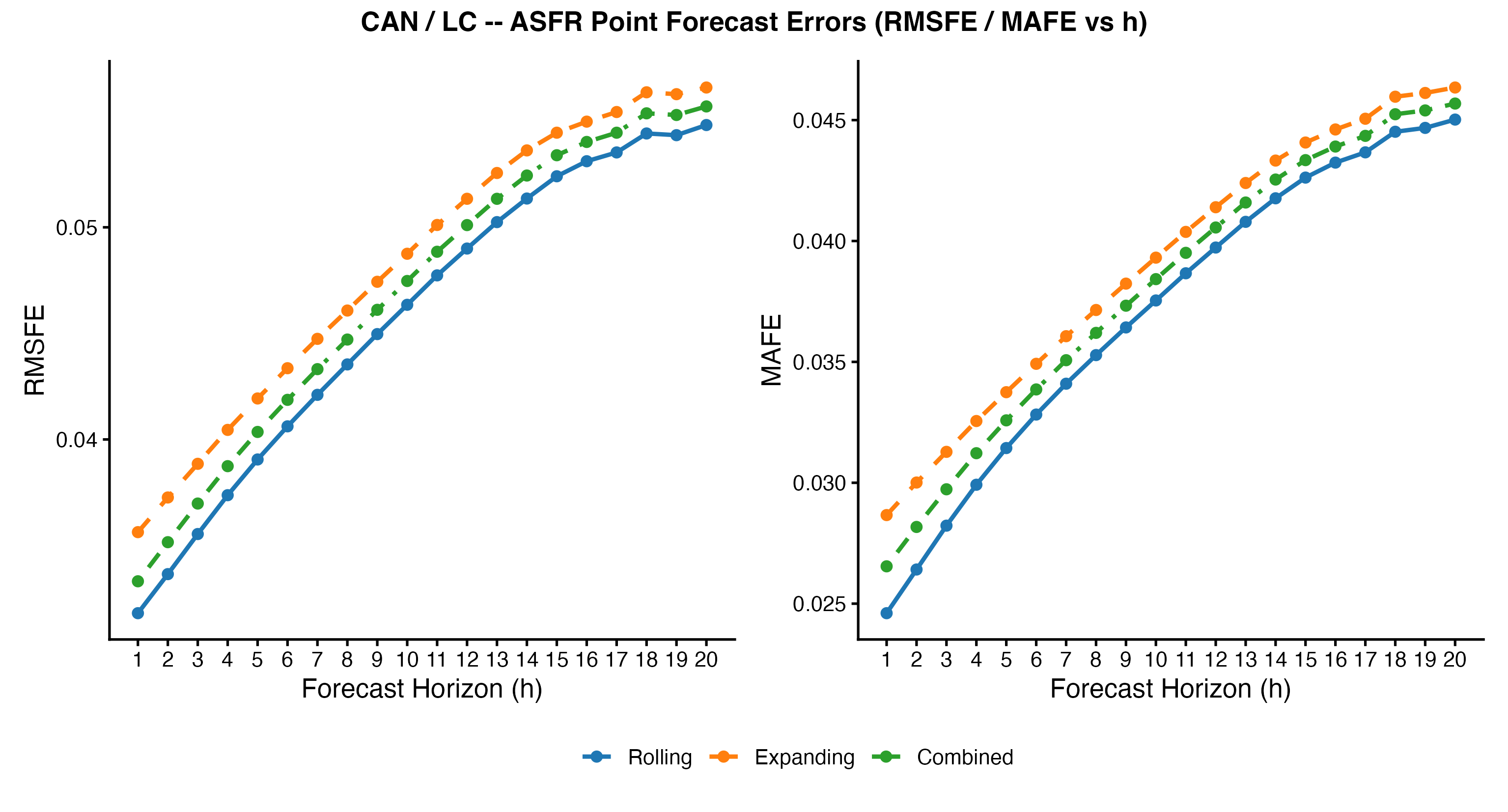}}
\caption{\small{CAN ASFR — LC Point Forecast Errors (RMSFE/MAFE vs Forecast Horizon ($h$)).}}
\label{fig:app-asfr-lc-can}
\end{figure}

\begin{figure}[!htb]
\centering
{\includegraphics[width=\textwidth]{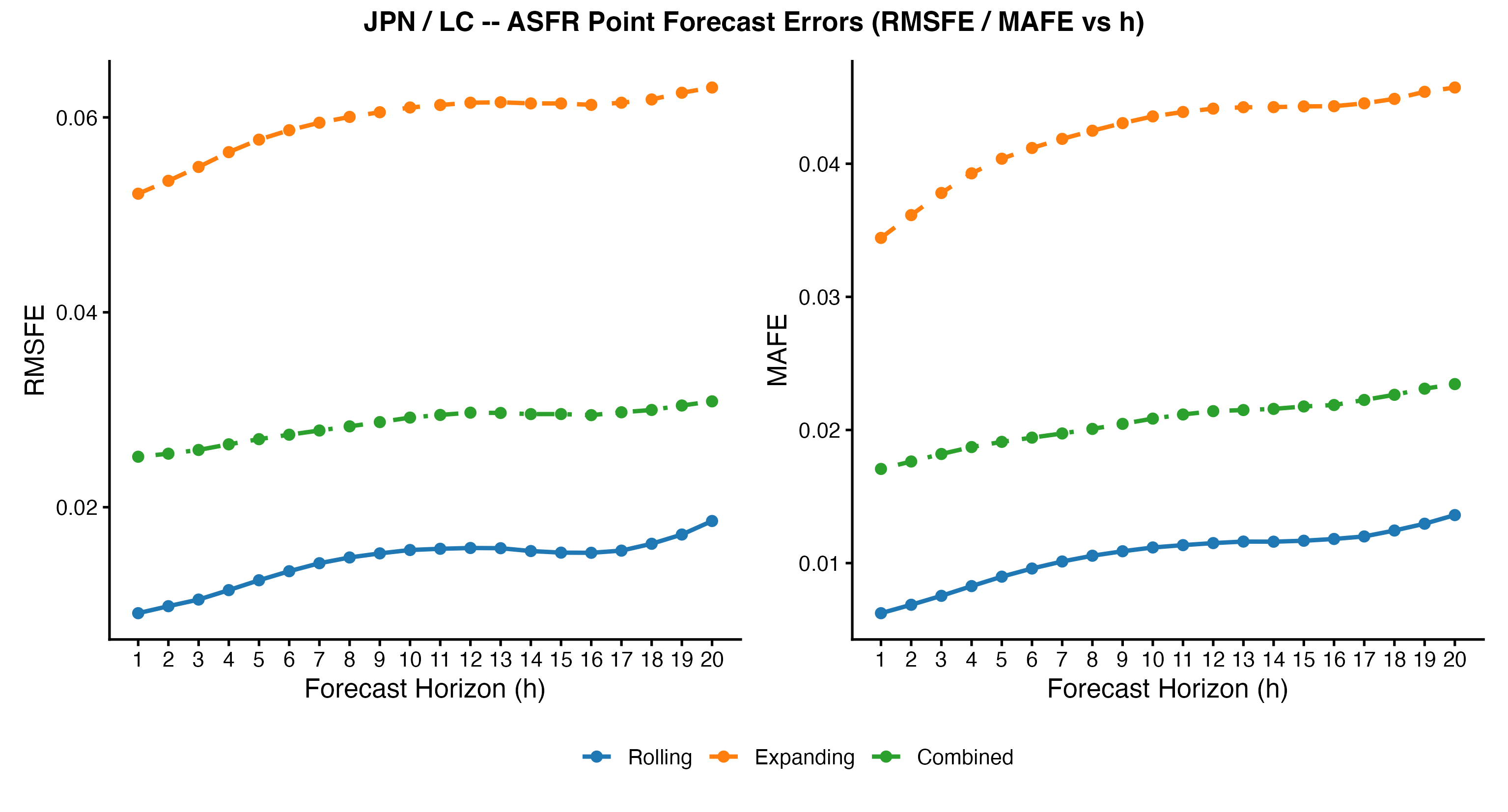}}
\caption{\small{JPN ASFR — LC point forecast errors (RMSFE/MAFE vs forecast horizon ($h$)).}}
\label{fig:app-asfr-lc-jpn}
\end{figure}

Figures~\ref{fig:app-asfr-lc-heatmap}--\ref{fig:app-asfr-lc-jpn} summarize the LC results for ASFR. The cross-country evidence strongly favors \texttt{Rolling}. In the RMSFE heatmap, \texttt{Rolling} wins, on average, 13.4 out of 16 countries across horizons, while \texttt{Combined} wins 2.5 and \texttt{Expanding} wins none. The MAFE heatmap is even more concentrated: \texttt{Rolling} wins, on average, 14.6 countries, \texttt{Combined} wins 1.4, and \texttt{Expanding} again records no average wins. The country-level panels for Canada and Japan show the same ordering. Under the LC specification, \texttt{Rolling} is consistently the lowest-error curve, \texttt{Expanding} is the highest-error curve, and \texttt{Combined} lies between them for both RMSFE and MAFE. This indicates that when ASFR is forecast with the one-factor LC structure, responsiveness to recent information is more valuable than using the full historical sample.

\begin{figure}[!htb]
\centering
{\includegraphics[width=0.92\textwidth]{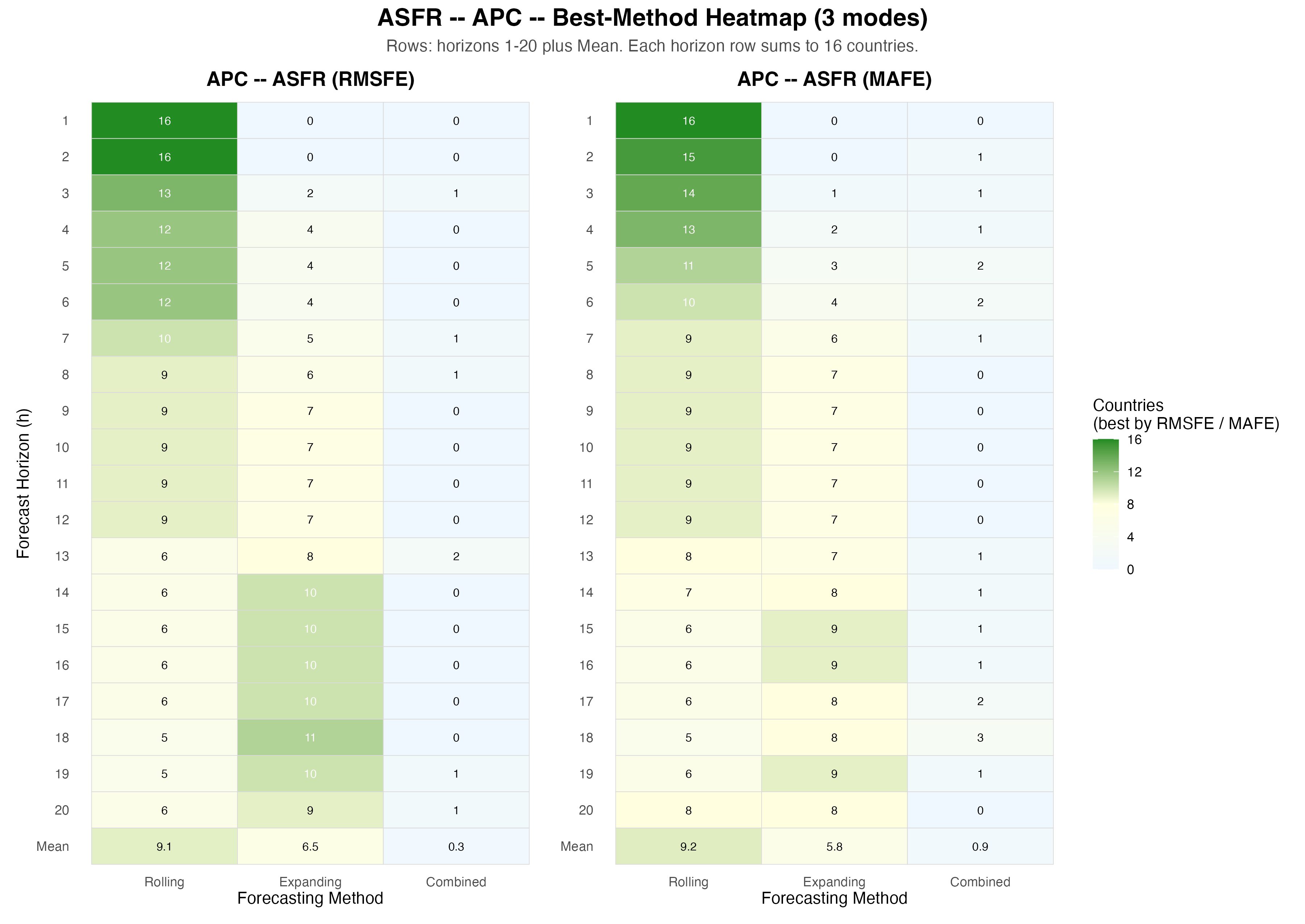}}
\caption{\small{Horizon-specific ranking heatmaps for the ASFR under the APC model, measured by the RMSFE and MAFE, for the 16 countries considered.}}
\label{fig:app-asfr-apc-heatmap}
\end{figure}

\begin{figure}[!htb]
\centering
{\includegraphics[width=0.92\textwidth]{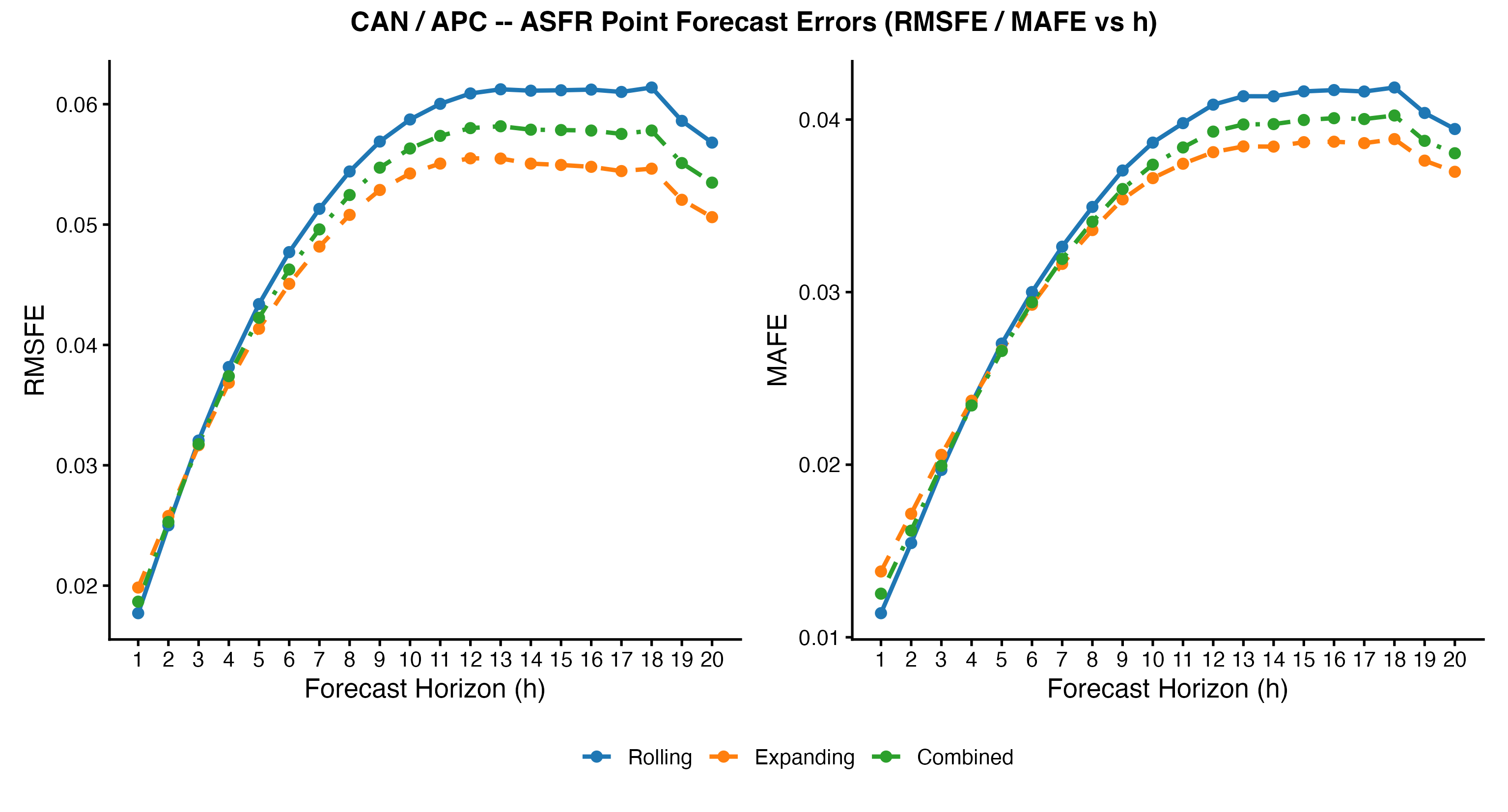}}
\caption{\small{CAN ASFR — APC point forecast errors (RMSFE/MAFE vs forecast horizon ($h$)).}}
\label{fig:app-asfr-apc-can}
\end{figure}

\begin{figure}[!htb]
\centering
{\includegraphics[width=0.92\textwidth]{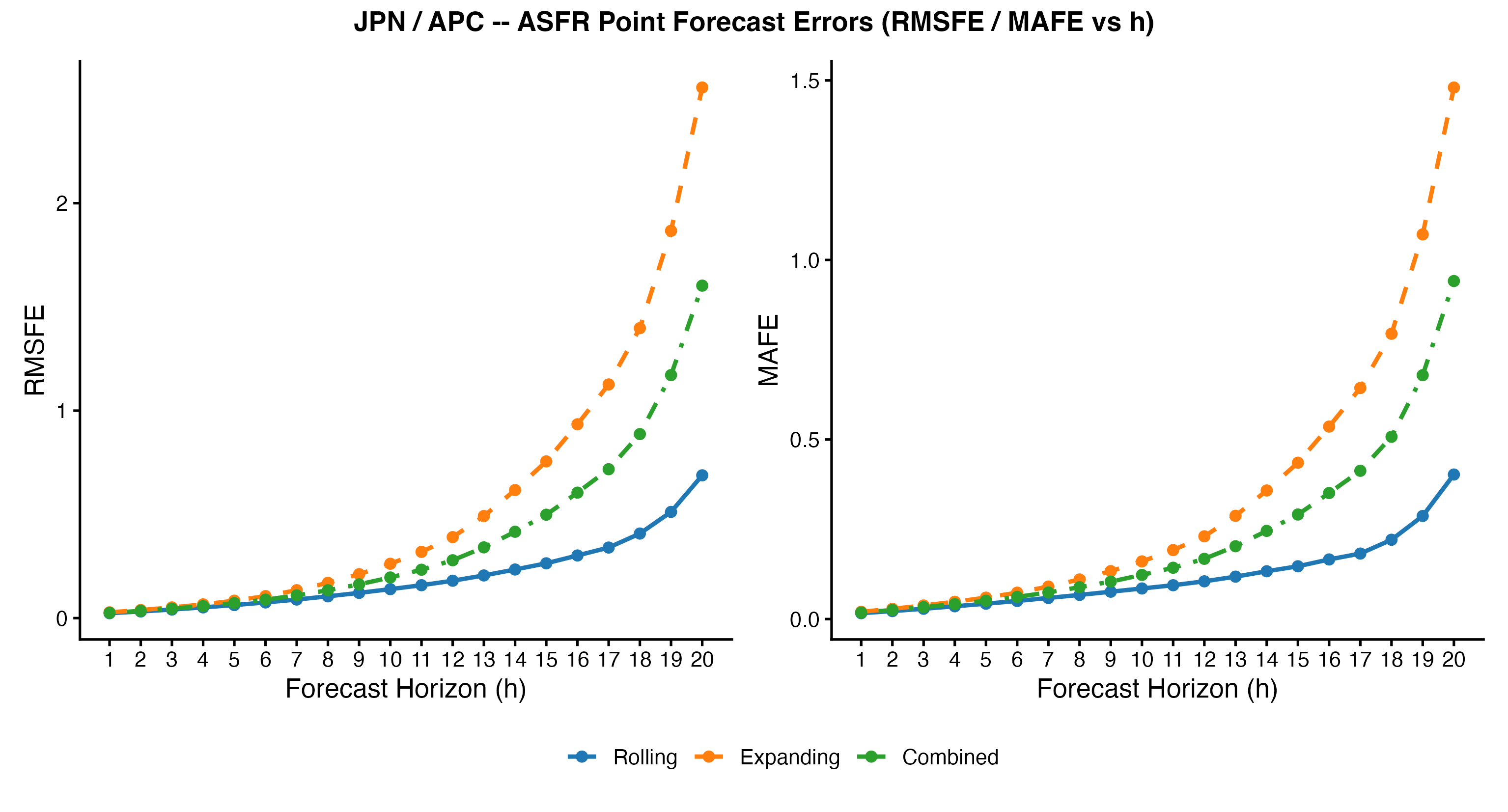}}
\caption{\small{JPN ASFR — APC point forecast errors (RMSFE/MAFE vs forecast horizon ($h$)).}}
\label{fig:app-asfr-apc-jpn}
\end{figure}

Figures~\ref{fig:app-asfr-apc-heatmap}--\ref{fig:app-asfr-apc-jpn} present the corresponding APC results for ASFR. Compared with LC, the APC model produces a more horizon-dependent ranking. \texttt{Rolling} remains the leading method on average, winning 9.1 countries under RMSFE and 9.2 countries under MAFE, while \texttt{Expanding} wins 6.5 and 5.8 countries, respectively. The heatmaps show that \texttt{Rolling} dominates the shortest horizons, especially $h=1$ and $h=2$, whereas \texttt{Expanding} gains more wins at longer horizons. The \texttt{Combined} scheme has relatively few heat-map wins under APC, with average counts of fewer than one country for both metrics.

The two country examples clarify why the APC heatmap is less uniform than the LC heatmap. For Canada, \texttt{Expanding} is generally more accurate at medium and long horizons, with \texttt{Combined} remaining close and \texttt{Rolling} becoming the least accurate method toward the end of the horizon range. For Japan, the opposite pattern occurs: \texttt{Rolling} is stable and clearly the best, while \texttt{Expanding} and \texttt{Combined} increase sharply at long horizons. Thus, APC-ASFR forecasts are more sensitive to country-specific cohort dynamics and to the stability of the estimated cohort component. The equal-weight average dampens the extreme long-horizon behavior of \texttt{Expanding} in Japan, but it does not frequently become the best method in the cross-country rankings.

The sensitivity analysis shows that the best-fitting window scheme depends not only on the demographic component and forecast horizon, but also on the forecasting model. Under the LC and APC benchmarks, \texttt{Rolling} is the dominant method for mortality point forecasts and for LC-based ASFR point forecasts. The main exception is APC-based ASFR, where \texttt{Rolling} remains the average winner, but \texttt{Expanding} gains a substantial share of wins at longer horizons and can be clearly preferable for countries such as Canada. Japan illustrates the opposite case, where long-horizon APC forecasts based on the expanding window become unstable, and \texttt{Rolling} is strongly preferred.

Compared with the functional time-series results in Section~\ref{sec:4}, the LC/APC sensitivity analysis gives a larger role to recent-information weighting. This is consistent with the more restrictive structure of LC and APC models: when the age-time surface is summarized into one period component and possibly one cohort component, fitting the model to the most recent observations can reduce the influence of older regimes that no longer reflect current demographic dynamics. The \texttt{Combined} method remains useful as a simple hedge, particularly when the single-window forecasts diverge at long horizons, but in these LC/APC point forecast experiments, it is less often the direct winner than in the main functional time-series analysis. Overall, the appendix reinforces the need to report horizon-specific and model-specific evidence rather than relying on a single aggregate ranking.

\subsection*{Comparison of interval forecast accuracy}\label{sec:appendix_2}

This section extends the LC/APC sensitivity analysis from point forecasts to interval forecasts. The interval forecasts are constructed using the same validated standard-deviation framework as in Section~\ref{sec:5.1}. Calibration is assessed by the empirical coverage probability (ECP), with the nominal target fixed at $0.80$, and sharpness-adjusted interval performance is assessed by the interval score of \citet{GR07}. A smaller absolute deviation of ECP from $0.80$ indicates better calibration, while a smaller interval score indicates a better balance between calibration and interval width. Because the interval procedure requires a validation segment for estimating the standard deviation adjustment, the interval-forecast comparison is reported for horizons $h=1,\ldots,19$.

In the heatmaps below, the ECP winner is the method whose ECP is closest to the nominal target, while the interval-score winner is the method with the smallest interval score. For ASFR, each horizon row sums to 16 countries; for mortality, each horizon row sums to 23 countries. Fractional entries occur when ties are split across methods. The country-level interval scores are reported in Table~\ref{tab:interval_score_mortality} for the Australian and Canadian mortality examples and in Table~\ref{tab:interval_score_asfr} for ASFR.

\tabcolsep 0.1in
\renewcommand{\arraystretch}{0.79}
\begin{longtable}{@{}lrrrrrrrrrrrr@{}}
\caption{Interval scores for mortality interval forecasts}
\label{tab:interval_score_mortality}\\
\toprule
& \multicolumn{4}{c}{Rolling} 
& \multicolumn{4}{c}{Expanding} 
& \multicolumn{4}{c}{Combined} \\
\cmidrule(lr){2-5}\cmidrule(lr){6-9}\cmidrule(lr){10-13}
& \multicolumn{2}{c}{AUS} & \multicolumn{2}{c}{CAN} & \multicolumn{2}{c}{AUS} & \multicolumn{2}{c}{CAN} \\
$h$ & F & M & F & M & F & M & F & M & F & M & F & M \\
\midrule
\endfirsthead
\caption[]{Interval scores for mortality interval forecasts (continued)}\\
\toprule
& \multicolumn{4}{c}{Rolling} 
& \multicolumn{4}{c}{Expanding} 
& \multicolumn{4}{c}{Combined} \\
\cmidrule(lr){2-5}\cmidrule(lr){6-9}\cmidrule(lr){10-13}
 & \multicolumn{2}{c}{AUS} & \multicolumn{2}{c}{CAN} & \multicolumn{2}{c}{AUS} & \multicolumn{2}{c}{CAN} \\
$h$ & F & M & F & M & F & M & F & M & F & M & F & M \\
\midrule
\endhead
\midrule
\multicolumn{13}{r}{Continued on next page}
\endfoot
\bottomrule
\endlastfoot
\multicolumn{13}{l}{\hspace{-.1in}{\textit{Panel A: LC}}} \\
1 & 0.50 & 0.54 & 0.51 & 0.63 & 0.95 & 0.83 & 1.31 & 1.14 & 0.58 & 0.66 & 0.85 & 0.82 \\
2 & 0.53 & 0.60 & 0.58 & 0.70 & 0.95 & 0.90 & 1.42 & 1.29 & 0.61 & 0.73 & 0.93 & 0.91 \\
3 & 0.57 & 0.67 & 0.65 & 0.78 & 0.98 & 0.96 & 1.53 & 1.39 & 0.64 & 0.80 & 1.02 & 1.00 \\
4 & 0.60 & 0.76 & 0.71 & 0.87 & 0.96 & 1.04 & 1.66 & 1.53 & 0.65 & 0.88 & 1.11 & 1.10 \\
5 & 0.65 & 0.86 & 0.76 & 0.92 & 0.97 & 1.16 & 1.68 & 1.60 & 0.70 & 1.00 & 1.16 & 1.17 \\
6 & 0.70 & 0.94 & 0.84 & 1.02 & 1.10 & 1.18 & 1.80 & 1.65 & 0.77 & 1.05 & 1.26 & 1.25 \\
7 & 0.76 & 1.19 & 0.94 & 1.10 & 1.02 & 1.45 & 1.88 & 1.75 & 0.79 & 1.31 & 1.36 & 1.34 \\
8 & 0.77 & 1.16 & 1.01 & 1.18 & 1.12 & 1.40 & 1.91 & 1.82 & 0.85 & 1.27 & 1.41 & 1.43 \\
9 & 0.85 & 1.40 & 1.10 & 1.21 & 1.08 & 1.65 & 1.96 & 1.91 & 0.89 & 1.52 & 1.47 & 1.50 \\
10 & 0.80 & 1.56 & 1.24 & 1.19 & 1.16 & 1.69 & 2.04 & 1.90 & 0.86 & 1.62 & 1.60 & 1.48 \\
11 & 0.85 & 1.69 & 1.31 & 1.19 & 1.28 & 1.57 & 2.09 & 1.91 & 0.94 & 1.63 & 1.66 & 1.48 \\
12 & 0.99 & 2.12 & 1.36 & 1.21 & 1.26 & 1.76 & 2.16 & 1.98 & 1.06 & 1.93 & 1.72 & 1.53 \\
13 & 0.95 & 1.84 & 1.54 & 1.29 & 1.27 & 1.77 & 2.29 & 2.09 & 1.02 & 1.81 & 1.89 & 1.63 \\
14 & 1.05 & 2.62 & 1.86 & 1.43 & 1.34 & 2.36 & 2.61 & 2.23 & 1.10 & 2.49 & 2.21 & 1.76 \\
15 & 1.16 & 2.73 & 1.82 & 1.54 & 1.58 & 2.44 & 2.56 & 2.39 & 1.29 & 2.58 & 2.15 & 1.89 \\
16 & 1.20 & 2.75 & 1.64 & 1.64 & 1.43 & 2.62 & 2.40 & 2.42 & 1.27 & 2.68 & 1.97 & 1.95 \\
17 & 1.48 & 3.15 & 1.93 & 49.29 & 91.74 & 3.23 & 2.57 & 2.30 & 46.60 & 3.19 & 2.23 & 25.16 \\
18 & 1.34 & 2.95 & 2.31 & 9.41 & 1.66 & 2.89 & 2.94 & 2.43 & 1.46 & 2.91 & 2.61 & 5.51 \\
19 & 105.25 & 12.96 & 2.13 & 2.74 & 2.10 & 4.00 & 20.47 & 3.43 & 53.62 & 8.31 & 10.86 & 3.07 \\
\midrule
\multicolumn{13}{l}{\hspace{-.1in}{\textit{Panel B: APC}}} \\
1 & 0.47 & 0.49 & 1.06 & 1.16 & 0.55 & 0.78 & 1.21 & 1.82 & 0.48 & 0.58 & 1.07 & 1.46 \\
2 & 0.48 & 0.53 & 1.25 & 1.32 & 0.58 & 0.77 & 1.37 & 2.00 & 0.50 & 0.59 & 1.25 & 1.63 \\
3 & 0.50 & 0.54 & 1.36 & 1.46 & 0.60 & 0.83 & 1.49 & 2.15 & 0.52 & 0.62 & 1.37 & 1.78 \\
4 & 0.53 & 0.58 & 1.48 & 1.59 & 0.67 & 0.87 & 1.66 & 2.36 & 0.57 & 0.66 & 1.51 & 1.94 \\
5 & 0.56 & 0.60 & 1.62 & 1.72 & 0.70 & 0.89 & 1.80 & 2.49 & 0.60 & 0.69 & 1.64 & 2.07 \\
6 & 0.61 & 0.65 & 1.70 & 1.84 & 0.76 & 0.99 & 1.94 & 2.60 & 0.65 & 0.77 & 1.75 & 2.19 \\
7 & 0.64 & 0.69 & 1.82 & 1.94 & 0.81 & 0.99 & 2.03 & 2.75 & 0.69 & 0.79 & 1.85 & 2.30 \\
8 & 0.71 & 0.74 & 1.86 & 2.10 & 0.89 & 1.07 & 2.08 & 3.01 & 0.76 & 0.86 & 1.89 & 2.52 \\
9 & 0.75 & 0.76 & 1.86 & 2.16 & 0.95 & 1.10 & 2.12 & 3.09 & 0.82 & 0.88 & 1.90 & 2.58 \\
10 & 0.79 & 0.82 & 1.87 & 2.31 & 1.02 & 1.23 & 2.22 & 3.26 & 0.87 & 0.99 & 1.98 & 2.76 \\
11 & 0.86 & 0.87 & 1.79 & 2.49 & 1.09 & 1.34 & 2.22 & 3.37 & 0.94 & 1.05 & 1.94 & 2.91 \\
12 & 0.94 & 0.88 & 1.95 & 2.68 & 1.18 & 1.30 & 2.35 & 3.54 & 1.03 & 1.03 & 2.10 & 3.09 \\
13 & 0.97 & 1.04 & 2.13 & 2.84 & 1.23 & 1.47 & 2.59 & 3.80 & 1.05 & 1.21 & 2.31 & 3.30 \\
14 & 1.10 & 1.03 & 2.43 & 3.12 & 1.30 & 1.50 & 2.85 & 4.03 & 1.18 & 1.21 & 2.59 & 3.54 \\
15 & 1.29 & 1.12 & 2.51 & 3.41 & 1.48 & 1.67 & 2.89 & 4.32 & 1.36 & 1.34 & 2.67 & 3.83 \\
16 & 1.15 & 1.15 & 71.31 & 3.65 & 1.33 & 1.54 & 65.75 & 4.54 & 1.20 & 1.29 & 68.53 & 4.03 \\
17 & 1.19 & 1.20 & 3.55 & 3.92 & 1.37 & 1.65 & 60.49 & 4.62 & 1.25 & 1.38 & 31.12 & 4.21 \\
18 & 1.40 & 72.91 & 7.47 & 41.10 & 1.62 & 64.52 & 11.76 & 4.83 & 1.44 & 68.71 & 9.59 & 21.37 \\
19 & 1.31 & 1.53 & 4.68 & 3.71 & 1.76 & 1.73 & 4.49 & 4.41 & 1.51 & 1.57 & 4.49 & 3.99 \\
\end{longtable}

\begin{table}[!htb]
\centering
\renewcommand{\arraystretch}{0.96}
\caption{Interval scores for ASFR interval forecasts}
\label{tab:interval_score_asfr}
\scriptsize
\resizebox{\textwidth}{!}{%
\begin{tabular}{@{}rrrrrrrcrrrrrr@{}}
\toprule
& \multicolumn{6}{c}{\textit{Panel A: LC}} & & \multicolumn{6}{c}{\textit{Panel B: APC}} \\
\cmidrule(lr){2-7} \cmidrule(lr){9-14}
& \multicolumn{2}{c}{Rolling} & \multicolumn{2}{c}{Expanding} & \multicolumn{2}{c}{Combined} & & \multicolumn{2}{c}{Rolling} & \multicolumn{2}{c}{Expanding} & \multicolumn{2}{c}{Combined} \\
\cmidrule(lr){2-3}\cmidrule(lr){4-5}\cmidrule(lr){6-7}
\cmidrule(lr){9-10}\cmidrule(lr){11-12}\cmidrule(lr){13-14}
$h$ & CAN & JPN & CAN & JPN & CAN & JPN & & CAN & JPN & CAN & JPN & CAN & JPN \\
\midrule
1  & 0.12 & 0.06 & 0.13 & 0.13 & 0.13 & 0.07 & & 0.09  & 0.06   & 0.10 & 0.08   & 0.09 & 0.07   \\
2  & 0.13 & 0.07 & 0.14 & 0.14 & 0.14 & 0.08 & & 0.10  & 0.09   & 0.12 & 0.11   & 0.11 & 0.09   \\
3  & 0.14 & 0.08 & 0.15 & 0.15 & 0.14 & 0.09 & & 0.11  & 0.11   & 0.13 & 0.15   & 0.12 & 0.12   \\
4  & 0.15 & 0.10 & 0.15 & 0.15 & 0.15 & 0.10 & & 0.13  & 0.14   & 0.15 & 0.21   & 0.14 & 0.17   \\
5  & 0.17 & 0.12 & 0.16 & 0.17 & 0.16 & 0.11 & & 0.15  & 0.18   & 0.18 & 0.28   & 0.17 & 0.21   \\
6  & 0.18 & 0.13 & 0.17 & 0.18 & 0.18 & 0.12 & & 0.18  & 0.21   & 0.20 & 0.39   & 0.19 & 0.28   \\
7  & 0.20 & 0.15 & 0.19 & 0.19 & 0.19 & 0.13 & & 0.21  & 0.26   & 0.24 & 0.52   & 0.22 & 0.36   \\
8  & 0.20 & 0.16 & 0.20 & 0.20 & 0.20 & 0.14 & & 0.22  & 0.31   & 0.27 & 0.66   & 0.24 & 0.43   \\
9  & 0.21 & 0.18 & 0.20 & 0.21 & 0.21 & 0.16 & & 0.24  & 0.37   & 0.30 & 0.80   & 0.27 & 0.51   \\
10 & 0.22 & 0.20 & 0.21 & 0.22 & 0.21 & 0.17 & & 0.25  & 0.41   & 0.30 & 0.94   & 0.27 & 0.58   \\
11 & 0.24 & 0.20 & 0.23 & 0.23 & 0.23 & 0.18 & & 0.27  & 0.47   & 0.33 & 1.12   & 0.30 & 0.68   \\
12 & 0.24 & 0.18 & 0.24 & 0.24 & 0.24 & 0.17 & & 0.31  & 0.54   & 0.28 & 1.10   & 0.29 & 0.72   \\
13 & 0.22 & 0.17 & 0.22 & 0.24 & 0.22 & 0.17 & & 0.27  & 0.59   & 0.25 & 1.55   & 0.26 & 0.86   \\
14 & 0.17 & 0.17 & 0.17 & 0.24 & 0.17 & 0.17 & & 0.26  & 0.66   & 0.17 & 1.81   & 0.21 & 0.94   \\
15 & 4.93 & 0.24 & 0.19 & 0.26 & 2.50 & 0.21 & & 0.26  & 0.76   & 0.16 & 1.83   & 0.21 & 0.98   \\
16 & 0.18 & 0.23 & 0.19 & 5.79 & 0.18 & 3.01 & & 0.29  & 0.92   & 0.17 & 1.92   & 0.23 & 1.09   \\
17 & 3.62 & 7.91 & 3.70 & 4.80 & 3.66 & 6.36 & & 0.33  & 1.20   & 0.19 & 2.03   & 0.26 & 1.32   \\
18 & 2.66 & 4.62 & 2.72 & 2.63 & 2.69 & 3.63 & & 0.48  & 1.63   & 0.21 & 2.57   & 0.33 & 1.73   \\
19 & 1.46 & 0.72 & 0.37 & 2.67 & 0.91 & 1.69 & & 10.40 & 151.19 & 5.30 & 160.99 & 7.85 & 156.09 \\
\bottomrule
\end{tabular}%
}
\vspace{0.3em}
\end{table}

\begin{figure}[!htb]
\centering
{\includegraphics[width=0.81\textwidth]{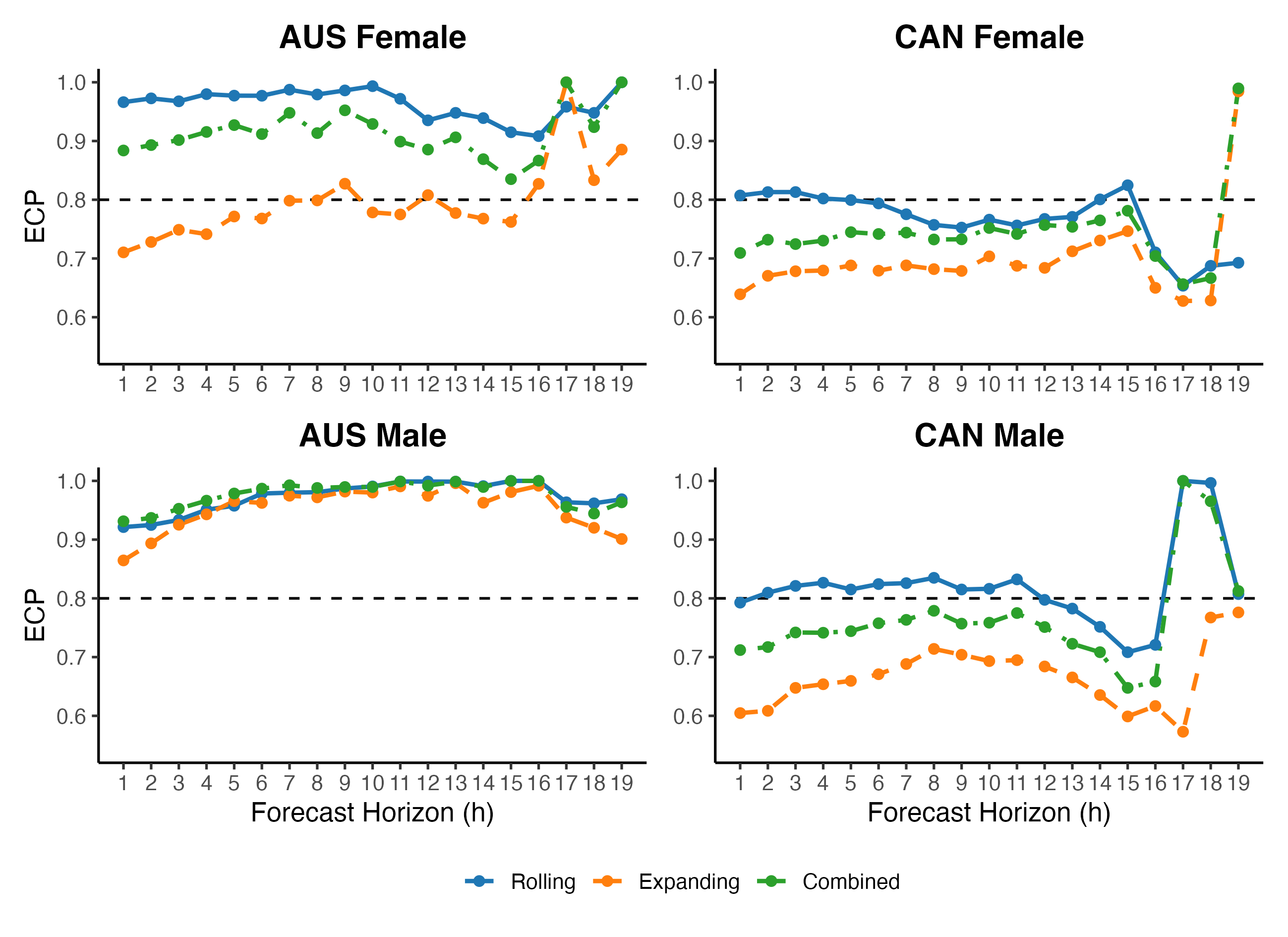}}
\caption{\small{Australia and Canada mortality -- LC interval forecasts: ECP by sex and forecast horizon ($h$).}}
\label{fig:app-mortality-lc-ecp}
\end{figure}

\begin{figure}[!htb]
\centering
{\includegraphics[width=0.81\textwidth]{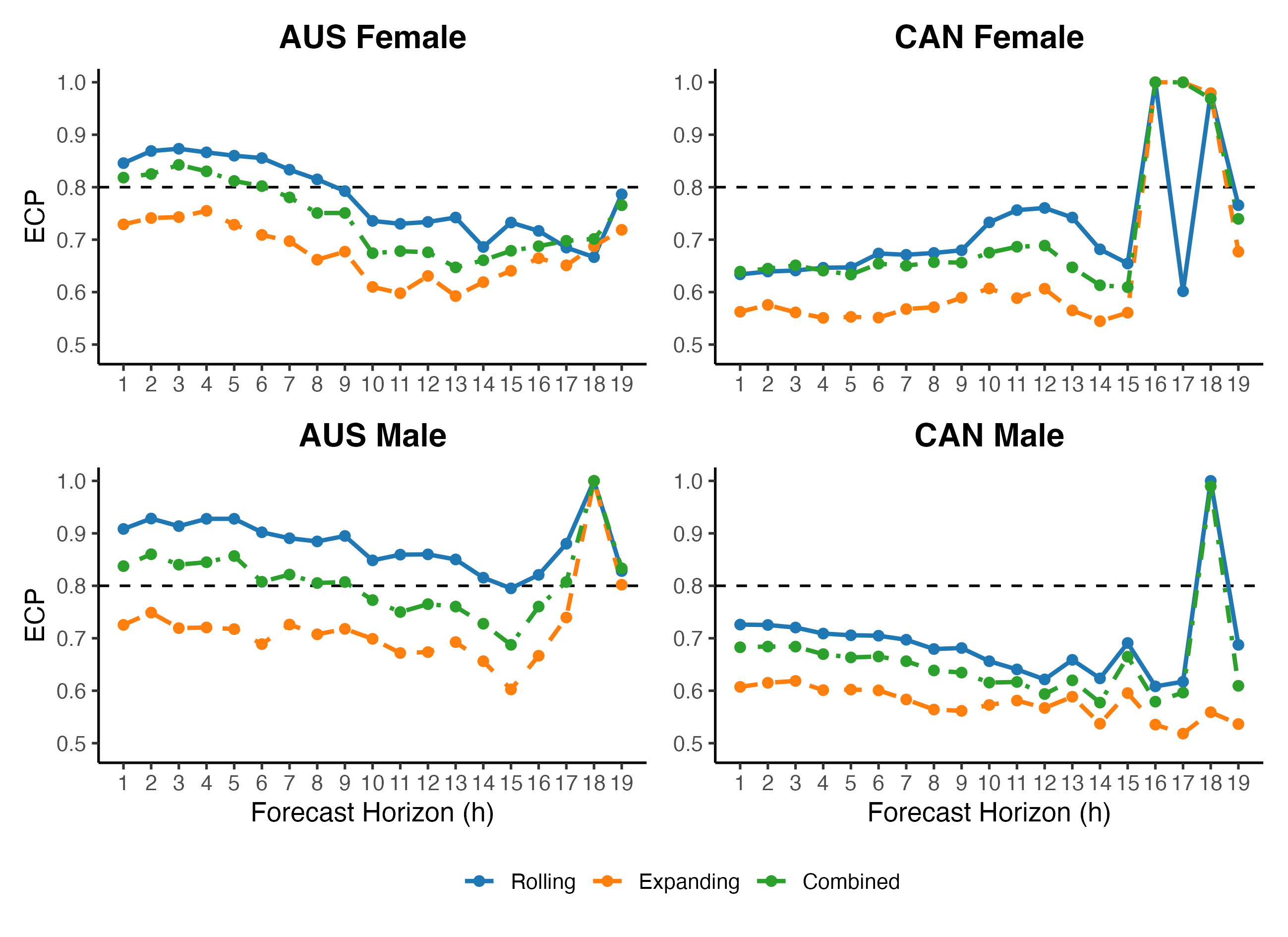}}
\caption{\small{Australia and Canada mortality -- APC interval forecasts: ECP by sex and forecast horizon ($h$).}}
\label{fig:app-mortality-apc-ecp}
\end{figure}

\begin{figure}[!htb]
\centering
{\includegraphics[width=0.77\textwidth]{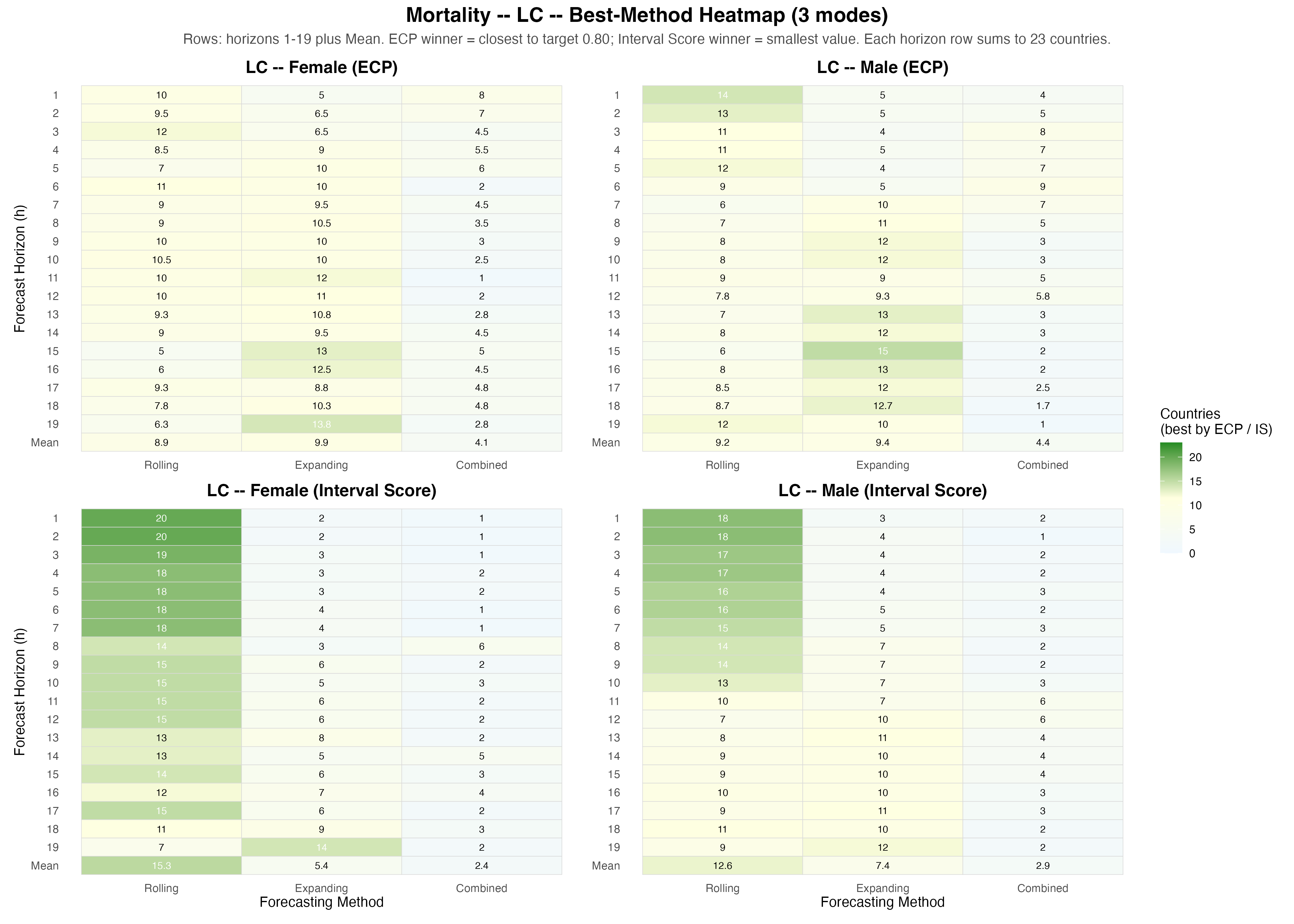}}
\caption{\small{Horizon-specific ranking heatmaps for the female and male mortality under the LC model, as measured by the CPD and score, for the 23 countries considered.}}
\label{fig:app-mortality-lc-interval-heatmap}
\end{figure}

\begin{figure}[!htb]
\centering
{\includegraphics[width=0.77\textwidth]{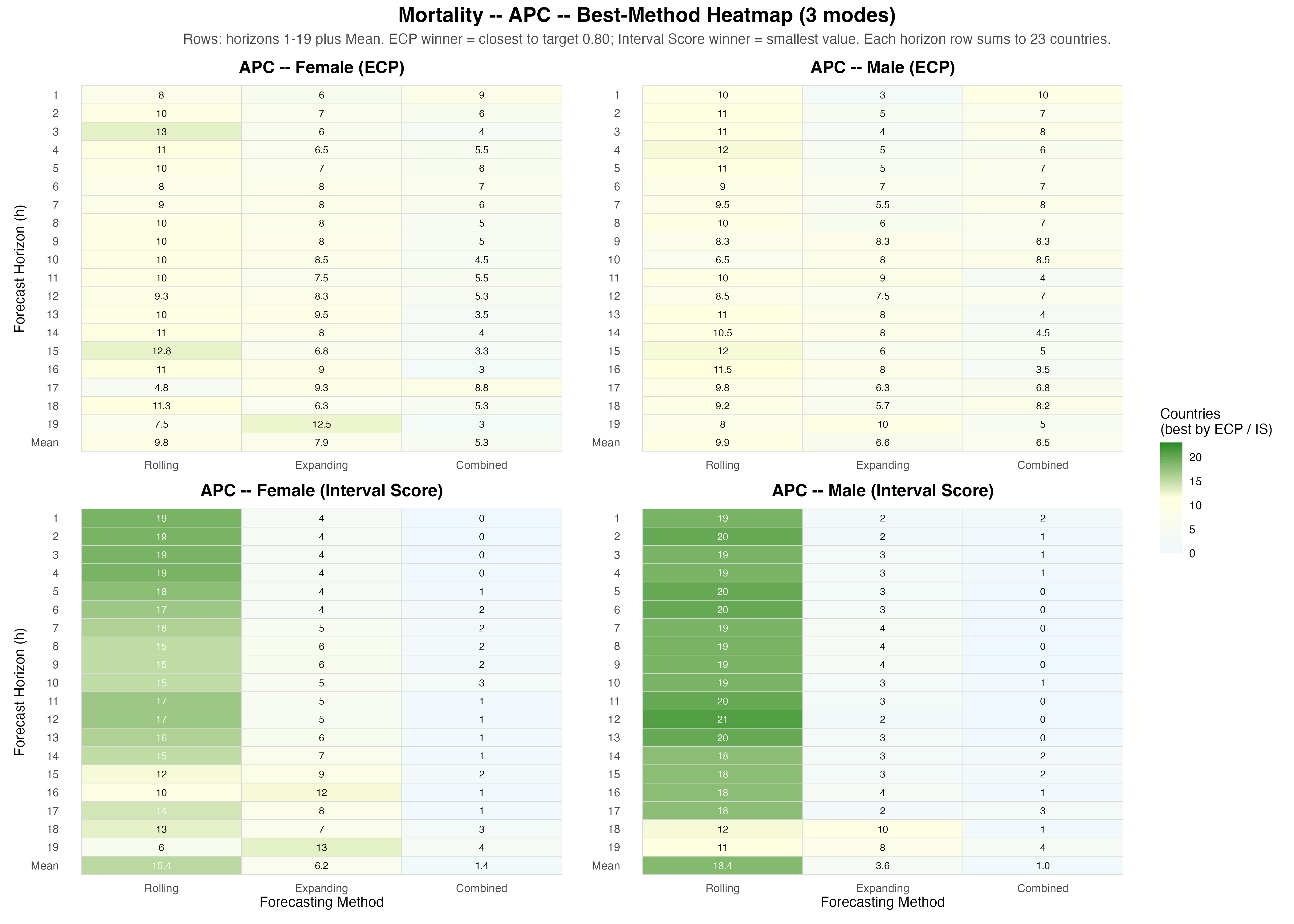}}
\caption{\small{Horizon-specific ranking heatmaps for the female and male mortality under the APC model, as measured by the CPD and score, for the 23 countries considered.}}
\label{fig:app-mortality-apc-interval-heatmap}
\end{figure}

Figures~\ref{fig:app-mortality-lc-ecp}--\ref{fig:app-mortality-apc-interval-heatmap} report the mortality interval-forecast results. For LC mortality, the ECP heatmaps show a less one-sided ranking than the point-forecast results. In the female panel, \texttt{Expanding} has the highest mean ECP win count, while \texttt{Rolling} is very close; in the male panel, the two single-window schemes are again close on average. This indicates that, judged only by closeness to the nominal coverage level, both \texttt{Rolling} and \texttt{Expanding} can be competitive under LC. However, the interval-score panels provide a clearer ranking. \texttt{Rolling} wins on average 15.3 countries for females and 12.6 countries for males, making it the strongest method once both interval width and miscoverage penalties are considered. The Australian and Canadian ECP plots also show that some individual series experience pronounced horizon-end instability, especially for Canadian mortality at longer horizons. The interval scores in Table~\ref{tab:interval_score_mortality} confirm that these late-horizon irregularities can generate large penalties even when the ECP trajectory appears close to or above the nominal level.

For APC mortality, the evidence in favor of \texttt{Rolling} is stronger. In the ECP heatmaps, \texttt{Rolling} has the largest mean win count for both females and males. The advantage is particularly visible in the male panel, where \texttt{Rolling} wins on average 9.9 countries, compared with 6.6 for \texttt{Expanding} and 6.5 for \texttt{Combined}. The interval-score heatmaps are even more concentrated in the rolling column: \texttt{Rolling} wins on average 15.4 countries for females and 18.4 countries for males. The country-level APC ECP plots show that Australia is generally better calibrated than Canada at shorter horizons, while Canada exhibits stronger late-horizon fluctuations, including sharp jumps around the final horizons. The interval-score tables show the same pattern: most short- and medium-horizon scores are moderate, but a small number of long-horizon cases produce very large penalties. These outlying scores are important because they show why ECP alone is insufficient for evaluating interval forecasts: a method can achieve high empirical coverage by producing overly wide or unstable intervals, which the interval score penalizes.

\begin{figure}[!htb]
\centering
{\includegraphics[width=0.87\textwidth]{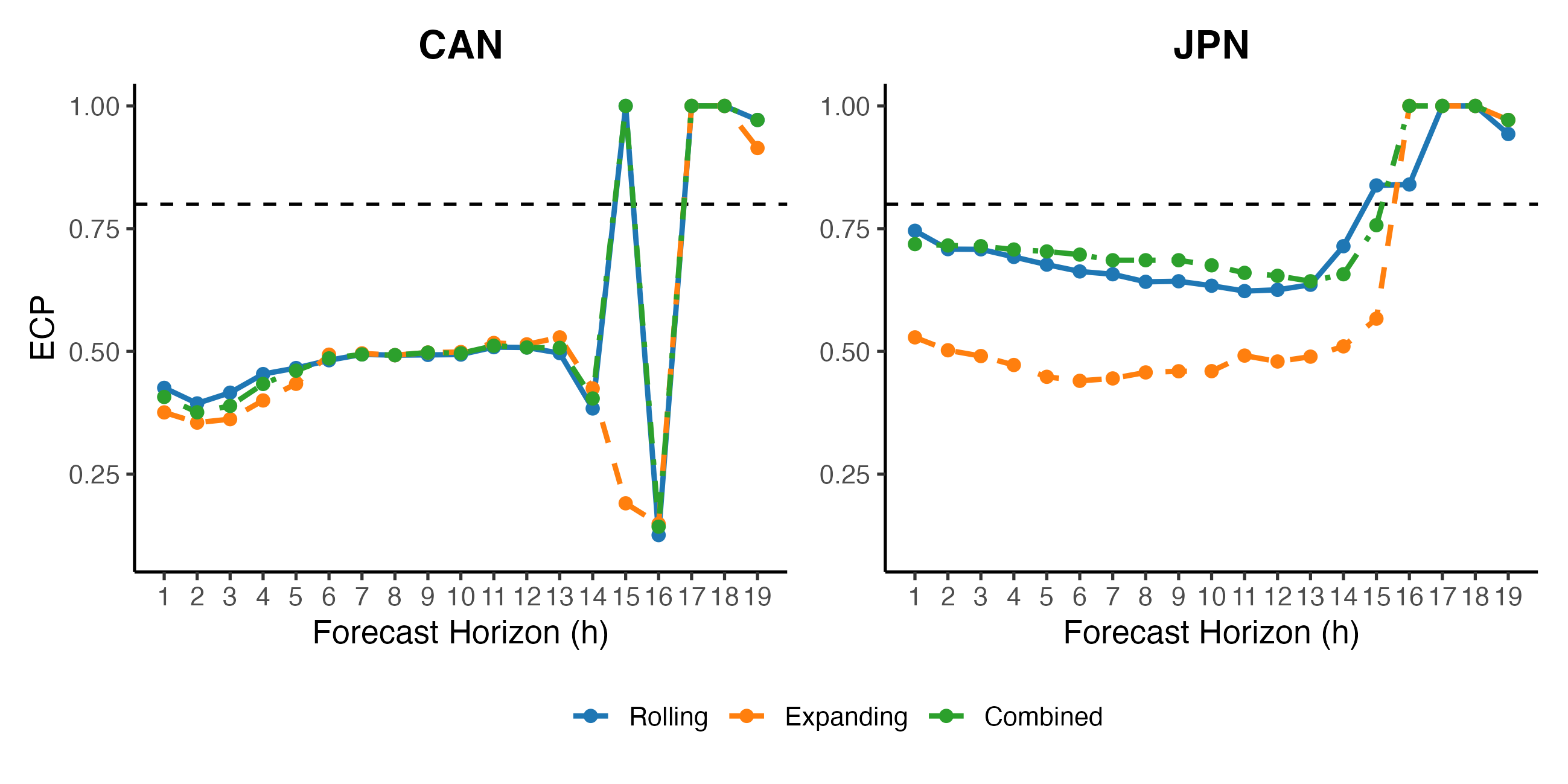}}
\caption{\small{Canada and Japan ASFR -- LC interval forecasts: ECP by forecast horizon ($h$).}}
\label{fig:app-asfr-lc-ecp}
\end{figure}

\begin{figure}[!htb]
\centering
{\includegraphics[width=0.87\textwidth]{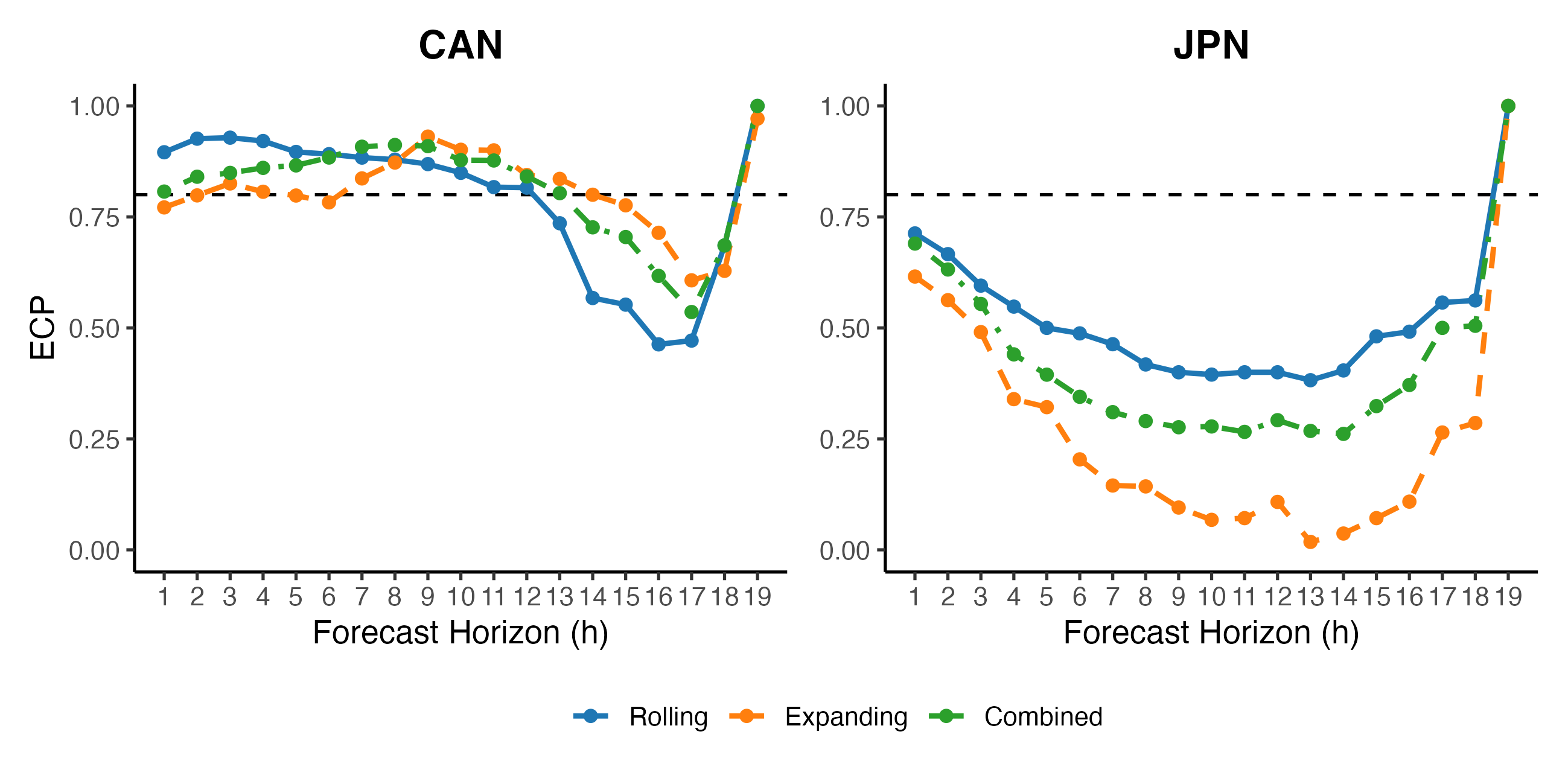}}
\caption{\small{Canada and Japan ASFR -- APC interval forecasts: ECP by forecast horizon ($h$).}}
\label{fig:app-asfr-apc-ecp}
\end{figure}

\begin{figure}[!htb]
\centering
{\includegraphics[width=0.87\textwidth]{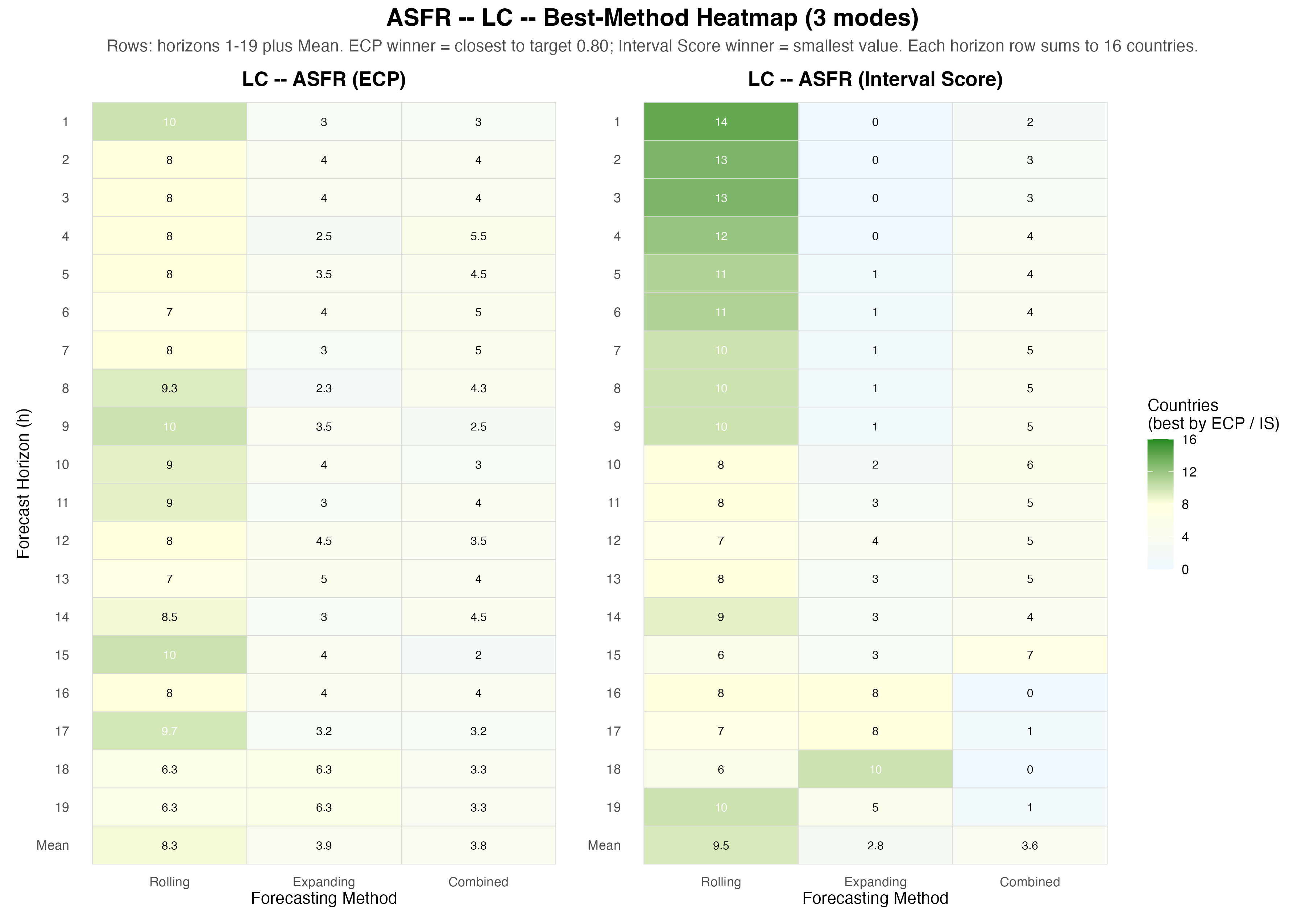}}
\caption{\small{Horizon-specific ranking heatmaps for the ASFR under the LC model, as measured by the CPD and score, for the 16 countries considered.}}
\label{fig:app-asfr-lc-interval-heatmap}
\end{figure}

\begin{figure}[!htb]
\centering
{\includegraphics[width=\textwidth]{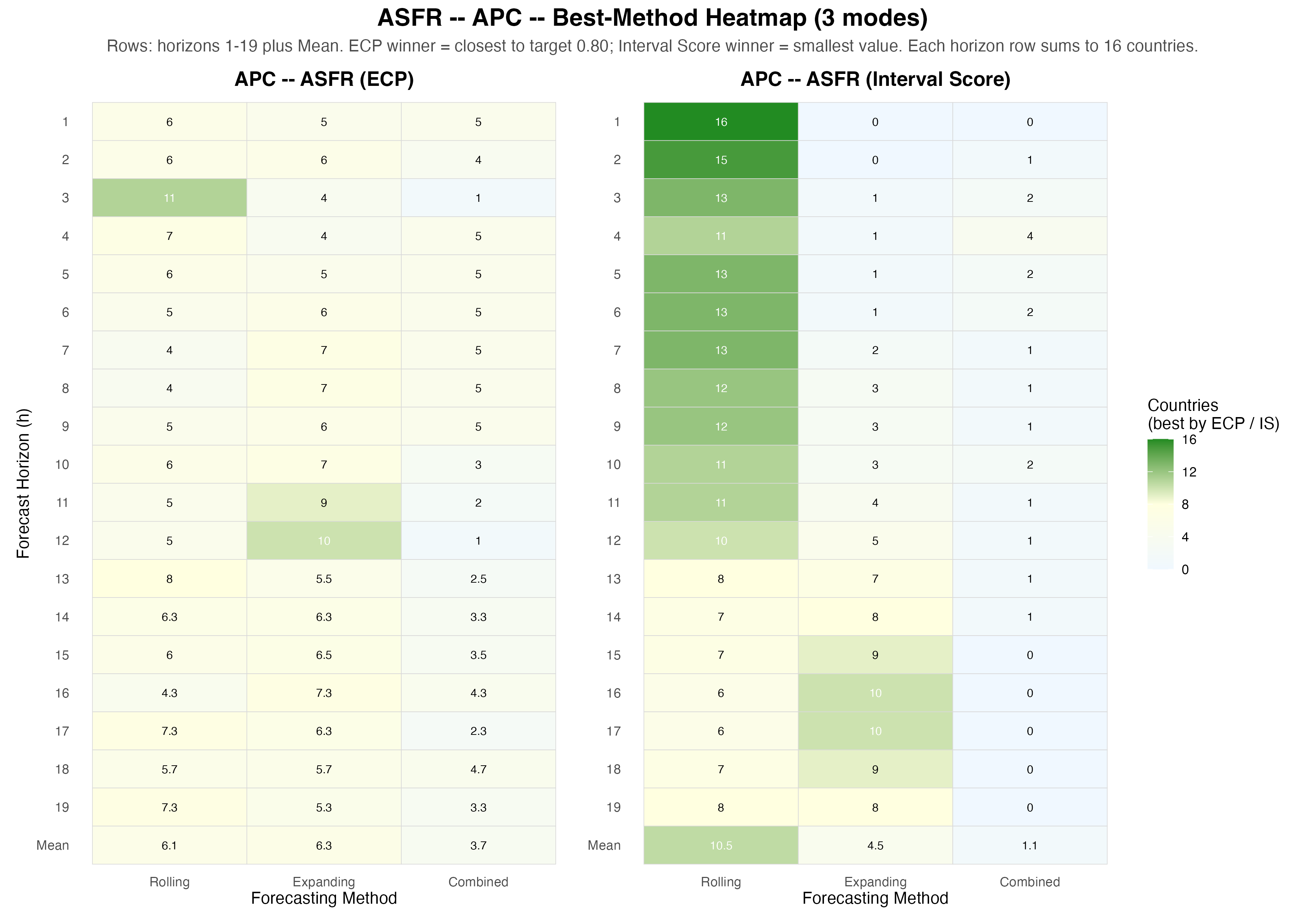}}
\caption{\small{Horizon-specific ranking heatmaps for the ASFR under the APC model, as measured by the CPD and score, for the 16 countries considered.}}
\label{fig:app-asfr-apc-interval-heatmap}
\end{figure}

Figures~\ref{fig:app-asfr-lc-ecp}--\ref{fig:app-asfr-apc-interval-heatmap} summarize the ASFR interval-forecast results under the LC and APC benchmarks. Under the LC specification, the heatmaps show a clear advantage for \texttt{Rolling}. In the ECP panel, \texttt{Rolling} wins on average 8.3 out of 16 countries, compared with 3.9 for \texttt{Expanding} and 3.8 for \texttt{Combined}. In the interval-score panel, \texttt{Rolling} again has the largest mean win count, at 9.5 countries. The country-level ECP plots show that the LC intervals are not uniformly calibrated: Canada exhibits substantial under-coverage over many horizons, while Japan remains below the nominal level for most horizons before improving at the end of the horizon range. Nevertheless, the cross-country heatmap indicates that, relative to the competing schemes, the rolling window most often delivers the closest coverage and the most favorable interval-score performance under LC.

The APC-ASFR results are more horizon-dependent. In the ECP heatmap, \texttt{Rolling} and \texttt{Expanding} are close on average, with mean win counts of 6.1 and 6.3 countries, respectively. This suggests that, for APC-based ASFR intervals, calibration alone does not identify a single dominant fitting scheme. However, the interval-score heatmap is more decisive: \texttt{Rolling} wins on average 10.5 countries, compared with 4.5 for \texttt{Expanding} and 1.1 for \texttt{Combined}. The Canada and Japan ECP panels explain this pattern. For Canada, the three APC schemes are reasonably close to the nominal level over short and medium horizons, although all methods become unstable near the longest horizons. For Japan, \texttt{Expanding} displays persistent under-coverage over most horizons, while \texttt{Rolling} stays closer to the target but still falls below nominal coverage until the final horizons. The interval-score values in Table~\ref{tab:interval_score_asfr} further show that APC-ASFR intervals can incur large end-horizon penalties, especially for Japan at $h=19$. Thus, although \texttt{Expanding} can be competitive for ECP under APC, \texttt{Rolling} is more reliable once sharpness and miscoverage penalties are considered jointly.

Overall, the LC/APC interval sensitivity analysis leads to three main conclusions. First, the ranking based on ECP is more variable than the ranking based on interval score, because ECP measures calibration only and does not account for interval width. Second, when calibration and sharpness are evaluated jointly through the interval score, \texttt{Rolling} is the most frequent winner for both LC and APC, especially for mortality. Third, the equal-weight \texttt{Combined} method remains useful as a stabilizing hedge in some country-level trajectories, but it is less often the direct winner in the LC/APC interval experiments than in the main functional time-series analysis. These findings are consistent with the point-forecast sensitivity results: under the more restrictive LC and APC model structures, recent-information weighting often improves forecast robustness, while expanding windows can become sensitive to older demographic regimes and long-horizon cohort effects.

\section*{Disclosure Statement}

The authors declare no conflict of interest.

\section*{Data Availability Statement}

The data used in this article are freely available upon registration in the Human Mortality Database \url{https://www.mortality.org/} and the Human Fertility Database \url{https://www.humanfertility.org/}.

\newpage
\bibliographystyle{apalike3}
\bibliography{averaging.bib}

\end{document}